\documentclass[journal,transmag]{IEEEtran}
\usepackage{graphicx}
\usepackage{multirow}
\usepackage{subfigure}
\usepackage{longtable}
\usepackage{diagbox}
\usepackage{amsfonts,amssymb}
\usepackage{array}
\usepackage{amsmath}
\usepackage{lineno,hyperref}

\newcolumntype{L}[1]{>{\raggedright\let\newline\\\arraybackslash\hspace{0pt}}m{#1}}
\newcolumntype{C}[1]{>{\centering\let\newline\\\arraybackslash\hspace{0pt}}m{#1}}
\newcolumntype{R}[1]{>{\raggedleft\let\newline\\\arraybackslash\hspace{0pt}}m{#1}}

\begin{document}

\title{Steganalysis of 3D Objects Using Statistics of Local Feature Sets}

\author{\IEEEauthorblockN{Zhenyu Li and Adrian G. Bors}

\IEEEauthorblockA{Department of Computer Science, University of York, York, UK YO10 5GH\\
Email: \{zl991, adrian.bors\}@york.ac.uk}
}

\maketitle

\begin{abstract}
3D steganalysis aims to identify subtle invisible changes produced in graphical objects through digital watermarking or
steganography.
Sets of statistical representations of 3D features, extracted from both cover and stego 3D mesh objects,
are used as inputs into machine learning classifiers in order to decide whether any information was hidden
in the given graphical object. According to previous studies, sets of local geometry features can be used to define the differences between stego and cover-objects. The features proposed in this paper include those representing the local object
curvature, vertex normals, the local geometry representation in the spherical coordinate system and are considered in various
combinations with others. We also analyze the effectiveness of various 3D feature sets applied for steganalysis based on the Pearson correlation coefficient. The classifiers proposed in this study for discriminating the 3D stego and cover-objects include
Support Vector Machine and the Fisher Linear Discriminant ensemble. Three different watermarking
and steganographic methods are used for hiding information in the 3D objects used for testing the performance of the proposed steganalysis methodology.
\end{abstract}


\IEEEpeerreviewmaketitle

\section{Introduction}
Computer graphics are becoming a more and more important medium in the interactive techniques such as Virtual Reality (VR) and 3D printing. In order to protect the copyright of the 3D objects, the digital watermarks maybe embedded into the 3D objects. 3D watermarking and steganographic algorithms seek to hide information in 3D objects, which can then
be used for a variety of applications. Information is hidden in 3D objects for various purposes: copyright protection, defining behaviour characteristics similar to the DNA in beings, or hiding certain information for marketing purpose, etc. Moreover, the 3D objects can be the carriers of a covert communication channel when the 3D steganography is applied. While watermarking seeks to robustly embed rather smaller codes, steganography would
embed larger payload messages without enforcing robustness. All these approaches aim to hide information in such a way
that the changes they produce to the 3D objects are not visible. On the other hand steganalysis algorithms are being developed in order to find
whether information is embedded in a certain media data. Several steganalysis algorithms have been proposed for
audio signals \cite{ren2015amr,yan2013steganalysis,liu2009temporal}, digital images
\cite{qiao2015steganalysis,lu2014recognizing,li2013embedding,cogranne2011cover,gul2010svd}
and video \cite{wang2014video,cao2012video,budhia2006digital}. While 3D objects can be represented in various ways,
their most usual data representation is by means of meshes. Such irregular representations modelling complex
3D shapes are very different from the regular structural array representing audio, digital images or video.
Consequently, the existing image and video steganalysis algorithms cannot be successfully applied to 3D objects.

Most research on 3D watermarking involves modifying certain geometrical properties of the object, most often in
a statistical manner, in such a way that there are no visible changes. Research on 3D watermarking
started in 1997 when Ohbuchi {\em et al.} proposed two 3D information hiding algorithms using ratios of local
geometric measurements \cite{ohbuchi1997embedding}. Cho {\em et al.} \cite{cho2007oblivious} proposed
two blind robust watermarking algorithms based on modifying
the mean and variance of the distribution of the vertices' radial distances in the spherical coordinate system. Cayre and Macq proposed a steganographic approach for 3D triangle meshes in \cite{cayre2003data} whose
key idea is to consider each triangle from the mesh as a two-state geometrical object embedding a bit. Luo and Bors proposed changing the statistics of geodesic distance distributions in 3D objects \cite{luo2011surface}. The same authors proposed
minimizing the surface distortions in 3D watermarking by using an optimization algorithm in \cite{bors2013optimized}.
Among the information hiding algorithms, we mention a multi-layer 3D steganographic method \cite{chao2009high} which embeds large payloads using vertices' projections onto the principal axis of
the object. This steganographic method can make use of several layers for embedding the information increasing thus significantly the embedded payload. However, not all bits embedded through this method are retrievable and some are lost.
More recently, Yang {\em et al.} \cite{yang2016watermarking} proposed a steganalysis-resistant watermarking algorithm which embeds the payload by changing the
histogram of the radial coordinates of the vertices. This watermarking method produces less embedding distortion in the 3D objects when compared to the methods
proposed in \cite{cho2007oblivious}. A distortion-free steganographic algorithm \cite{bogomjakov2008distortion} embeds the information into the meshes by permuting the order in which faces and vertices are stored. However, this algorithm is not robust to the vertex-reordering attack.

3D steganalysis only received very recently the attention of the scientific community.
The 3D steganalysis approach proposed in \cite{yang2014mesh} considered various features including the norms of
vertices in the Cartesian and Laplacian coordinate systems \cite{yang2010polygonal}, the dihedral angle of
faces adjacent to the same edge, and the face normal. Parameters representing the statistics of these features were
used as inputs to a quadratic classifier. Yang {\em et al.} \cite{yang2014steganalytic}, proposed a new steganalysis algorithm, specifically designed for the mean-based watermarking algorithm from \cite{cho2007oblivious}.
This steganalysis algorithm first estimates the number of bins through exhaustive search and then
detects the presence of the secret message by a tailor-made normality test. A steganalytic approaches which is designed specially for addressing the cover source mismatch scenario in 3D steganalysis by selecting the features which are robust to the variations of the cover source was proposed in \cite{li2016RRFS}. Cryptanalysis aspects of 3D watermarking
in a larger context have been discussed in the review paper from \cite{Iter2014}.

The aims of a steganalyzer are difficult to achieve because the stego and cover-obejcts are supposed to be almost identical
under human visual observation assumptions. Moreover, the steganalyzer should be able to find such subtle changes, when
any hiding algorithm may have been used in a large diversity of 3D shapes. The ability of a steganalyzer to generalize
from the training set to a large testing set is a very demanding requirement as well. The steganalyzer proposed in this study has the following
processing stages: extracting and combining multiple 3D features, statistical modelling of the features and the classifier. We propose some new features to be used for 3D steganalysis such as the
local curvatures, normal vectors calculated at vertex locations, as well as using the vertex representation in the spherical coordinate system
for 3D steganalytic feature representation. The statistics of sets of 3D features are then fed into machine learning algorithms.
Fisher Linear Discriminant (FLD) ensemble \cite{kodovsky2012ensemble} is the most used classifier for image steganalysis because of its powerful ability to find quickly the non-linear separation boundaries between features characterizing the
cover- and stego-images. In this research study we propose to use FLD ensemble as well as the Support Vector Machine
(SVM) in 3D steganalysis. The discriminating ability for 3D steganalysis of these classifiers is then compared against
the quadratic classifier proposed in \cite{yang2014mesh}. The description of the 3D steganalysis framework formulated
in this study is provided in Section~\ref{Framew}. The 3D feature set, used by the steganalyzer
is presented in detail in Section~\ref{Features} and the algorithms used for training the steganalyzers
in Section~\ref{Learning}. The experimental results are provided in Section~\ref{Exper}, while the conclusions
of this study are outlined in Section~\ref{Con}.

\section{3D Steganalysis Framework}
\label{Framew}

\begin{figure}[!htbp]
  \centering
  \includegraphics[width=8.5cm]{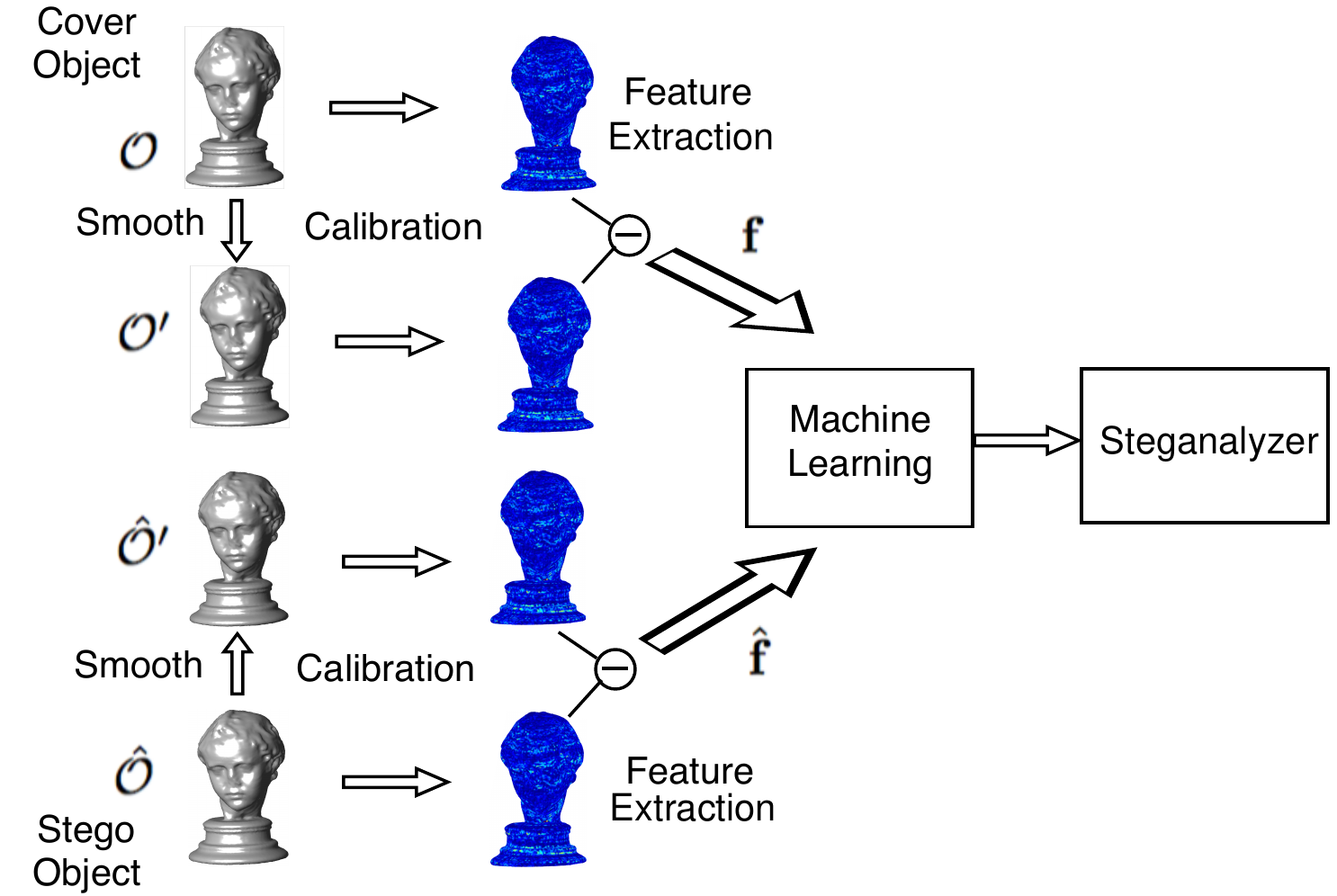}
  \caption{The 3D steganalysis framework based on learning from statistics of the local feature sets and classification by means
  of machine learning methods.}
  \label{framework}
\end{figure}

In this section, we will give a brief introduction of the 3D steganalysis framework. Let us assume that we have a mesh
representing the shape of a 3D object, considered as a cover mesh $\cal{O}=\{ \emph{V}, \emph{F}, \emph{E}\}$
and its corresponding watermarked stego mesh $\hat{\cal{O}}=\{  \hat{\emph{V}}, \hat{\emph{F}}, \hat{\emph{E}} \}$, containing
the vertex sets $\emph{V}=\{{\emph{v}}(i) | i = 1,2, \ldots, |V| \}$ and $\hat{\emph{V}}=\{ \hat{\emph{v}}(i) | i = 1,2, \ldots,  | \hat{V} | \}$,
where $| V |$ and $| \hat{V} |$ represent the number of vertices in the cover-object $\cal{O}$ and stego-object $\hat{\cal{O}}$, their face sets $\emph{F}$, $ \hat{\emph{F}}$, and their edge sets $\emph{E}$ and $\hat{\emph{E}}$, respectively.
The steganalysis framework is treated as a machine learning problem, consisting of training and testing stages.
The training of the steganalyzer has the following processing steps: calibration, feature extraction and learning, as illustrated
in  Figure~\ref{framework}. These processing steps produce a parameter set discriminating between the 3D objects carrying
hidden information and those that are not. The testing stage includes the same calibration and feature extraction steps as in
the training stage, while applying the parameters learnt during the training on the features extracted from sets of test objects.

Firstly, a series of preprocessing steps are used in order to ensure that the cover-object $\cal{O}$ and stego-object $\hat{\cal{O}}$
are normalized such that their size is constrained within a cube with sides of one. In the case of image steganalyzers, it was
observed that the difference between the stego-image and its smoothed version is more significant than
the difference between the cover-image and its corresponding smoothed version \cite{Frid2002,Kod2009}.
Similarly, it is expected that the difference between a mesh and its smoothed version is larger for a stego mesh than for a cover mesh.
In most 3D watermarking algorithms, the changes produced to the stego-object, following the watermark embedding can be associated
to noise. Consequently, when smoothing a cover mesh, the resulting modifications will be smaller than those obtained when smoothing
its corresponding stego mesh. We consider Laplacian smoothing, for both the cover-object $\cal{O}$ and the stego-object $\hat{\cal{O}}$,
resulting in their smoothed versions $\cal{O}'$ and stego-object $\hat{\cal{O}}'$.

Features, characterizing the local geometry of 3D objects are extracted before and after smoothing the stego-objects and
cover-objects, respectively. The discriminative features should be chosen such that they capture effectively the differences between the two
versions of mesh for a given object. In Section \ref{Features} we propose to use a set of new 3D features for steganalysis.
Then, statistics of differences between the features extracted from the cover-object $\cal{O}$ and the smoothed cover-object $\cal{O}'$,
are compared with those of differences between the features extracted from the stego-object $\hat{\cal{O}}$ and its smoothed
version $\hat{\cal{O}'}$, respectively. In order to model the differences, we consider the first four statistical moments, representing the mean,
variance, skewness and kurtosis, as in \cite{yang2014mesh}.  The statistics of the various combinations of 3D features are eventually used as inputs to machine learning algorithms.
The classifier separates the feature space defining the stego-objects from
that of the cover-objects. As in the case of supervised computational intelligence algorithms, we have a training stage, where a set of parameters
is estimated, and then a testing stage, where the classifier, using the learnt set of parameters is applied
on a different data set. Firstly, specific feature vectors are estimated from sets of cover- and stego-meshes, corresponding to the same set of 3D
objects.  It has been shown that breaking the cover-stego pairs correspondence may lead to a suboptimal performance \cite{schwamberger2010simple}.
The features should represent properties that would discriminate the cover-objects from their stego-objects counterparts.
Furthermore, choosing the appropriate machine learning algorithm and its training procedure are crucial, as steganalyzers trained by different machine
learning methods can provide different results on identical training sets.  In this research study we propose to use the FLD ensemble \cite{kodovsky2012ensemble}
and Support Vector Machine (SVM) methods for training the steganalyzer.

\section{Features for 3D Steganalysis}
\label{Features}

3D watermarking and steganographic methods are specifically designed to embed information in a way
that does not visibly alter the surface of the objects \cite{luo2011surface,bors2013optimized}. Nevertheless,
the changes produced in the 3D objects surface may be identified by steganalysis.
Depending on the specific algorithm used, such changes could be randomly distributed on the surface
of the 3D mesh \cite{bors2006} or they could be located specifically in certain regions of the object \cite{Alf2007}. Artefacts produced in objects, following the information hiding procedure,
could be assimilated to low level protuberances on mesh surfaces and consequently could be identified by feature detection algorithms.
In the following we outline some 3D local features which can be used for identifying whether objects
have been watermarked or not. Such feature detectors range from very simple vertex displacement
measurements to algorithms that take into account the local neighbourhoods and measure specific
shape characteristics.

\subsection{The YANG40 Features}
\label{Yang40}

The 40-dimensional feature vector YANG40 contains the most effective features from YANG208,
used in \cite{yang2014mesh}, which correspond to the statistics of features evaluated from
the vertices, edges and faces that make up the given meshes. For YANG40 we remove certain features, which provide lower performance, from YANG208 and abandon the strategy used in \cite{yang2014mesh} which treats the vertices with valence less, equal, or greater than six separately to reduce the dimensionality.

Let us denote by ${\bf \Phi}$, the feature vector
representing differences between the cover-object ${\cal O}$ and its smoothed version ${\cal O}'$.
Similarly, we evaluate $\hat{\bf \Phi}$, measuring the differences between the stego-object $\hat{\cal O}$ and
its smoothed version $\hat{\cal O}'$.
The first six components of ${\bf \Phi}$ represent the absolute distance, measured along each coordinate axis $x, y, z$
between the locations of vertices of the meshes $\cal{O}$ and $\cal{O}'$, in both the Cartesian and Laplacian coordinate
systems \cite{yang2010polygonal}:
\begin{equation}\label{v-position}
\begin{split}
 & \phi_1(i)= |{v_{x,c}}(i) - {{v'}_{x,c}}(i)|,\\
 & \phi_2(i)= |{v_{y,c}}(i) - {{v'}_{y,c}}(i)|,\\
 & \phi_3(i)= |{v_{z,c}}(i) - {{v'}_{z,c}}(i)|,
\end{split}
\end{equation}
\begin{equation}\label{v-lap}
\begin{split}
 &\phi_4 (i)= |{v_{x,l}}(i) - {{v'}_{x,l}}(i)|,\\
 &\phi_5 (i)= |{v_{y,l}}(i) - {{v'}_{y,l}}(i)|,\\
 &\phi_6 (i)= |{v_{z,l}}(i) - {{v'}_{z,l}}(i)|,
\end{split}
\end{equation}
where $v_{x,c}(i)$ and $v_{x,l}(i)$ represent the $x$-coordinate of $\emph{v}(i)$ in Cartesian and Laplacian coordinate systems, respectively, $i = 1,2, \ldots, |V|$.
Next, we evaluate the changes produced in the Euclidean distance between vertex locations and the center of the object, representing the
vertex norms. The absolute differences between the vertex norms of pairs of corresponding vertices in the meshes ${\cal O}$ and
 ${\cal O}'$ are calculated as:
 \begin{equation}
 \phi_7 (i) = |\| {\bf v}_c (i) \| -  \| {\bf v}'_c (i) \||
 \end{equation}
 \begin{equation}
 \phi_8 (i) = |\| {\bf v}_l (i) \| -  \| {\bf v}'_l (i) \||
 \label{v-cnorm}
 \end{equation}
 where $\| {\bf v}_c (i) \| $, $\| {\bf v}_l (i) \| $, represent the vector norms in Cartesian and Laplacian coordinates, respectively, for $i = 1,2, \ldots,  |V|$.

\begin{figure}[!htbp]
  \centering
  \includegraphics[width=5cm]{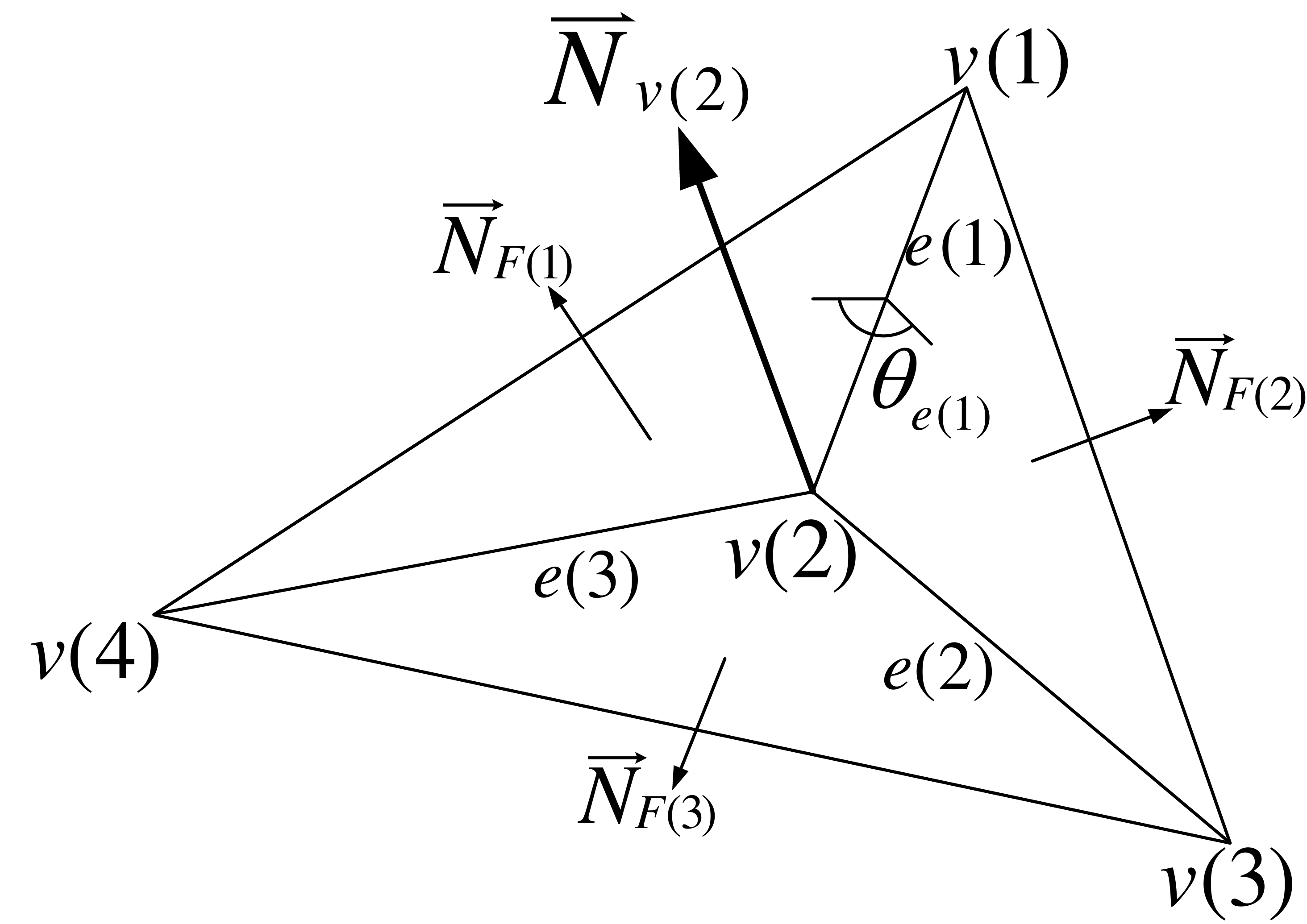}
  \caption{Dihedral angles and vertex-based normals for representing local geometry properties of the surface.}
  \label{normal}
\end{figure}

Another feature evaluates the local mesh surface variation by calculating the changes in the orientations of faces adjacent to the same edge.
This is measured by the absolute differences between the dihedral angles of neighbouring faces, calculated in the plane perpendicular on the
common edge $\{ e(i) \in {\cal O} | i = 1,2, \ldots, | E | \}$, where $|E |$ represents the number of edges part of the object ${\cal O}$ :
\begin{equation}
\phi_9(i)= | \theta_{e(i)} - \theta_{e(i)}' |,
\label{dihedral}
\end{equation}
where the calculation of the dihedral angle $\theta_{e(i)}$ is illustrated in Figure~\ref{normal}.
Changes in the local surface orientation are measured by calculating the angle between the surface normals ${\vec N}_{F(i)}$, of the faces from
the cover-object $F(i) \in {\cal O}$, and their correspondents  ${\vec N}_{F'(i)}$, from the smoothed cover-object $F'(i) \in {\cal O}'$:
\begin{equation}
\phi_{10}(i) =  {\rm{arccos}}\frac{{{{\vec N}_{F(i)}} \cdot {{\vec N}_{F'(i)}}}}{{\left\| {{{\vec N}_{F(i)}}} \right\| \cdot \left\| {{{\vec N}_{F'(i)}}} \right\|}}
\label{facenormal}
\end{equation}
where $i = 1,2, \ldots,  |F|$. The 40-dimensional feature vector YANG40 represents the first four statistical moments:  mean, variance, skewness and
kurtosis of the logarithm of the ten vectors $\{\phi_{i} | i=1,2, \ldots, 10\}$, described above.

\subsection{The Vertex Normal and Curvature Features}

In the following we propose to use some additional 3D features. The vertex normal is the weighted sum of the normals of all
faces that contain the vertex \cite{max1999weights}. A vertex normal is shown in Figure~\ref{normal} and is computed as:
\begin{equation}
{\vec N}_{v(i)} = \sum\limits_{F(j)} \frac{A(F(j)) \cdot {\vec N}_{F(j)}}{ \| e_{(v(i),1)} \|^2 \cdot  { \| e_{(v(i),2)} \|^2 }}
\label{vertexnormal}
\end{equation}
where $F(j)$ represents the $j$-th face that contains the vertex $ v(i)$, $A(F(j))$ represents its area, $e_{(v(i),1)}$ and $e_{(v(i),2)}$
are the two edges containing $ v(i)$ in the face $F(j)$. The change between the vertex normals is calculated as a dot product:
\begin{equation}
  \phi_{11} (i) =  {\rm{arccos}} \frac{{\vec N}_{v(i)} \cdot {\vec N}_{v'(i)}}{ \left\| {{\vec N}_{v(i)}} \right\| \cdot \left\| {{\vec N}_{v'(i)}} \right\|}
  \label{vnf}
\end{equation}
where ${\vec N}_{v'(i)}$ is the normal for a vertex from the smoothed object $ \{v'(i) \in {\cal O}' | i = 1,2, \ldots, |V| \}$ .

Next we consider the local shape curvatures, calculated according to the the Gaussian curvature and the curvature ratio formula
used in \cite{rugis2006scale}.
In differential geometry, the two principal curvatures of a surface are provided by the eigenvalues of the shape
operator, calculated at the location of a vertex using the vertices from its first neighbourhood. Such curvatures measure how the
local surface bends by different amounts in orthogonal directions at that point. The Gaussian curvature is defined as:
\begin{equation}
  K_G =K _1K _2,
  \label{gaussiancurveture}
\end{equation}
where $K_1$ is the minimum principal curvature and $K_2$ is the maximum principal curvature at a given point \cite{rusinkiewicz2004estimating}.
A special case is that of singularity in the shape operator, when we have a linear dependency in one direction or in both.
In this case we have locally a planar region, which is characterized by a linear relationship among its coordinates and consequently by zero curvature.
In our study we found that the curvature ratio proposed in \cite{rugis2006scale}, defined as
\begin{equation}
   K_r = \frac{{\min (|{K _1}|,|{K _2}|)}}{{\max (|{K _1}|,|{K _2}|)}},
\label{curvatureratio}
\end{equation}
is effective to be used as a feature when training steganalyzers. The Gaussian curvature from equation (\ref{gaussiancurveture}) and the curvature ratio from
(\ref{curvatureratio}) have been shown to be sensitive to very small mesh modifications and have been used to model 3D shape characteristics in various applications.
The two principal curvatures are evaluated at the location of each vertex in the cover-object ${\bf v}(i) \in {\cal O}$ and for its corresponding vertex from
the smoothed object ${\bf v}'(i) \in {\cal O}'$. Their absolute differences represent the features $\phi_{12}$ and $\phi_{13}$:
 \begin{equation}
\phi_{12} (i)= | K_G (\emph{v}(i)) - K_G (\emph{v}'(i)) |,
\label{fea12}
\end{equation}
\begin{equation}
\phi_{13}(i) = | K_r (\emph{v}(i)) - K_r (\emph{v}'(i)) |,
\label{fea13}
\end{equation}
for $i = 1,2, \ldots, |V|$.

\subsection{The Spherical Coordinate Features}

There are many information hiding algorithms would embed changes directly or indirectly in the spherical domain. Consequently,
in the following we consider the spherical coordinate system for defining characteristics that can be used by the steganalyzers.
We convert the 3D objects from the Cartesian coordinate system to the spherical coordinate system, considering the center
of the object as its reference.

\begin{figure}
  \centering
  \includegraphics[width=4cm]{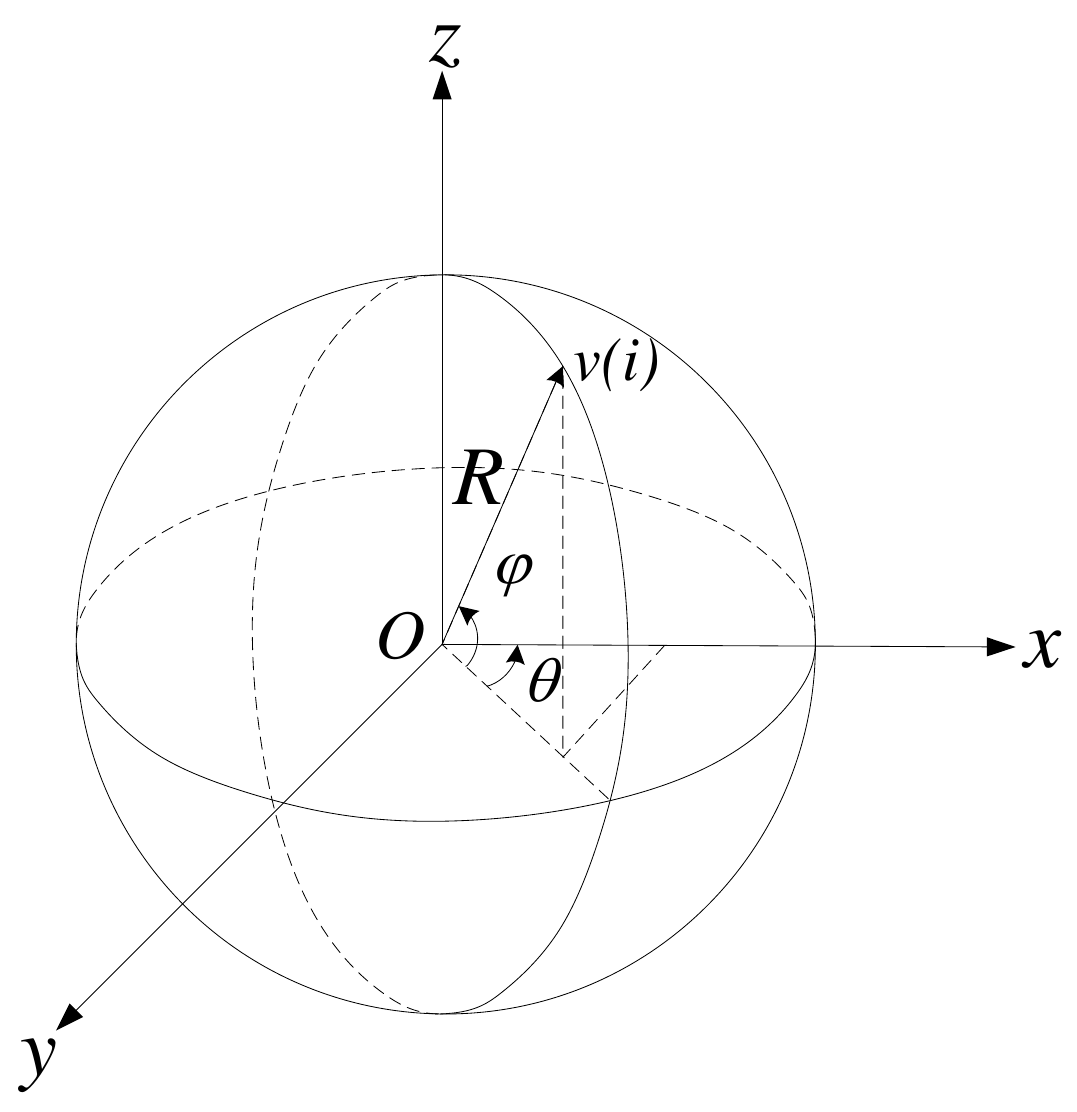}\\
  \caption{The spherical coordinate system, where $R$ is the radial distance of vertex $v(i)$, $\theta$ and $\varphi$ are its azimuth angle and elevation angle, respectively.}
  \label{scf}
\end{figure}

The spherical coordinate system specifies a point in the 3D space by a radius and two angles and the link to the Cartesian coordinate
system is given by:
\begin{equation}
\begin{split}
    &  v_x = R  \cos (\varphi) \cos (\theta) \\
    &  v_y = R  \cos (\varphi)  \sin (\theta)  \\
    &  v_z = R  \sin (\varphi)
\end{split}
\label{SpherCoord}
\end{equation}
where $\mathbf{v} = (v_x, v_y, v_z)$ represents the Cartesian coordinates of the vertex, and $(R, \theta, \varphi)$ its spherical coordinates, representing
$R$, the Euclidean norm from a fixed origin, $\theta$, the azimuth angle, while $\varphi$ is the elevation angle,
as illustrated in Figure~\ref{scf}. We compute the absolute differences of the spherical coordinates of all vertices,  $\{(R(i), \theta (i), \varphi(i))$
between the original object ${\cal O}$ and  the smoothed object ${\cal O}'$ in the spherical coordinate system:
\begin{equation}\label{SC}
\begin{split}
    & \phi_{14}(i)= | \theta(i) - \theta'(i) |, \\
    & \phi_{15}(i)= | \varphi(i) - \varphi'(i) |,\\
    & \phi_{16}(i)= | R(i) - R'(i) |
\end{split}
\end{equation}
where $i = 1,2, \ldots,  |V|$. The center of the spherical coordinate system is $O$, representing the center of the 3D object
calculated by averaging all the vertices in the object, as shown in Figure~\ref{scf}.

We also use statistics of the edges, defined in the spherical coordinate system. In this case, edges are defined by the differences in
the spherical coordinates of the two vertices that define the edge ends:
 \begin{equation}
\begin{split}
 & K_{\theta}(e_{(i,j)})= | \theta(i) - \theta(j) |,\\
 & K_{\varphi}(e_{(i,j)})= | \varphi(i) - \varphi(j) |,\\
 & K_{R}(e_{(i,j)})= | R(i) - R(j) |
\end{split}
\label{KSC}
\end{equation}
where $e_{(i,j)}$ is the edge connecting vertices $v(i)$ and $v(j)$, and $e_{(i,j)} \in E$. The corresponding features extracted from
both the original object and its smoothed version are
\begin{equation}
\label{FSC}
\begin{split}
 & \phi_{17}(i)= | K_{\theta}(i) - K_{\theta}'(i) |, \\
 & \phi_{18}(i)= | K_{\varphi}(i) - K_{\varphi}'(i) |,\\
 & \phi_{19}(i)= | K_{R}(i) - K_{R}'(i) |
\end{split}
\end{equation}
where, for example, $K_{\theta}(i)$ is obtained from the $i$-th edge of the original object, $K_{\theta}'(i)$ is its corresponding edge
from the smoothed object, for $i = 1,2, \ldots, |E|$, $|E|$ is the total number of edges in object ${\cal O}$.

Firstly, we apply the logarithm on all features in order to reduce the range of  their values and enforce a degree of evenness in their distribution. After applying the logarithm, we consider the four statistical moments, representing the mean, variance, skewness and kurtosis, of the logarithm of all the vertex normals,
Gaussian curvatures, curvature ratios, and the spherical coordinate features calculated as indicated above, as in the case of the feature set YANG40, defined in
Section~\ref{Yang40}.
In this way we define a vector set ${\bf \Phi}$  of 76 dimensions, which we call LFS76.
The four order moments capture almost entirely the statistical characteristics of the distribution of the features, representing their center and the
deviation from the center, as indicated by the mean and variance, respectively. The degree of symmetry in the logarithm of feature values is indicated by the skewness,
while the level of peakedness and the presence of specific values in the statistical distribution is indicated by the kurtosis.

A subset of the proposed features set, LFS52, was used in \cite{Li2016lfs}. That feature set did not include the 24-dimensional
feature vector extracted in the spherical coordinate system of 3D objects. A higher dimensional feature set, used in \cite{yang2014mesh}, is represented by
the 208-dimensional vector defined as YANG208. This feature set considers separately the statistics of the first eight features described above,
distinctly on vertex sets with valences less, equal, or greater than six. Moreover, YANG208 feature set considers the histogram differences of the ten
features defined in Section~\ref{Yang40}, as well.

\section{Training steganalyzers}
\label{Learning}

In the following we describe how we can use machine learning methods as 3D steganalyzers, as illustrated in Figure \ref{framework}.
We consider three machine learning methods: Quadratic Discriminant Analysis (QDA), Fisher Linear Discriminant (FLD) ensemble, and
Support Vector Machine (SVM), for training the steganalyzers using a training set of features extracted from pairs of stego-objects and
cover-objects. The machine learning algorithms estimate the parameters defining the nonlinear separation surfaces
between the spaces defined by the feature sets corresponding to the cover-objects, $\bf \Phi$, and the stego-objects, $\hat{\bf \Phi}$.

QDA fits mixtures of multivariate Gaussian distributions to the given feature data , \cite{krzanowski2000principles}:
\begin{equation}
  {{\bf M}(\bf \Phi)}=\sum_k \frac{1}{(2\pi)^{p/2}|{\bf \Sigma}_k|^{1/2}}e^{-\frac{1}{2}({\bf \Phi} - {\bf \mu}_k)^T {\bf \Sigma}_k^{-1}({\bf \Phi} - {\bf \mu}_k)}
  \label{multi-Gaussian}
\end{equation}
where ${\bf \mu}_k$,  ${\bf \Sigma}_k$, represent the mean and the covariance matrix of each Gaussian component.
These functions are then used for modelling boundaries between the classes of stego and cover-object data spaces by means of
a quadratic discriminative function:
\begin{equation}\label{QDA-function}
  \delta_{k({\bf \Phi})}=-\frac{1}{2}\log|{\bf \Sigma}_k|- \frac{1}{2}({\bf \Phi} - {\bf \mu}_k)^T\ {\bf \Sigma}_k^{-1}({\bf \Phi} - {\bf \mu}_k)+\log\pi_k
\end{equation}
where $\pi_k$ is the prior probability of class $k$, with $\sum\nolimits_{k=1}^2\pi_k=1$.

The FLD ensemble classifier was successfully used in image steganalysis \cite{kodovsky2012ensemble} and is characterized by a
high detection accuracy of stego-images with a relatively low computational cost. The FLD ensemble consists of a set of base
learners trained uniformly on randomly selected features from the feature space of $\bf \Phi$ and $\hat{\bf \Phi}$, corresponding to cover- and
stego-objects. The random subspace dimensionality and the number of base learners is found by minimizing the out-of-bag (OOB) error,
representing an estimate of the testing error calculated on bootstrap samples of the training set, \cite{duda2012pattern}.

Another steganalyzer considered in this study is the SVM with Gaussian kernels which is a well known classifier which can efficiently find
the best separation boundary providing the optimal separation margin between two classes. The training of SVMs in the kernel space
$G({\bf \Phi}_i,{\bf \Phi}_j)$  is performed by means of solving a convex optimization problem:
\begin{eqnarray}
 & \min\limits_{\mathbf{w},\xi_i,b}\left\{\frac{1}{2} \sum\limits_{i=1}^N \sum\limits_{j=1}^N y_i \xi_i G({\bf \Phi}_i,{\bf \Phi}_j) y_j \xi_j + C \sum\limits_{i=1}^N \xi_i \right\}  \\
 &  \mbox{subject to  } \; \sum\limits_{i=1}^N \xi_i y_i , \xi_i \ge 0, \forall i\\
 &  b = \sum\limits_{j=1}^N \xi_j y_j G({\bf \Phi}_i,{\bf \Phi}_j) - y_i
  \label{svm-1}
\end{eqnarray}
where $y_i$ represents the 3D object class  (cover-objects or stego-objects), $C$ is a regularization parameter, and $b$ represents the offset
to the origin of the coordinate system. The non-negative slack variables $\xi_i$ measure the degree of misclassification of data $\mathbf{\Phi}_i$ which are
estimated using quadratic programming. The regularization parameter $C$ controls the trade-off between the training error and model complexity \cite{cortes1995support}.
The kernel considered is the Gaussian radial basis function:
\begin{equation}
 G({\bf \Phi}_i, {\bf \Phi}_j) = \exp ( -\gamma \| {\bf \Phi}_i - {\bf \Phi}_i \|^2),
  \label{svm-kernel}
\end{equation}
where $\gamma$ is the kernel scale parameter which can be seen as the inverse of the radius of influence of samples selected as support vectors.
The selection of the kernel scale parameter $\gamma$ and the regularization parameter $C$ used in the training of the proposed SVM-based 3D
steganalyzers are estimated empirically, as described in the Appendix. A new data ${\bf z}$ is classified using SVM according to the following formula:
\begin{equation}
\mbox{sgn} \left[ \sum_{i=1}^N \xi_i y_i G({\bf \Phi}_i,{\bf z}) - b \right].
\end{equation}

An important property of the steganalyzers is their ability to avoid overfitting on the training set while generalizing well in the case of data sets which are not similar to those used for training.

\section{Experimental results}
\label{Exper}

\begin{figure}[htbp]
  \centering
    \includegraphics[width=3in]{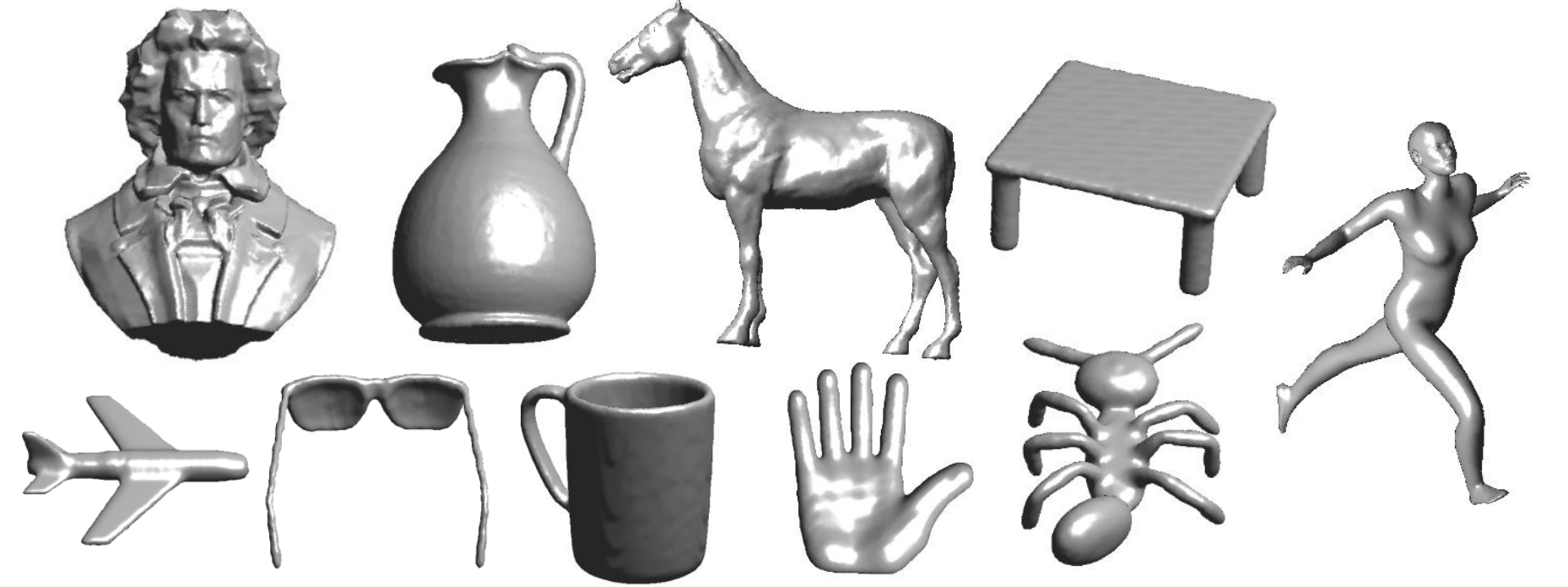}
  \caption{3D objects used in the steganalytic tests.}
  \label{objects}
\end{figure}

In the following we provide the experimental results of the proposed 3D steganalysis methodology.
During the tests we consider detecting the information embedded in 3D objects by three different steganographic methods:
the steganalysis-resistant 3D watermarking proposed in \cite{yang2016watermarking},
the mean-based watermarking method from \cite{cho2007oblivious}, and the high capacity embedding
method proposed in \cite{chao2009high}.
For the experiments we use the Princeton Mesh Segmentation project
\cite{chen2009benchmark} database, which consists of 354 3D objects represented as meshes.
The shapes of ten objects from this database are shown in Figure~\ref{objects}.
For each steganalyzer, we split the 354 pairs of cover- and stego-objects into 260 pairs, used for
training, and 94 pairs for testing. We consider 30 different splits of the given 3D object database,
into the training and testing data sets. The results are given by considering two measurements.
The first is the median value of the sum of false negatives (missed detections) and false positives
(false alarms) from all 30 trials, while the other one is the median value of the area under
the Receiver Operating Curves (ROC) of the detection results, evaluated over the 30 splits
of the data into training and testing sets.

During the pre-processing, we first apply three iterations of Laplacian smoothing on both cover- and stego-objects,
by considering the updating weight of 0.3. Next stage consists in feature extraction and their statistical modelling.
We consider the proposed feature set LFS76, discussed in Section \ref{Features} and compare their results
against YANG208, proposed in \cite{yang2014mesh}, and its simplified version, called YANG40. We also consider
the feature sets combining YANG40 and the vertex normal feature, VNF4, representing the mean, variance,
skewness and kurtosis of  $\phi_{11}$ from equation (\ref{vnf}), the combination of YANG40 and the curvature feature, CF8, representing
the mean, variance, skewness and kurtosis of  $\phi_{12}$ and $\phi_{13}$ from equations (\ref{fea12}) and (\ref{fea13}).
We also compare the LFS76 with the feature set proposed in our previous work \cite{Li2016lfs}, LFS52, which consists of YANG40, VNF4 and CF8 features.

\begin{figure}[htbp]
  \centering
  \subfigure[Cover]{
  \label{fig:subfig:a} 
  \includegraphics[width=1.5in]{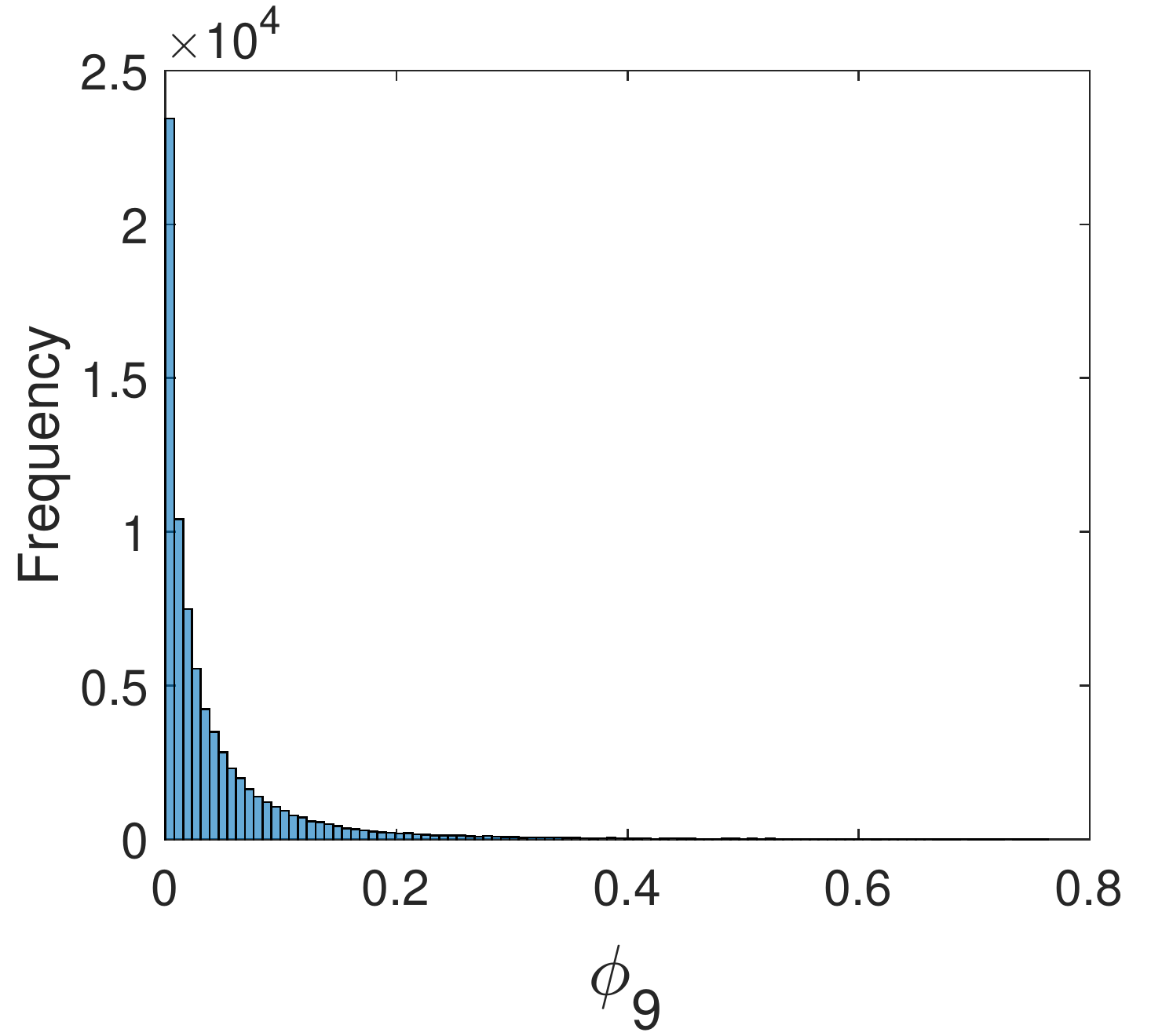}}
  \centering
  \subfigure[Stego]{
  \label{fig:subfig:b} 
  \includegraphics[width=1.5in]{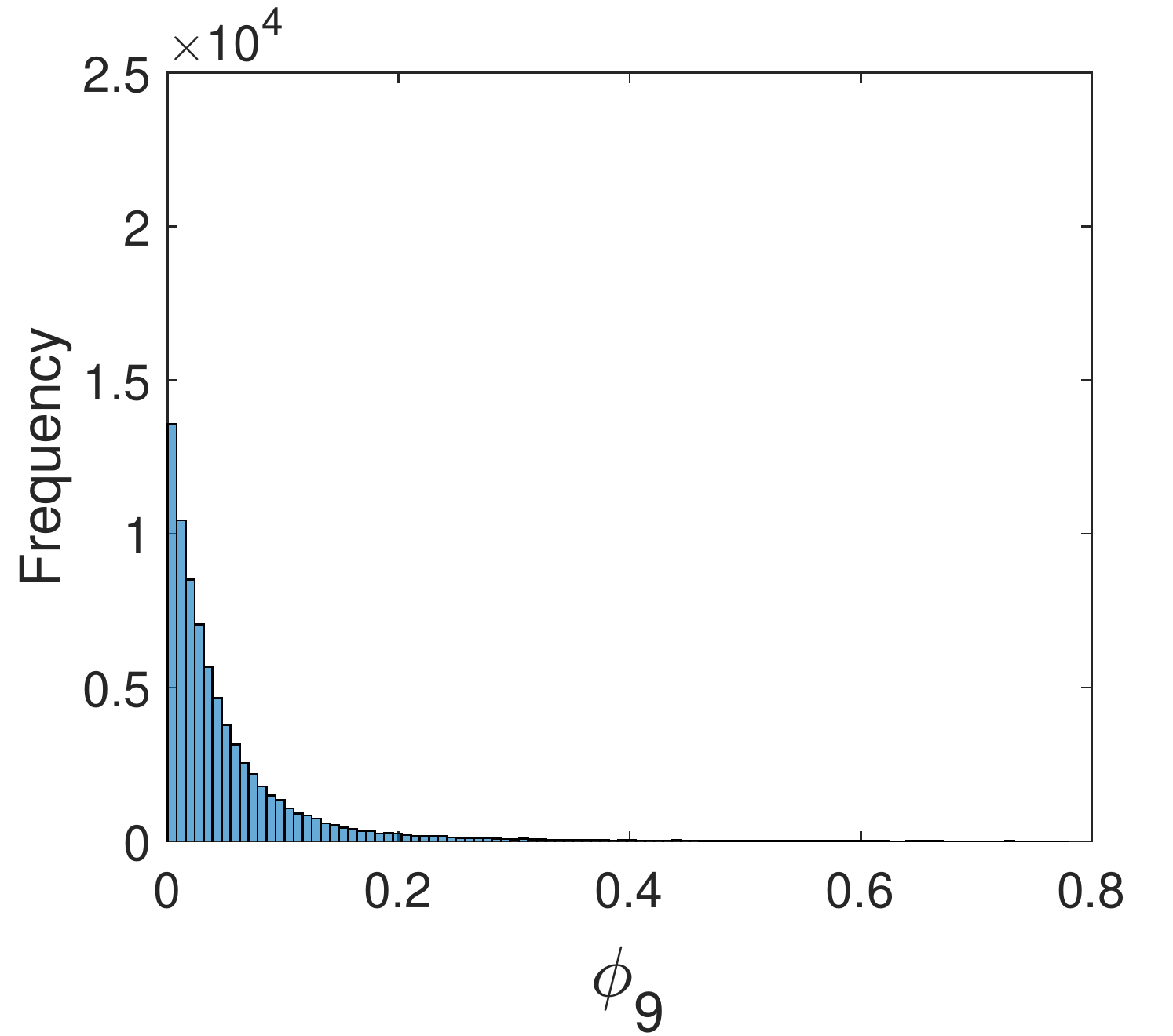}}
  \subfigure[Cover]{
  \label{fig:subfig:b} 
  \includegraphics[width=1.5in]{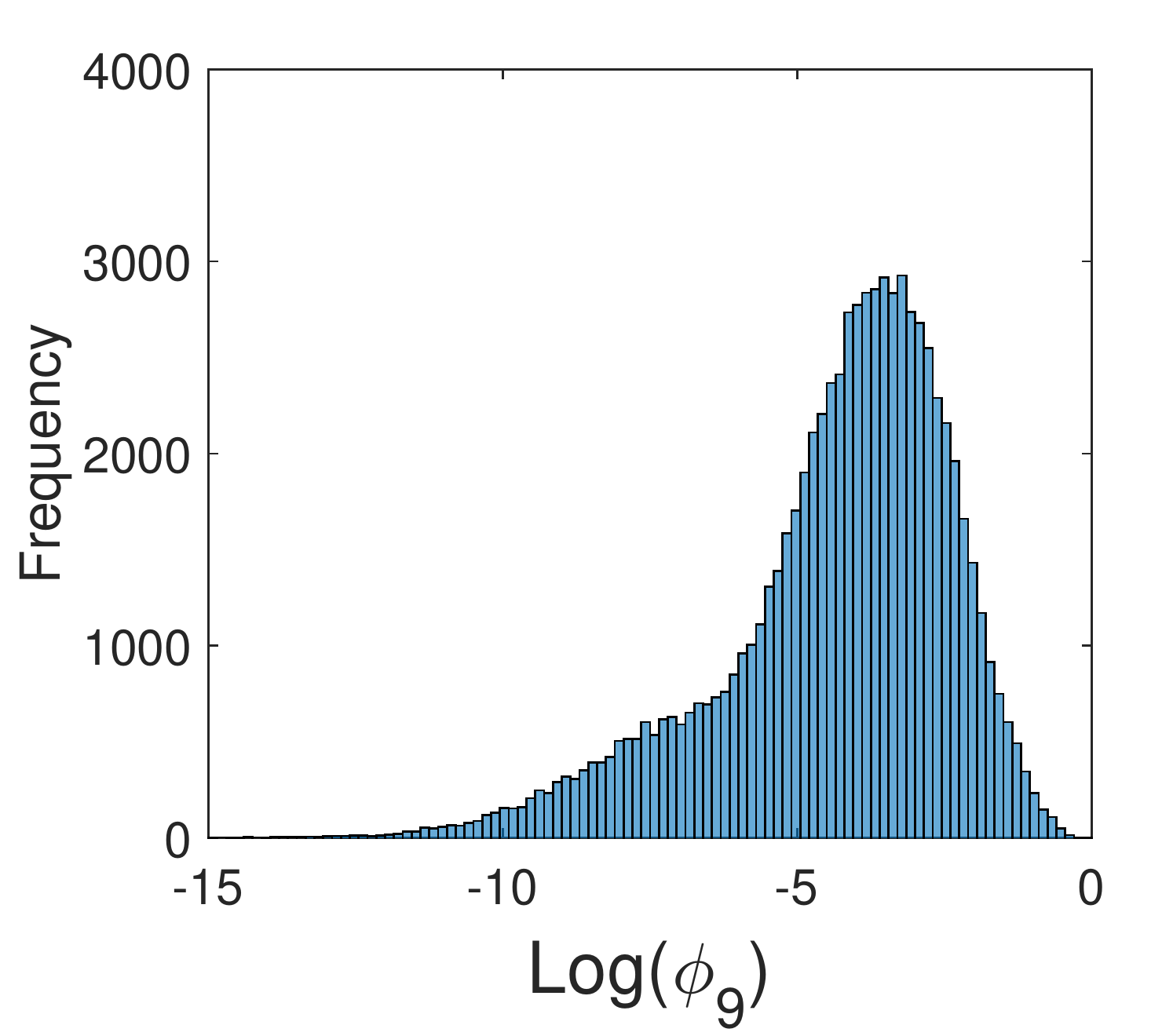}}
  \subfigure[Stego]{
  \label{fig:subfig:b} 
  \includegraphics[width=1.5in]{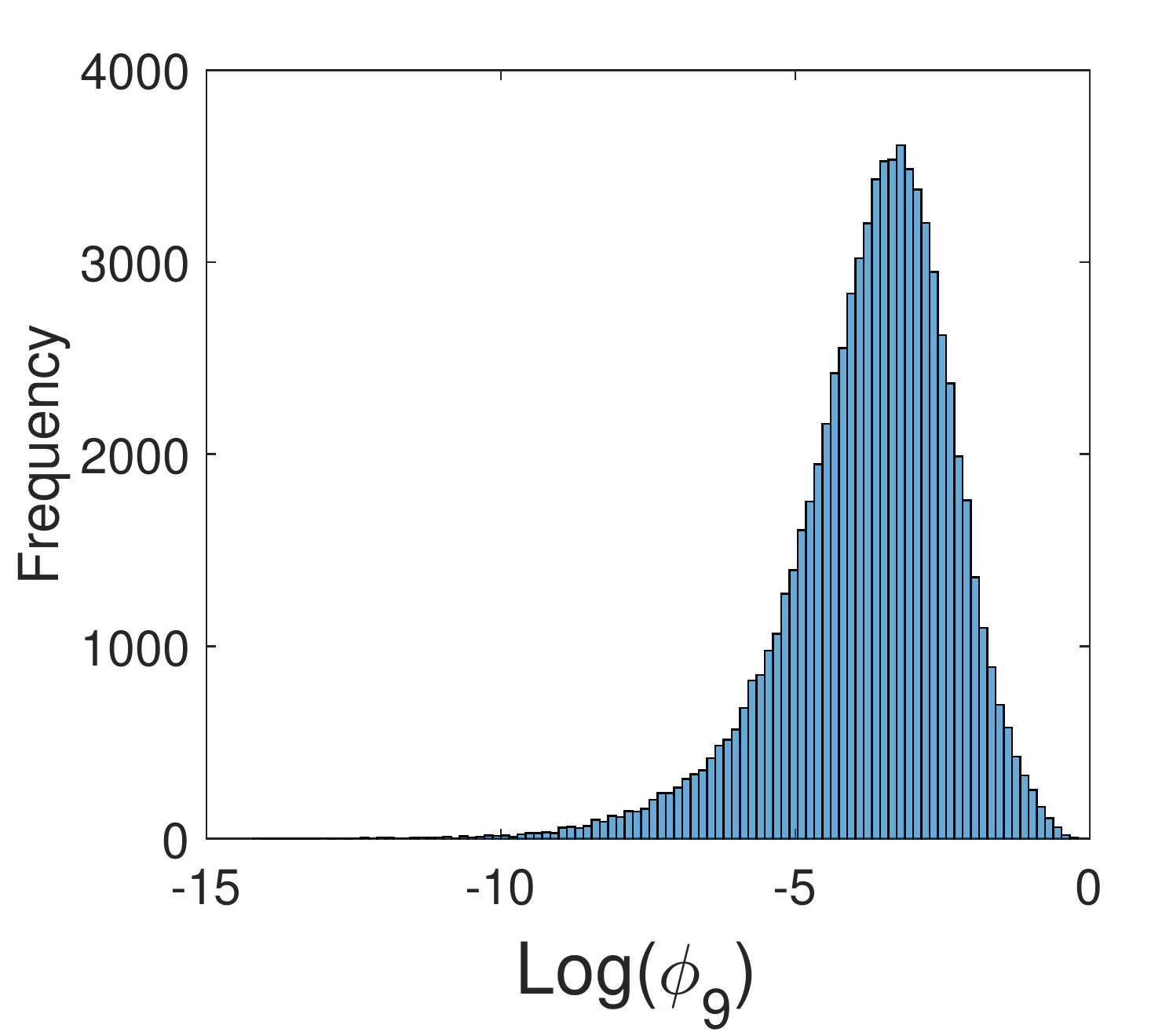}}
  \caption{Histograms of dihedral angle feature and its logarithm of the cover and stego versions of the ``Head statue" object from the database from \cite{chen2009benchmark}.}
  \label{311-di}
  \end{figure}

\begin{figure}[htbp]
  \centering
  \subfigure[Cover]{
  \label{fig:subfig:a} 
  \includegraphics[width=1.5in]{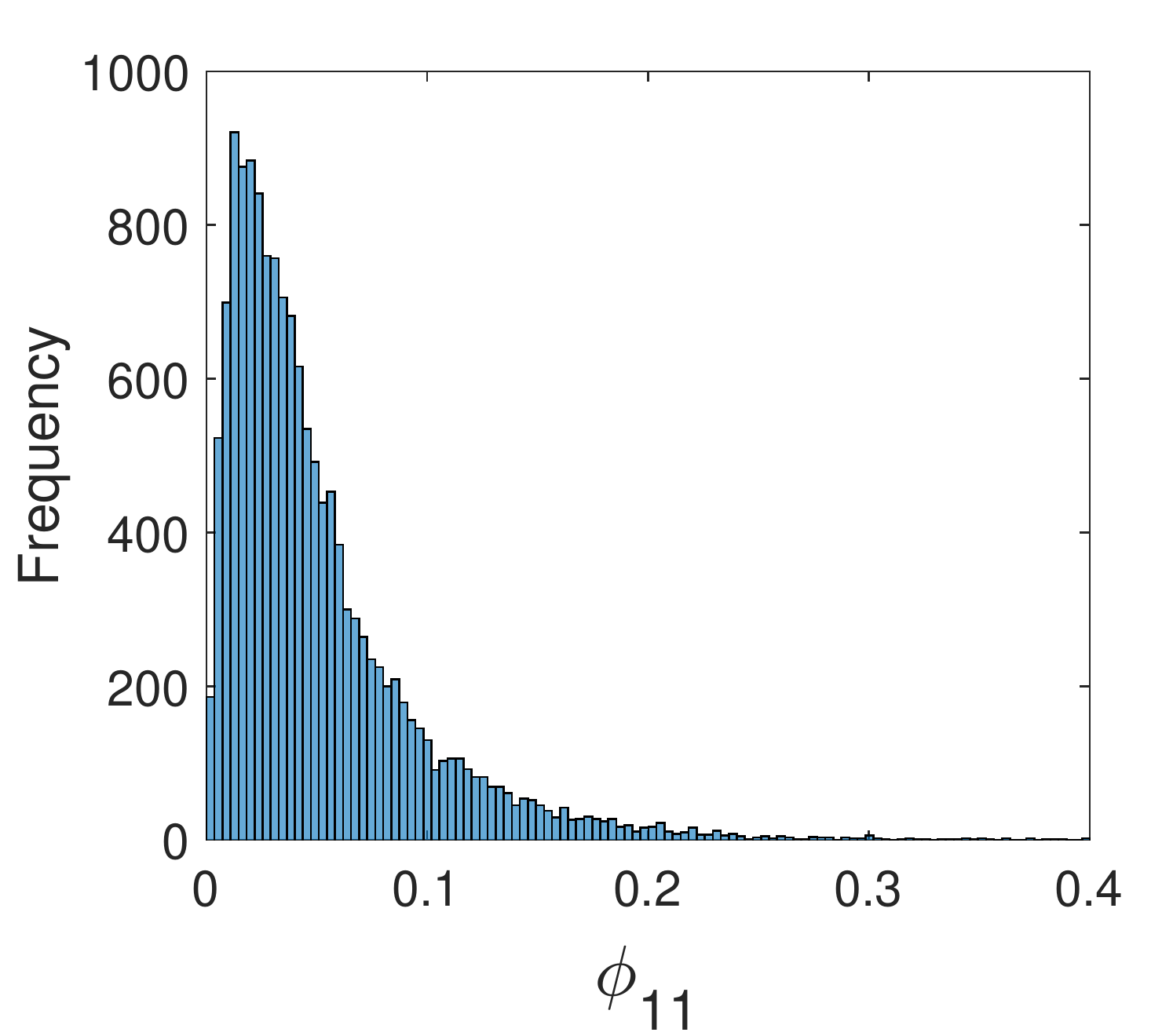}}
  \centering
  \subfigure[Stego]{
  \label{fig:subfig:b} 
  \includegraphics[width=1.5in]{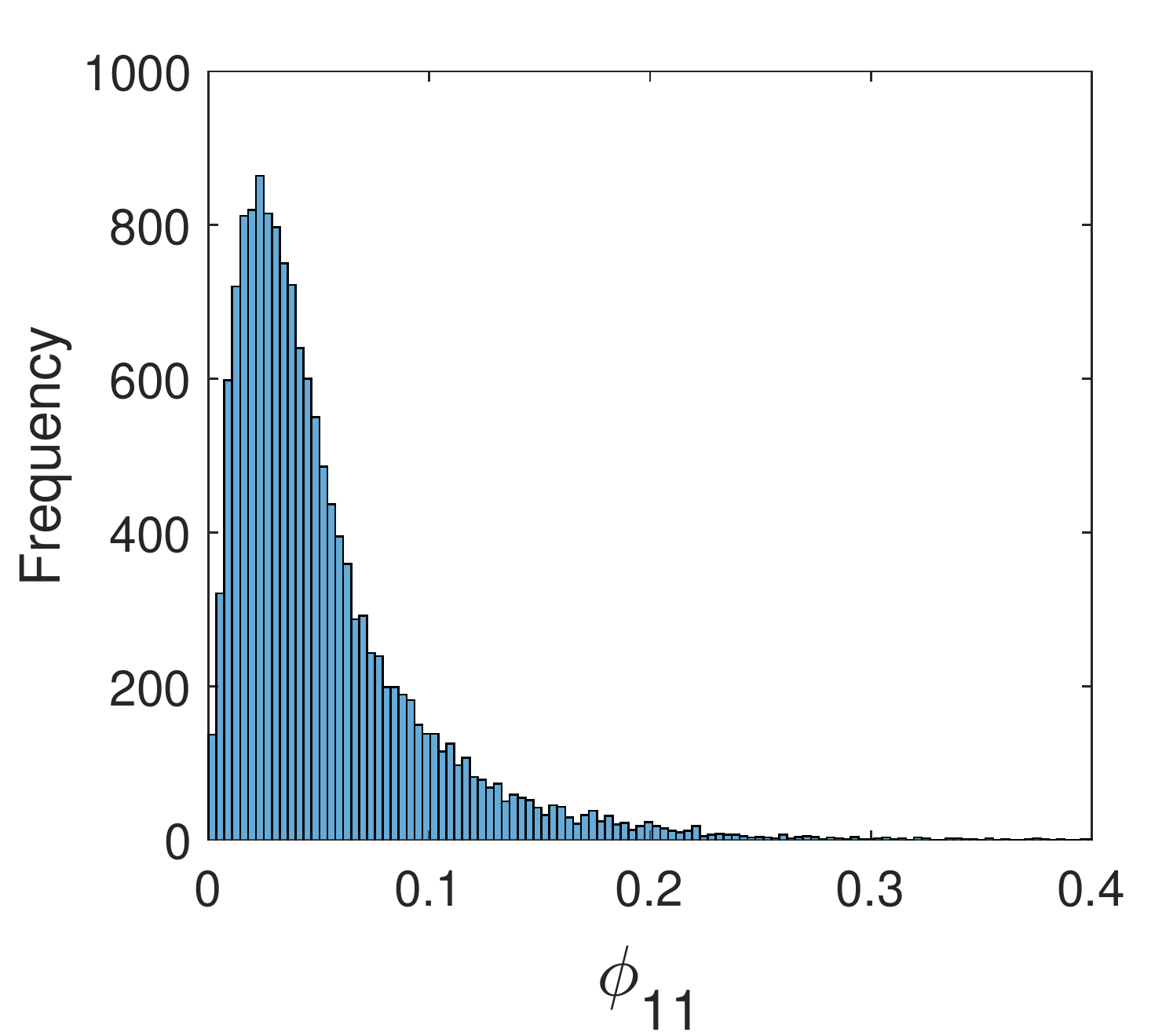}}
  \subfigure[Cover]{
  \label{fig:subfig:b} 
  \includegraphics[width=1.5in]{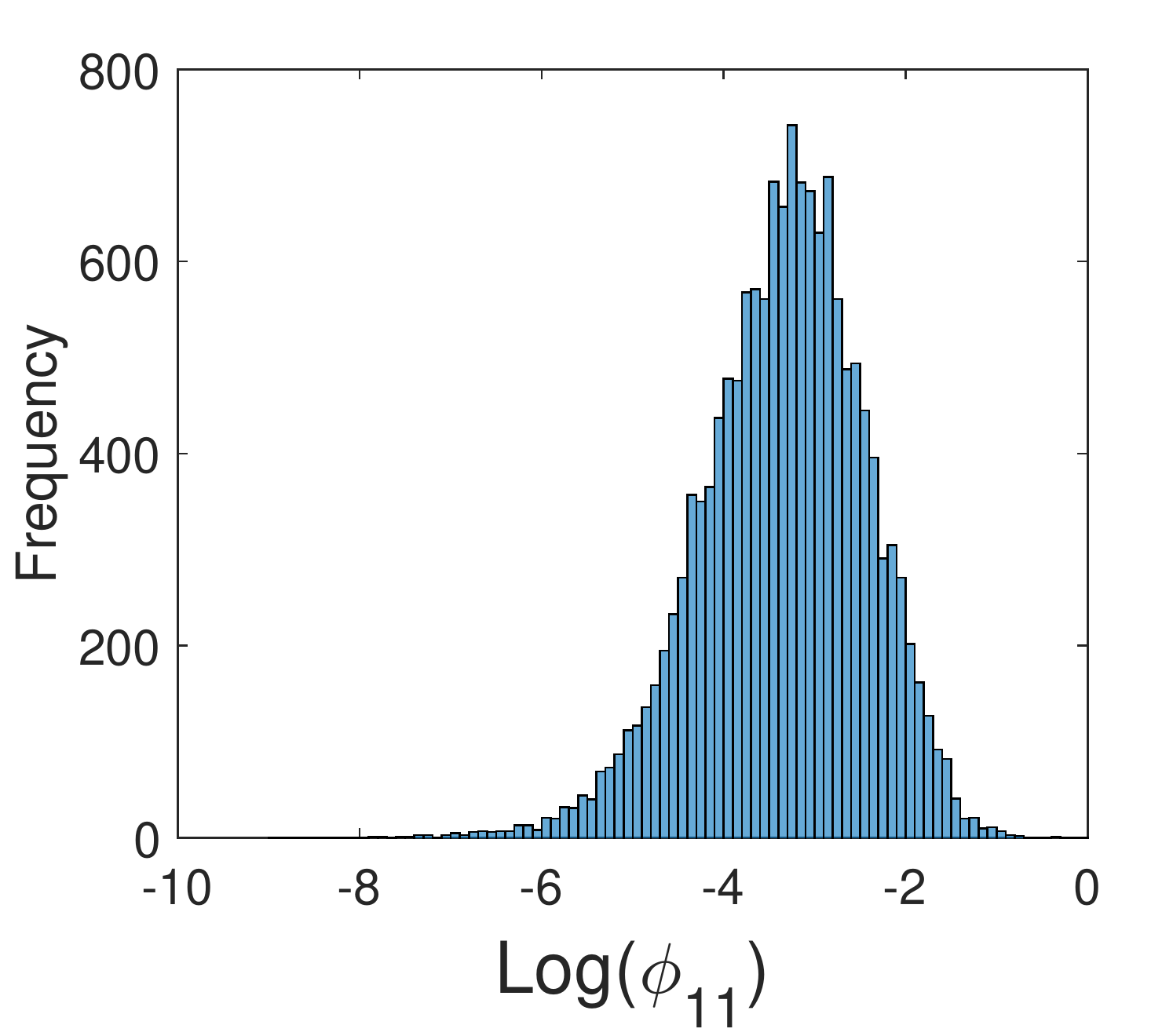}}
  \subfigure[Stego]{
  \label{fig:subfig:b} 
  \includegraphics[width=1.5in]{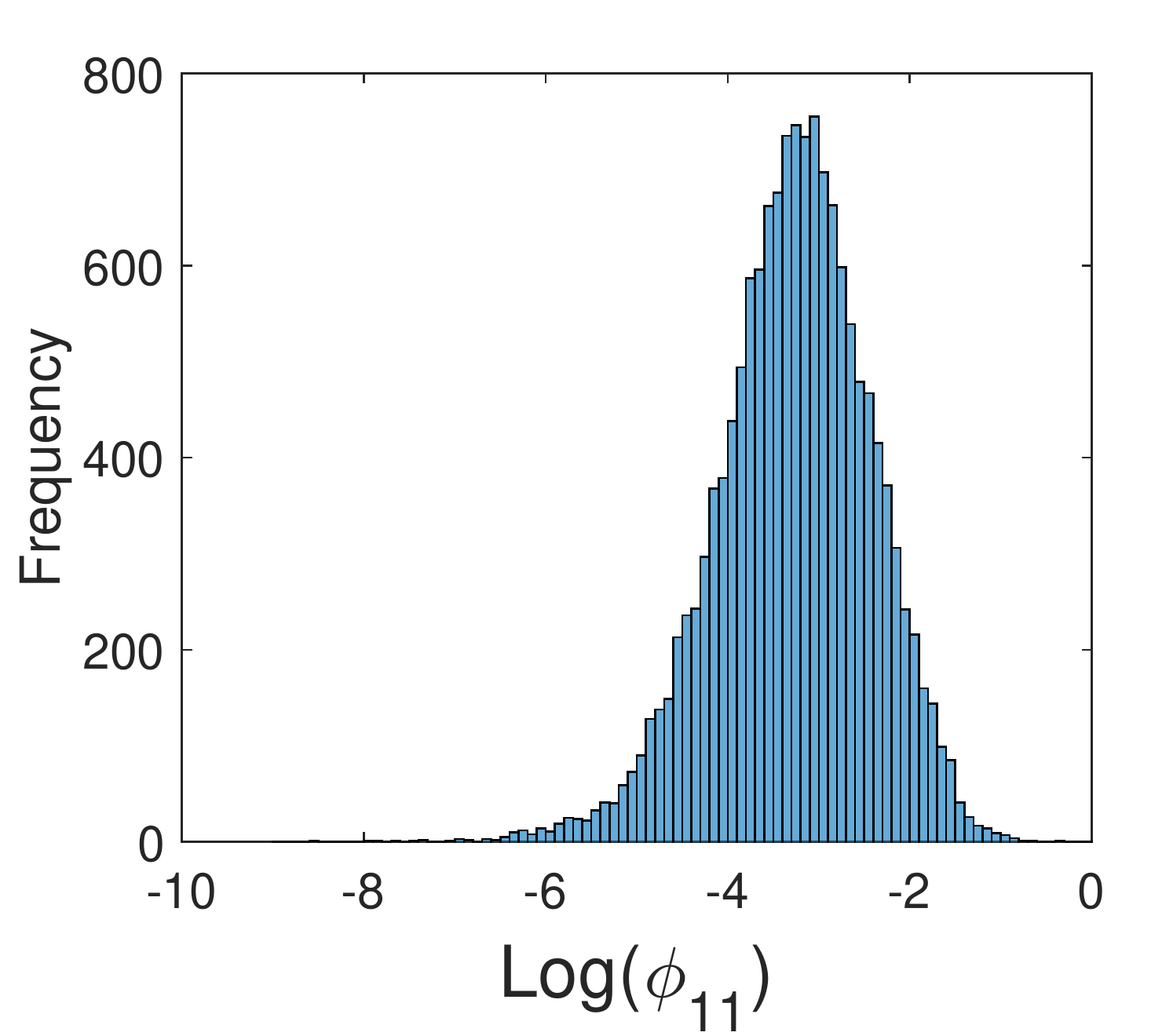}}
  \caption{Histograms of vertex normals feature and its logarithm of the cover and stego versions of `the `Horse" object from the database from \cite{chen2009benchmark}.}
  \label{387-vnf} 
\end{figure}

Figures~\ref{311-di} (a) and (b) show the histograms of the dihedral angles $\phi_9$, calculated according to equation (\ref{dihedral}),
for the cover- and stego-object, respectively, for the object ``Head statue'' shown in Figure~\ref{311-FeatureDiff}(f).
The histograms of the logarithm of these features are shown in Figures~\ref{311-di} (c) and (d), for the cover- and stego-objects, respectively.
Figures~\ref{387-vnf} (a) and (b) show the histograms of the vertex normal  $\phi_{11}$ calculated according to equation (\ref{vnf}),
while Figures~\ref{387-vnf} (c) and (d) show the corresponding histogram of logarithms for the cover-object ``Horse'' shown in Figure~\ref{objects},
and its corresponding stego-object embedded by the steganographic method from \cite{cho2007oblivious}.
From these figures, we can observe following the application of the logarithm, the distributions of feature components $\phi_9$ and  $\phi_{11}$ become similar to that of normal distributions where it is easier to model the differences between the
distributions of cover- and stego-objects when using the four statistical moments of mean, variance, skewness and kurtosis.

The steganalyzers are trained as binary classifiers implemented using three methods:
the Quadratic Discriminant Analysis (QDA), Fisher Linear Discriminant (FLD) ensemble and the
Support Vector Machine (SVM) with Gaussian kernel, described in Section~\ref{Learning}.
The quadratic discriminant that fits multivariate normal densities with covariance estimates
\cite{krzanowski2000principles} was used in \cite{yang2014mesh} as well. The implementation of
FLD ensemble is an extension of the version proposed in \cite{cogranne2015modeling}.
When training the SVM classifiers, the optimal values for the parameters
$C$ from (18) and $\gamma$  from (\ref{svm-kernel}), are found by grid-search using
five-fold cross validation, as detailed in the Appendix. The implementation of SVM is based
on LIBSVM \cite{CC01a}.

\subsection{Steganalysis of the information hiding methods}

\begin{figure*}[htbp]
  \centering
  \subfigure[]{
  \label{fig5:subfig:a} 
  \includegraphics[width=1.2in]{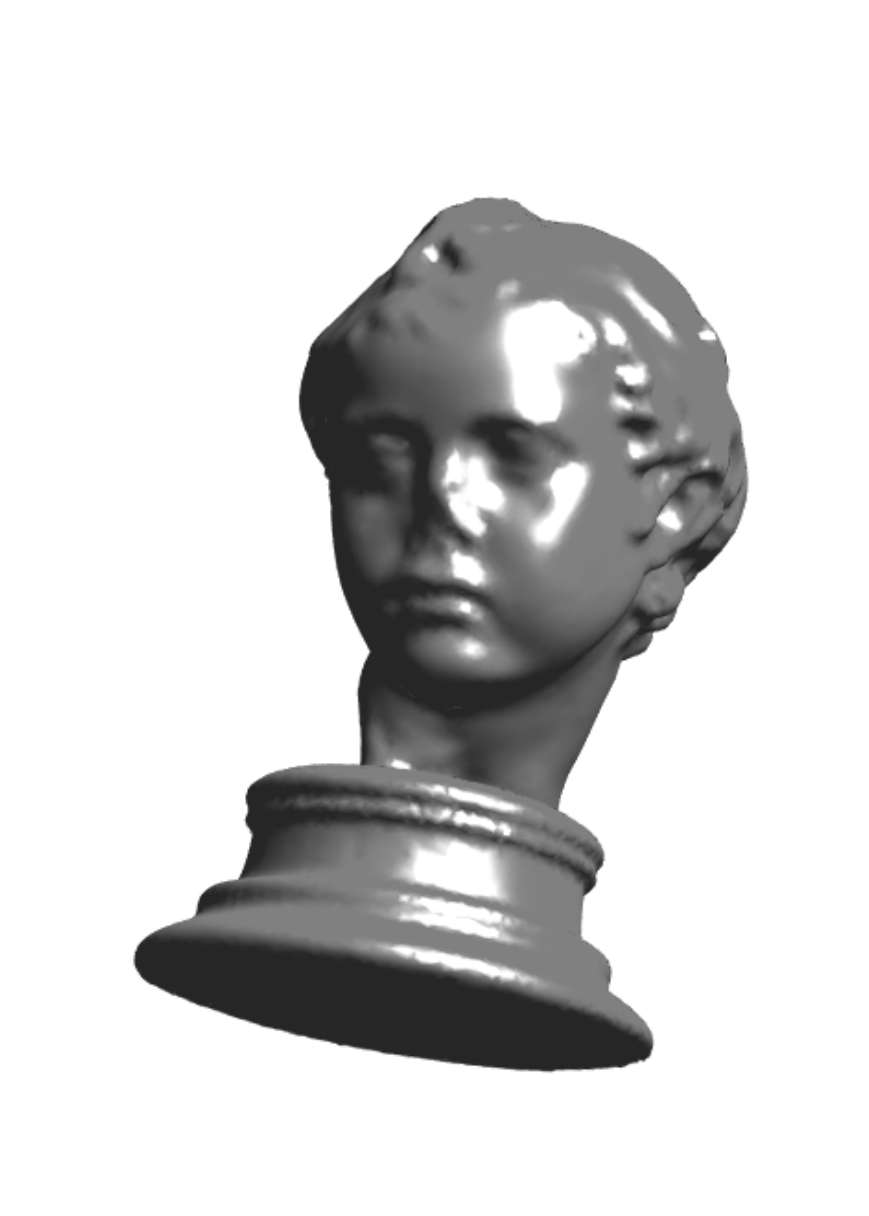}}
  \subfigure[]{
  \label{fig5:subfig:b} 
  \includegraphics[width=1.2in]{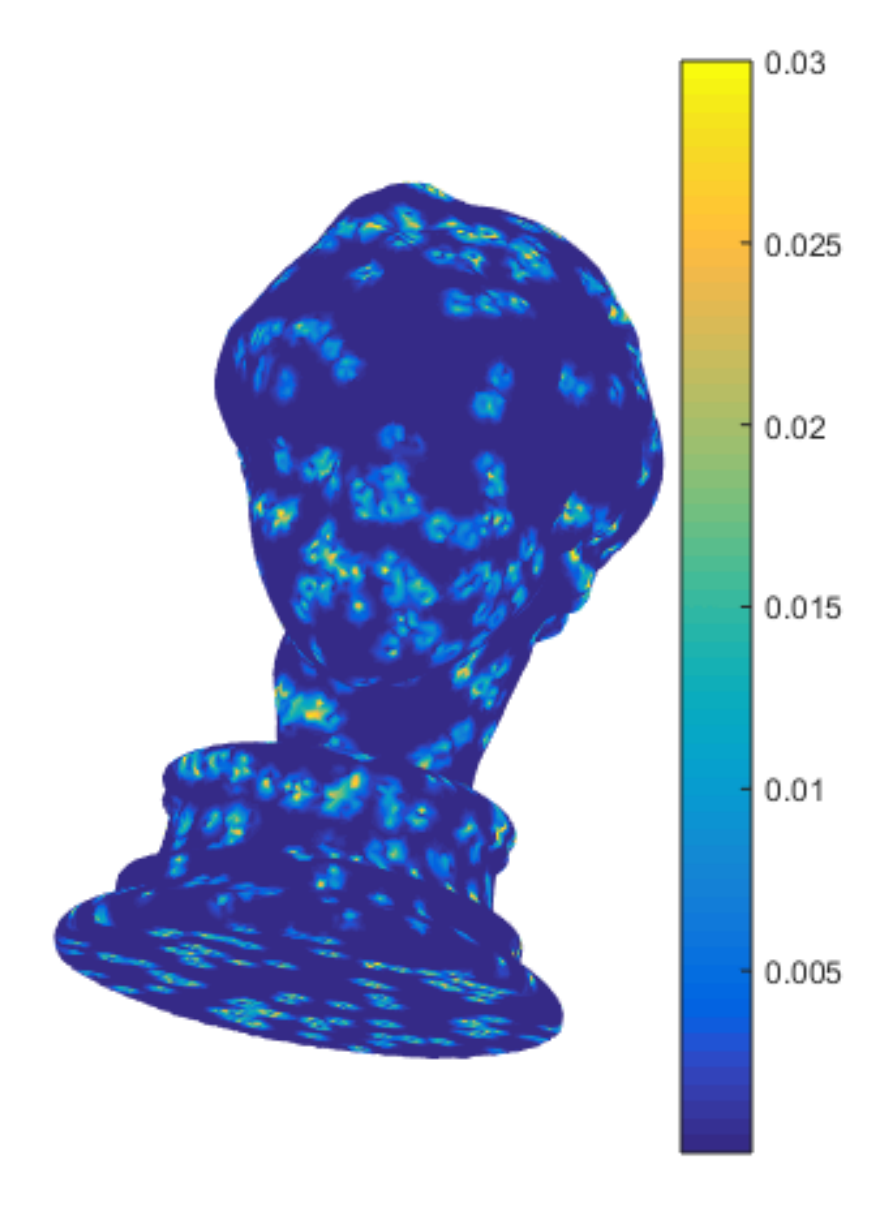}}
  \subfigure[]{
  \label{fig5:subfig:c} 
  \includegraphics[width=1.2in]{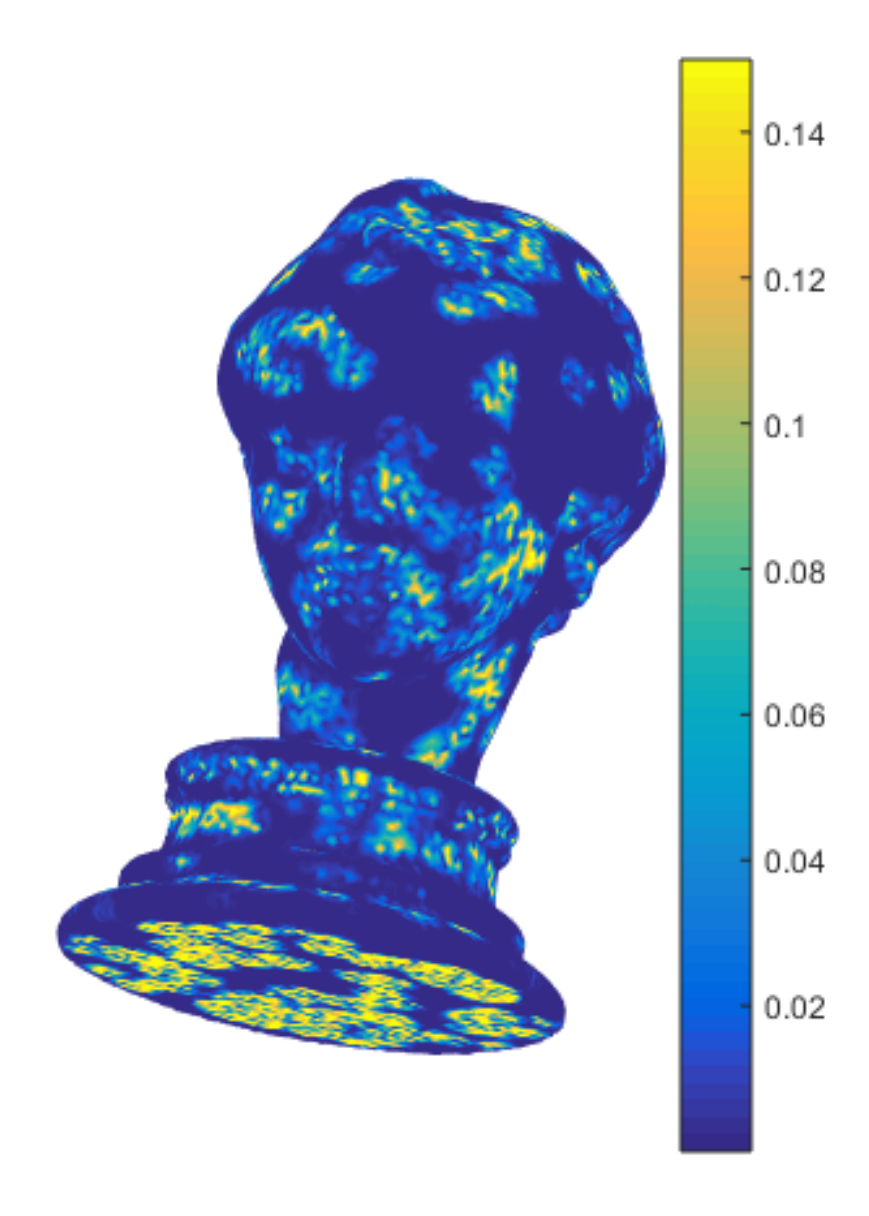}}
  \subfigure[]{
  \label{fig5:subfig:d} 
  \includegraphics[width=1.2in]{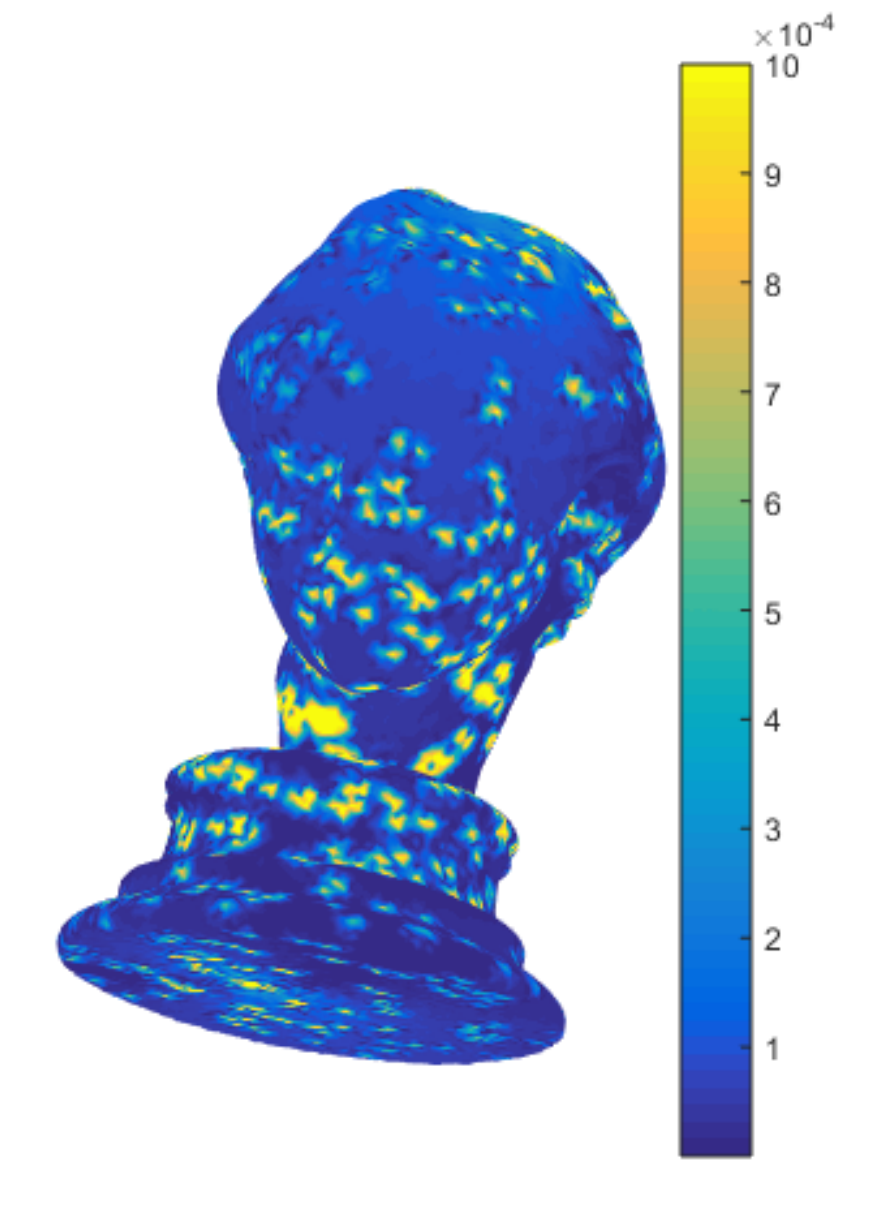}}
  \subfigure[]{
  \label{fig5:subfig:e} 
  \includegraphics[width=1.2in]{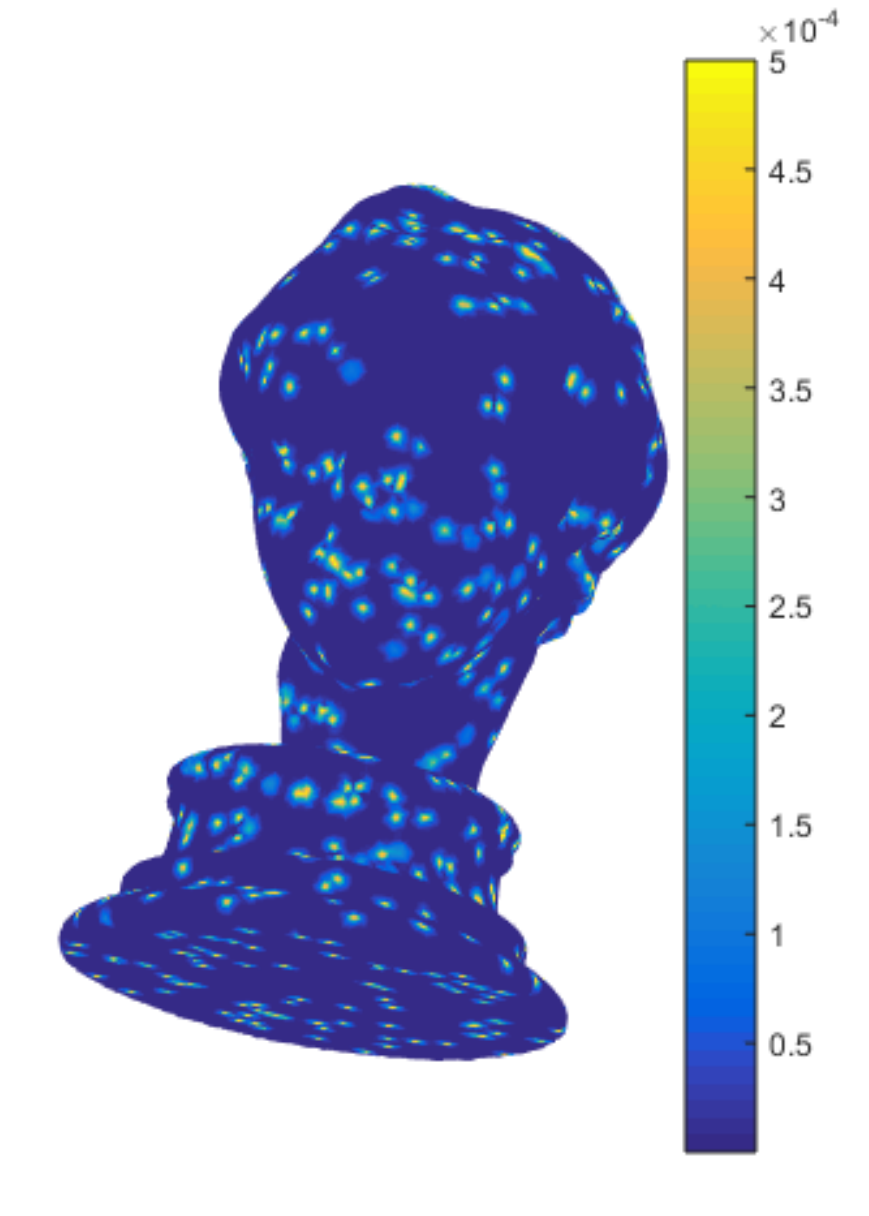}}
  \subfigure[]{
  \label{fig5:subfig:f} 
  \includegraphics[width=1.2in]{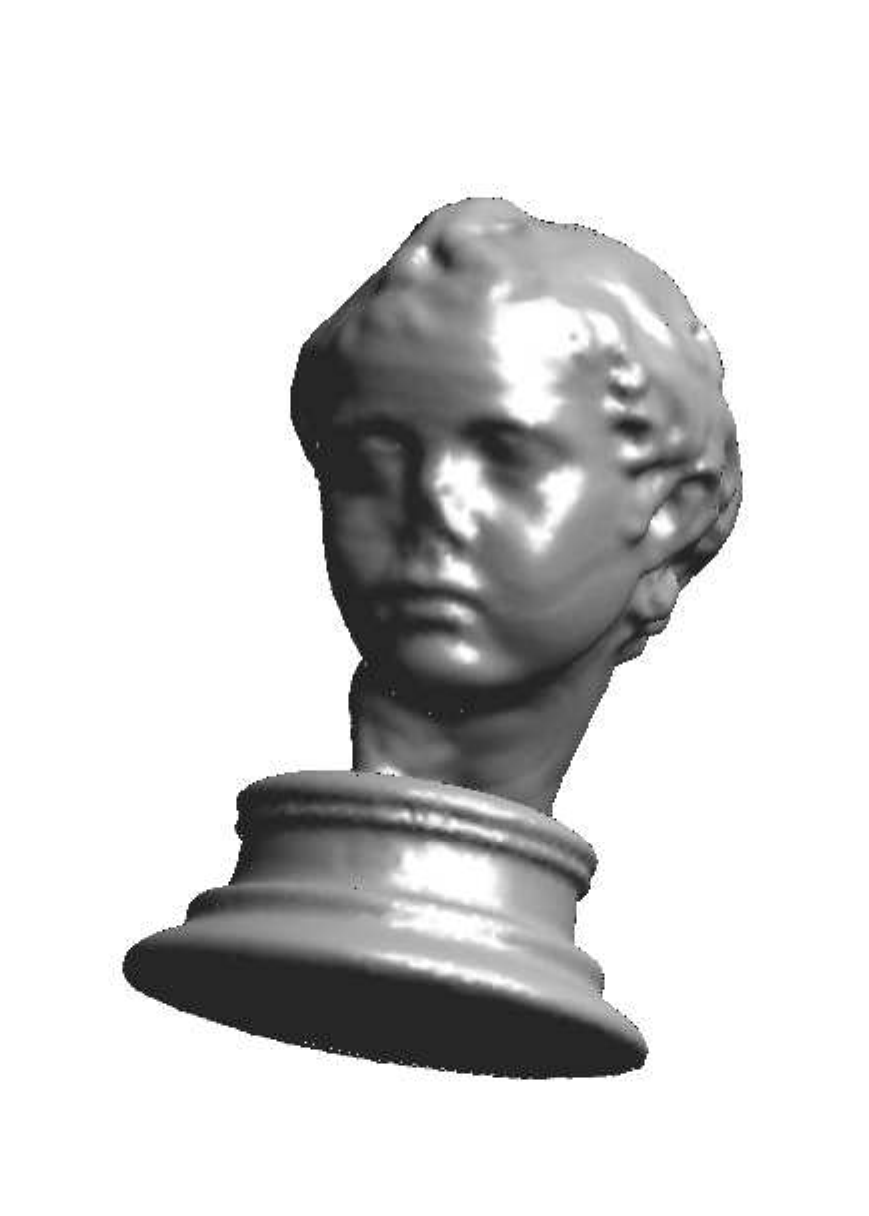}}
  \subfigure[]{
  \label{fig5:subfig:g} 
  \includegraphics[width=1.2in]{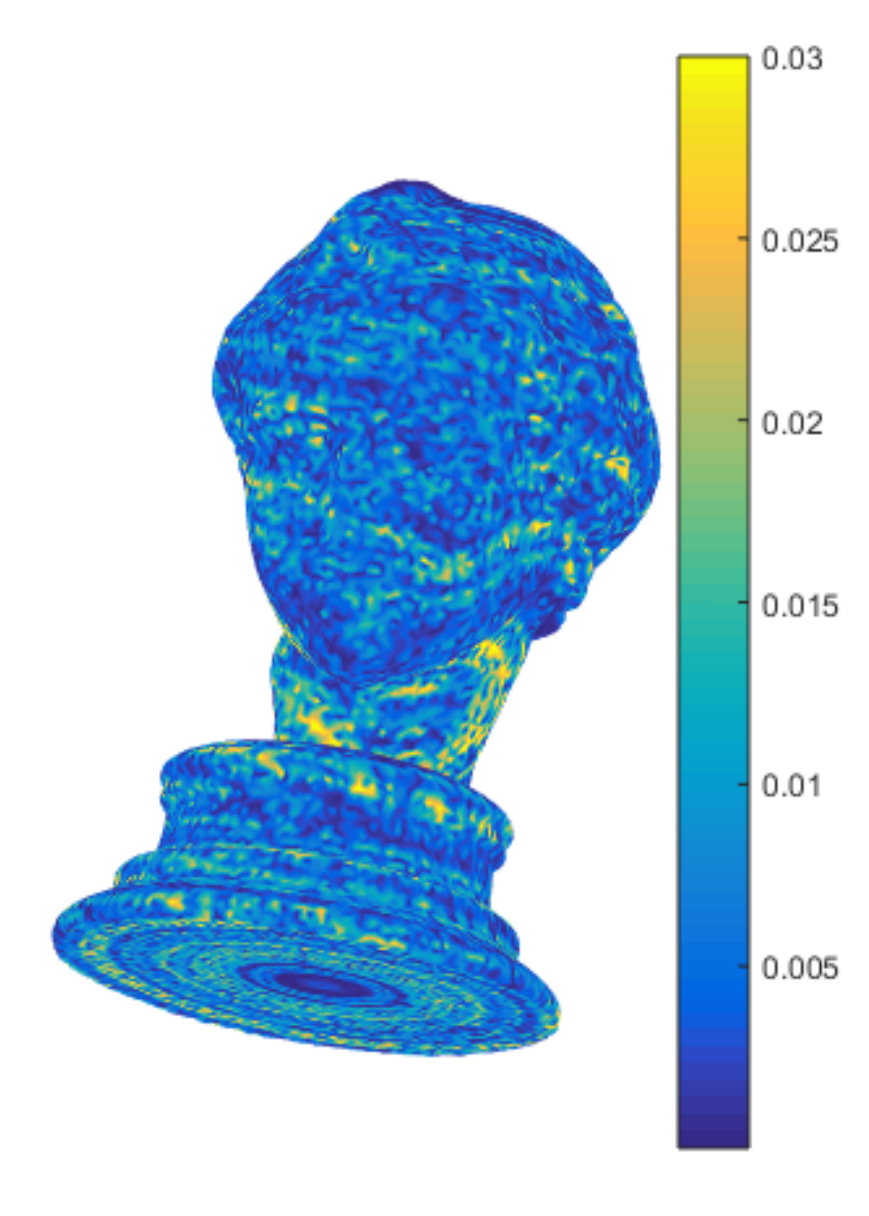}}
  \subfigure[]{
  \label{fig5:subfig:h} 
  \includegraphics[width=1.2in]{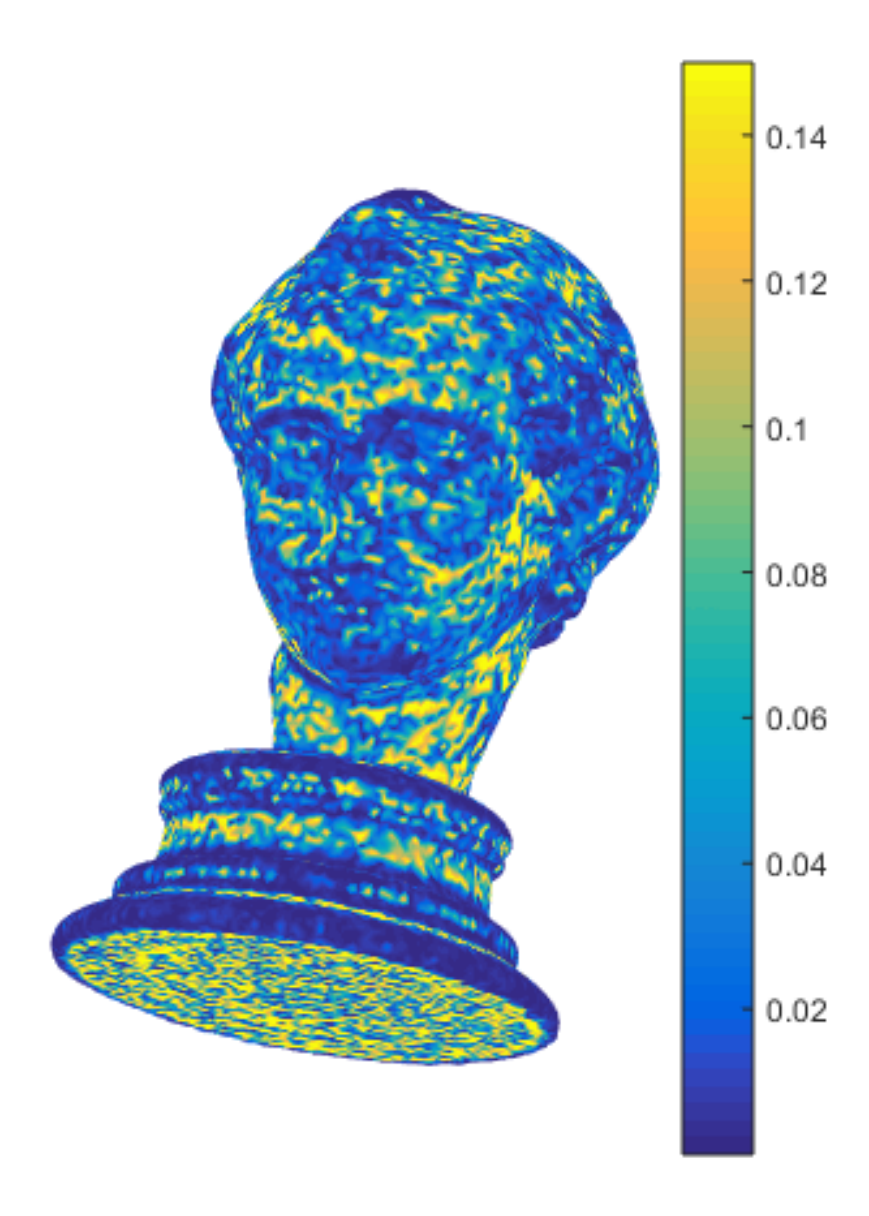}}
  \subfigure[]{
  \label{fig5:subfig:i} 
  \includegraphics[width=1.2in]{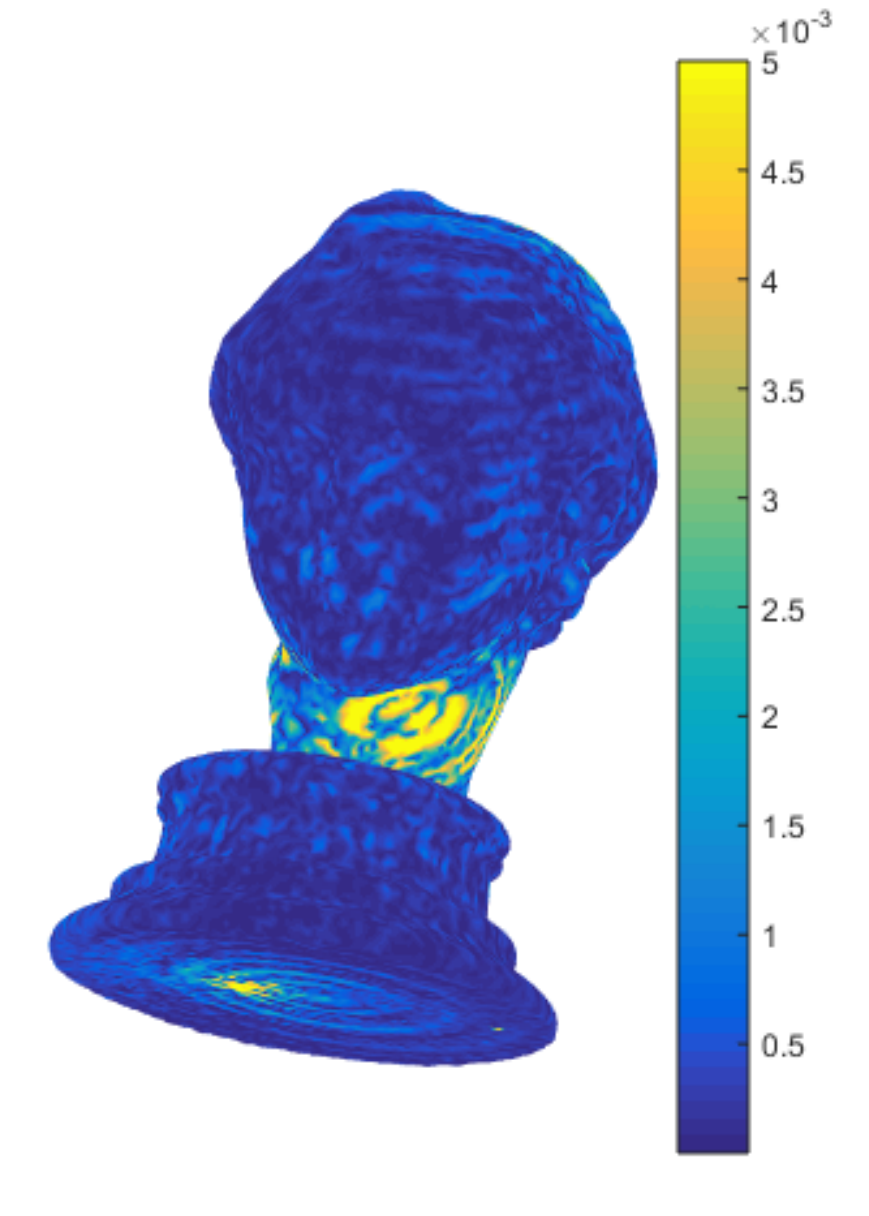}}
  \subfigure[]{
  \label{fig5:subfig:j} 
  \includegraphics[width=1.2in]{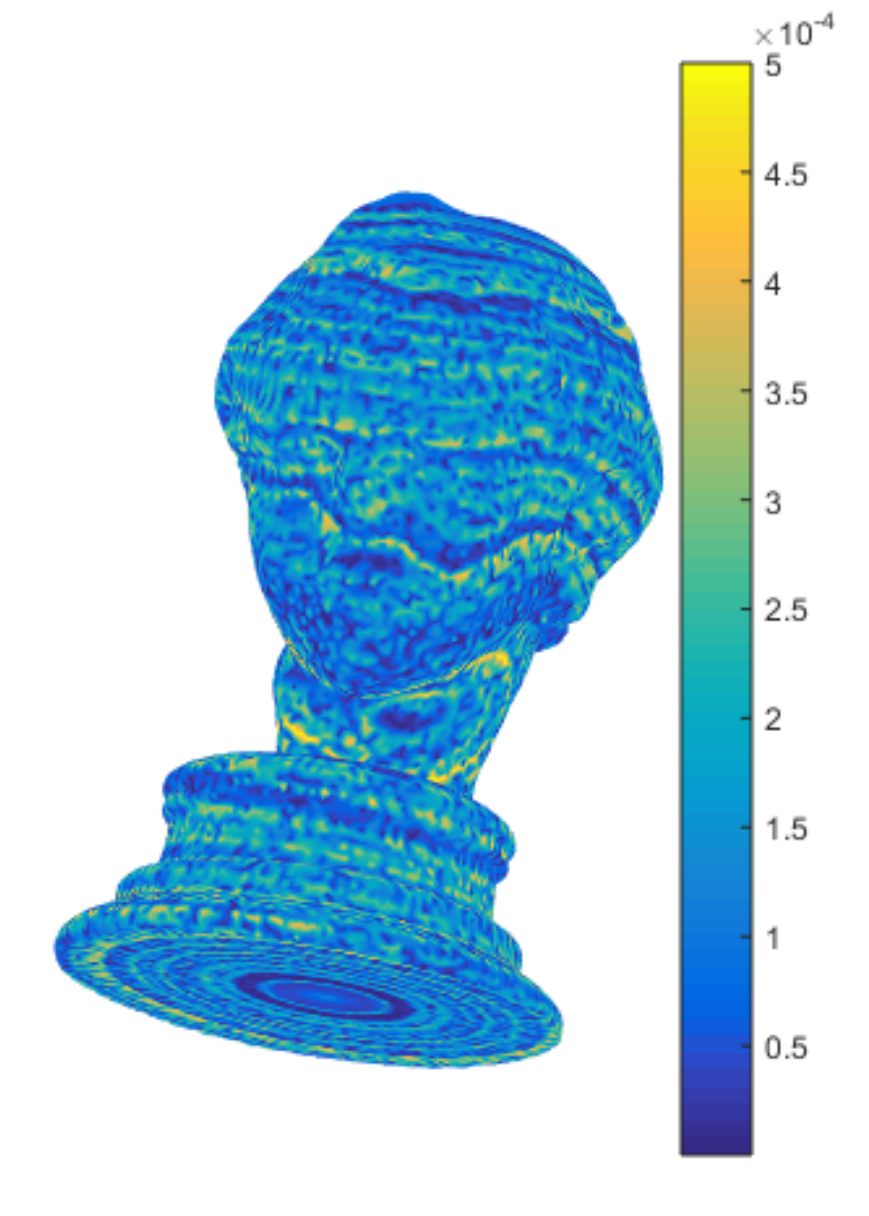}}
  \subfigure[]{
  \label{fig5:subfig:k} 
  \includegraphics[width=1.2in]{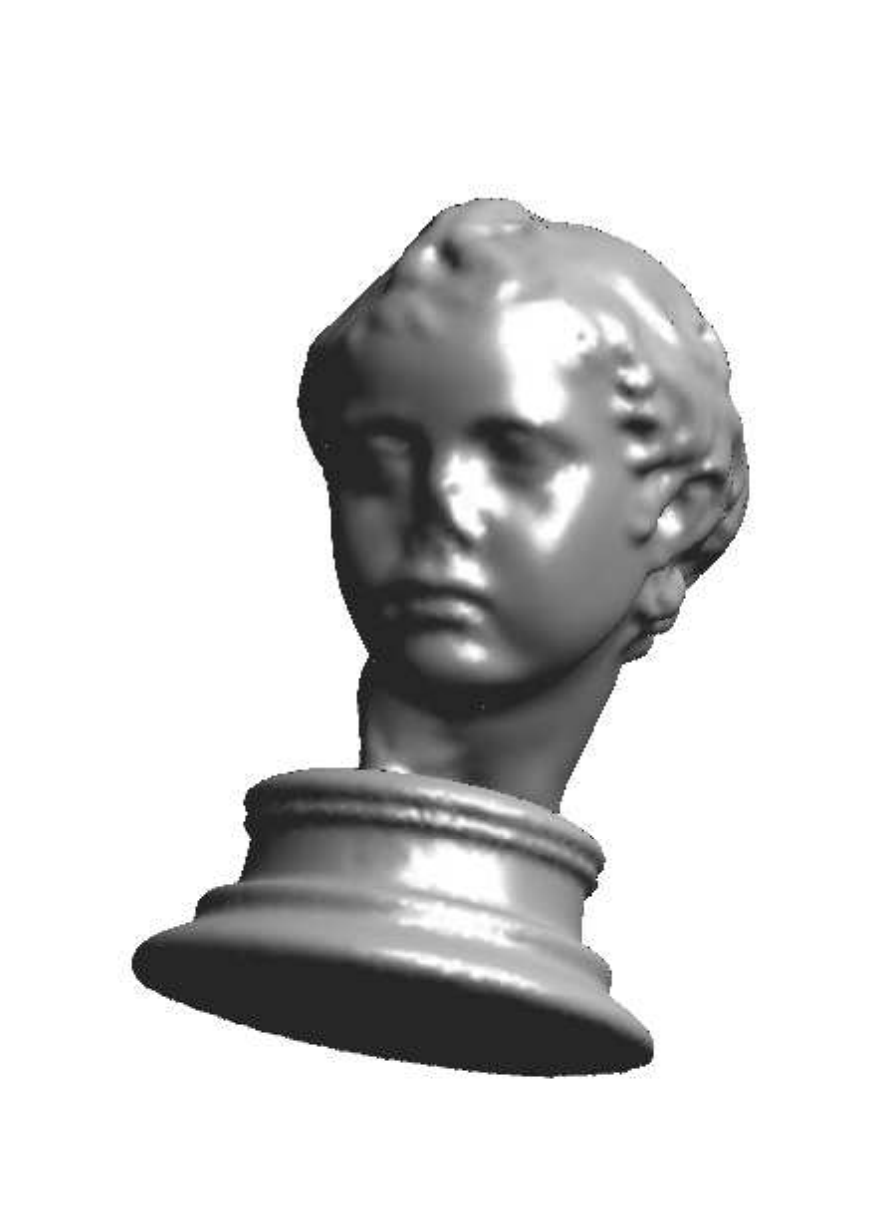}}
  \subfigure[]{
  \label{fig5:subfig:l} 
  \includegraphics[width=1.2in]{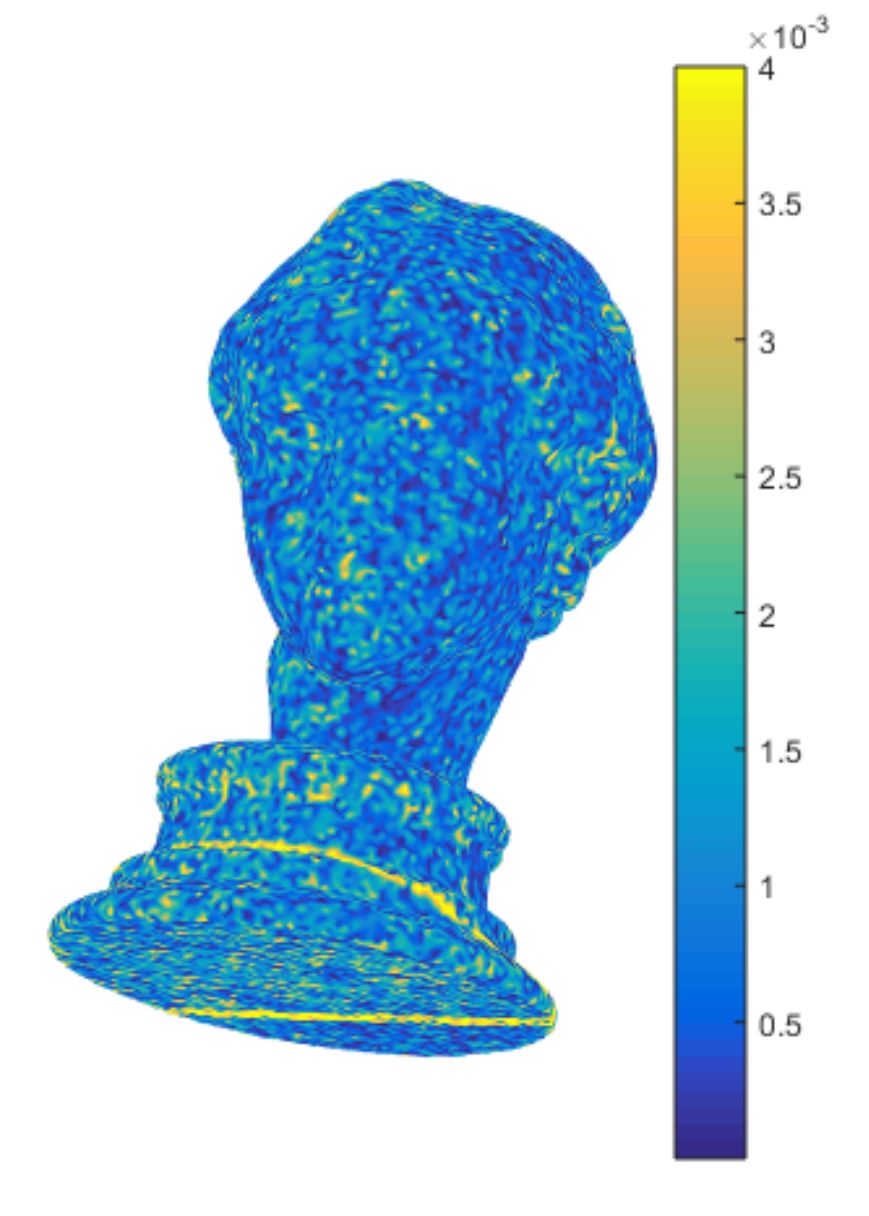}}
  \subfigure[]{
  \label{fig5:subfig:m} 
  \includegraphics[width=1.2in]{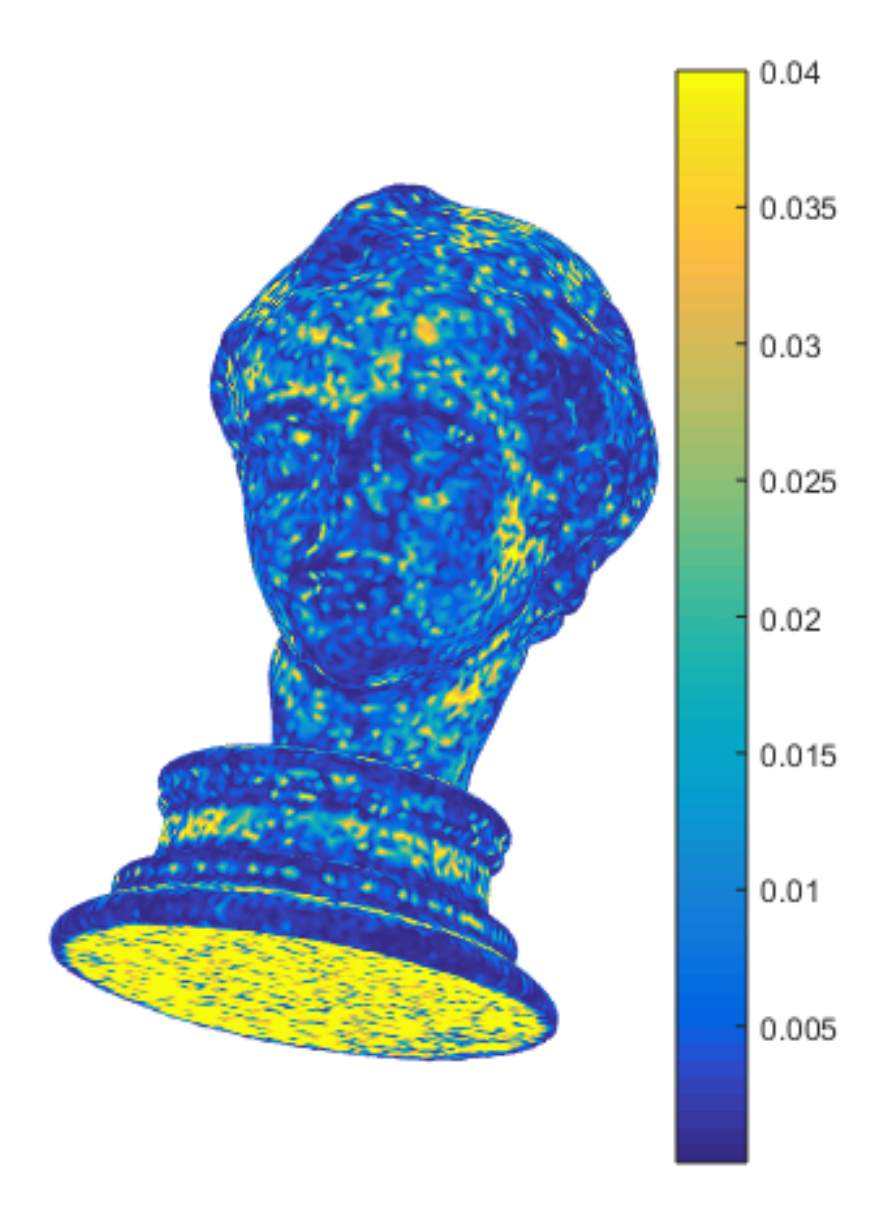}}
  \subfigure[]{
  \label{fig5:subfig:n} 
  \includegraphics[width=1.2in]{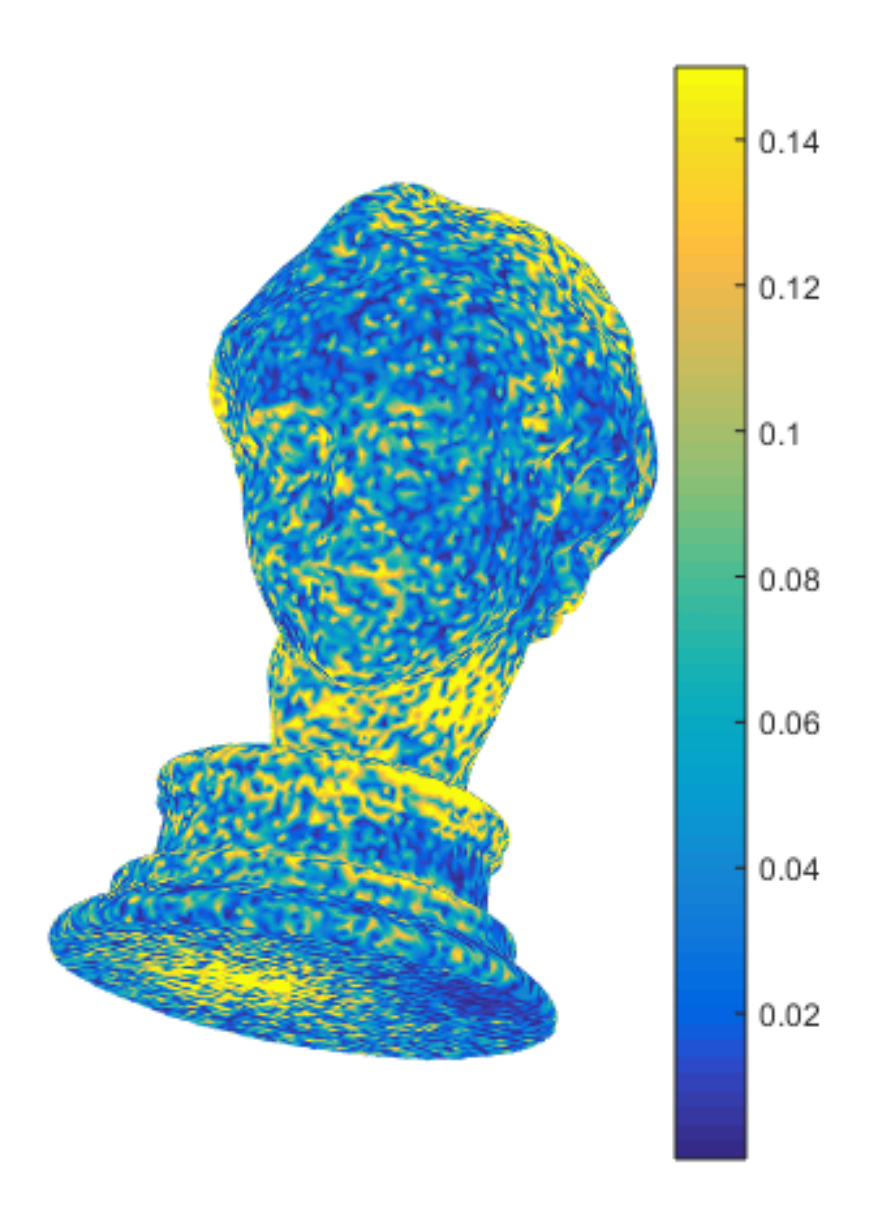}}
  \subfigure[]{
  \label{fig5:subfig:o} 
  \includegraphics[width=1.2in]{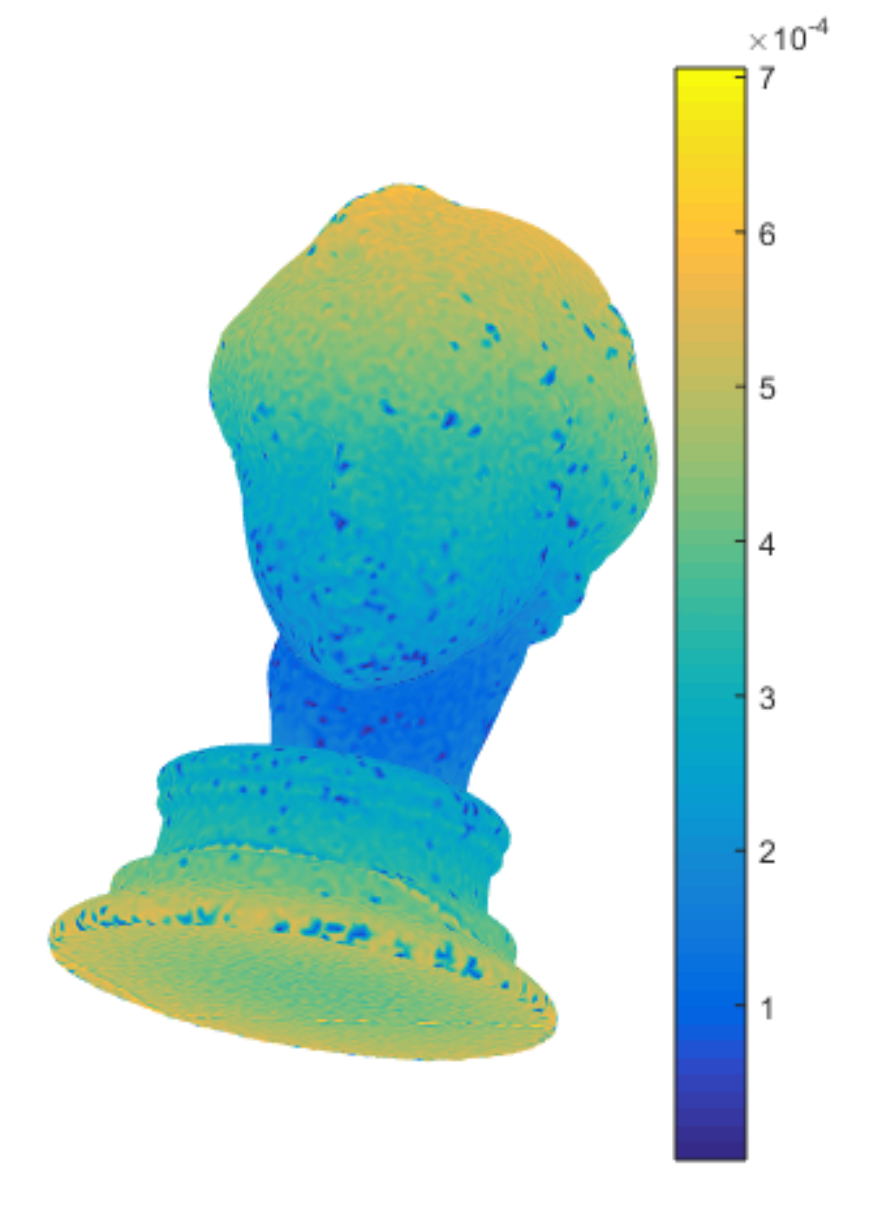}}
  \caption{Stego-objects and the visualization of differences in the detection of features used for steganalysis.
  (a), (f) and (k) are the stego-objects obtained after using the information hiding
  algorithms described in \cite{yang2016watermarking}, \cite{cho2007oblivious} and \cite{chao2009high}, respectively; (b), (g) and (l) show the absolute differences of vertex normals $\phi_{11}$ between those stego-objects and their corresponding cover-object, respectively; (c), (h) and (m) for the curvature
  ratios $\phi_{13}$; (d), (i) and (n) for the azimuth angle $\phi_{14}$; (e), (j) and (o) for the radial distance $\phi_{16}$.}
  \label{311-FeatureDiff}
\end{figure*}

\begin{figure*}[htbp]
  \centering
  \subfigure[]{
  \label{fig5:subfig:a} 
  \includegraphics[width=1.2in]{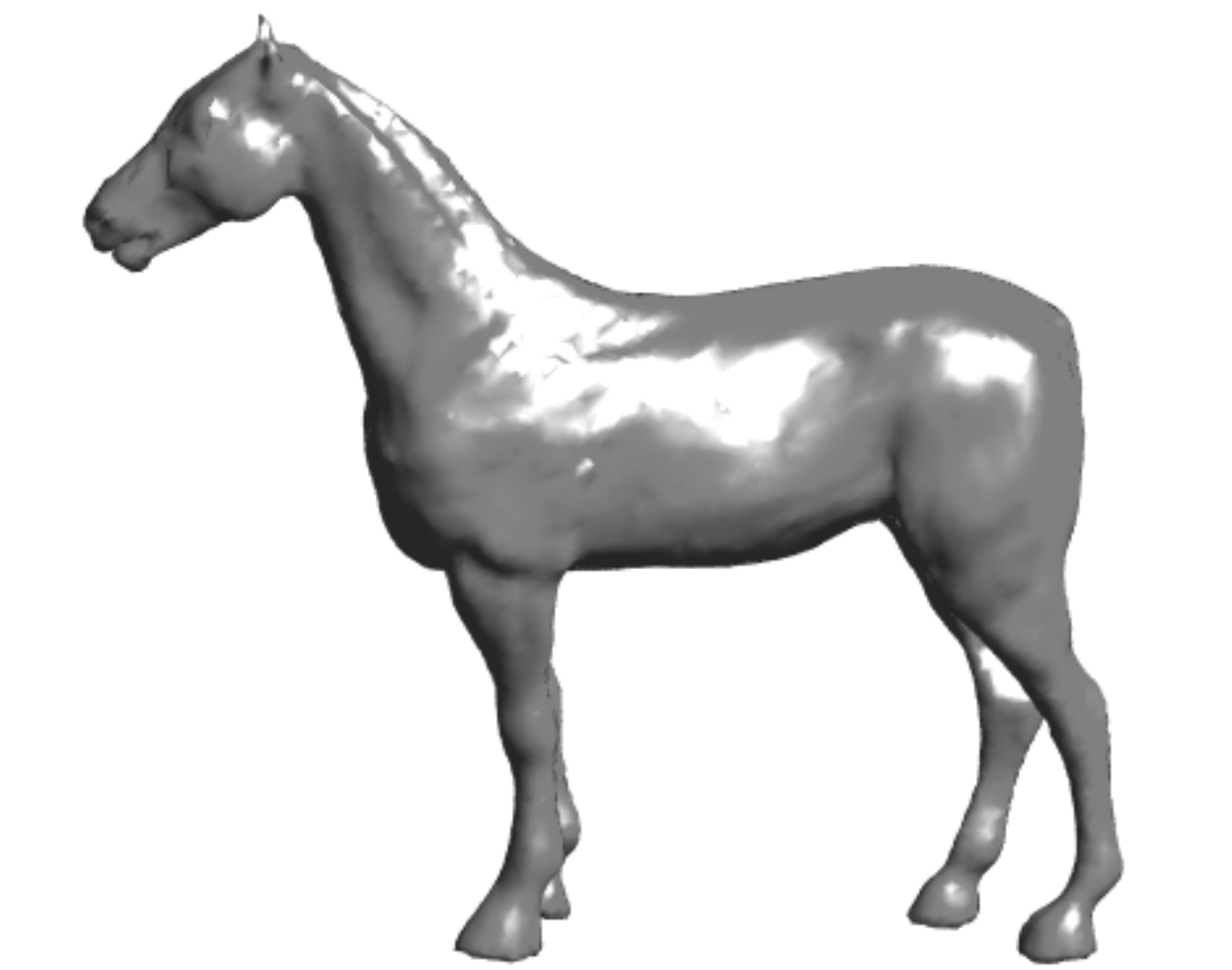}}
  \subfigure[]{
  \label{fig5:subfig:b} 
  \includegraphics[width=1.2in]{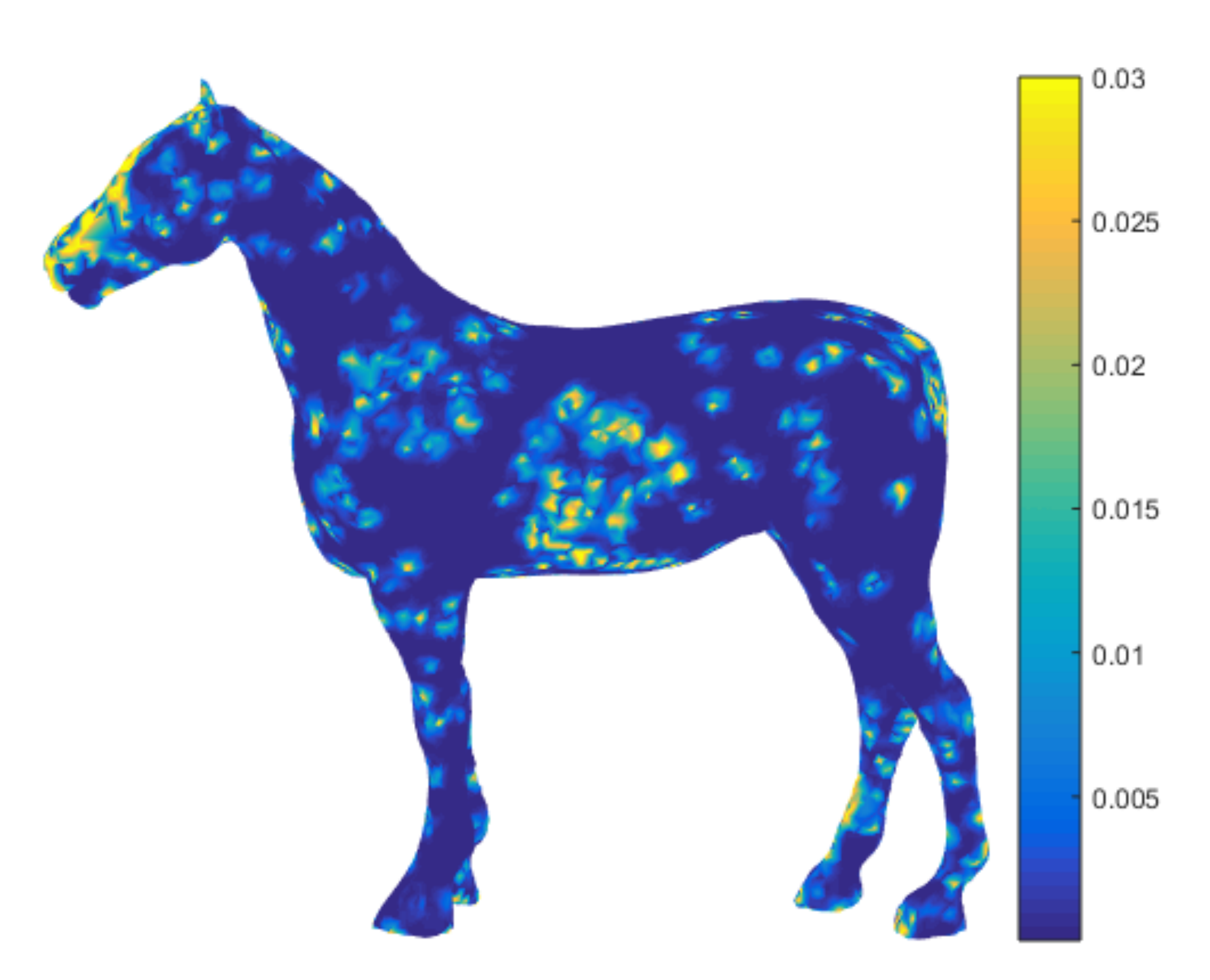}}
  \subfigure[]{
  \label{fig5:subfig:c} 
  \includegraphics[width=1.2in]{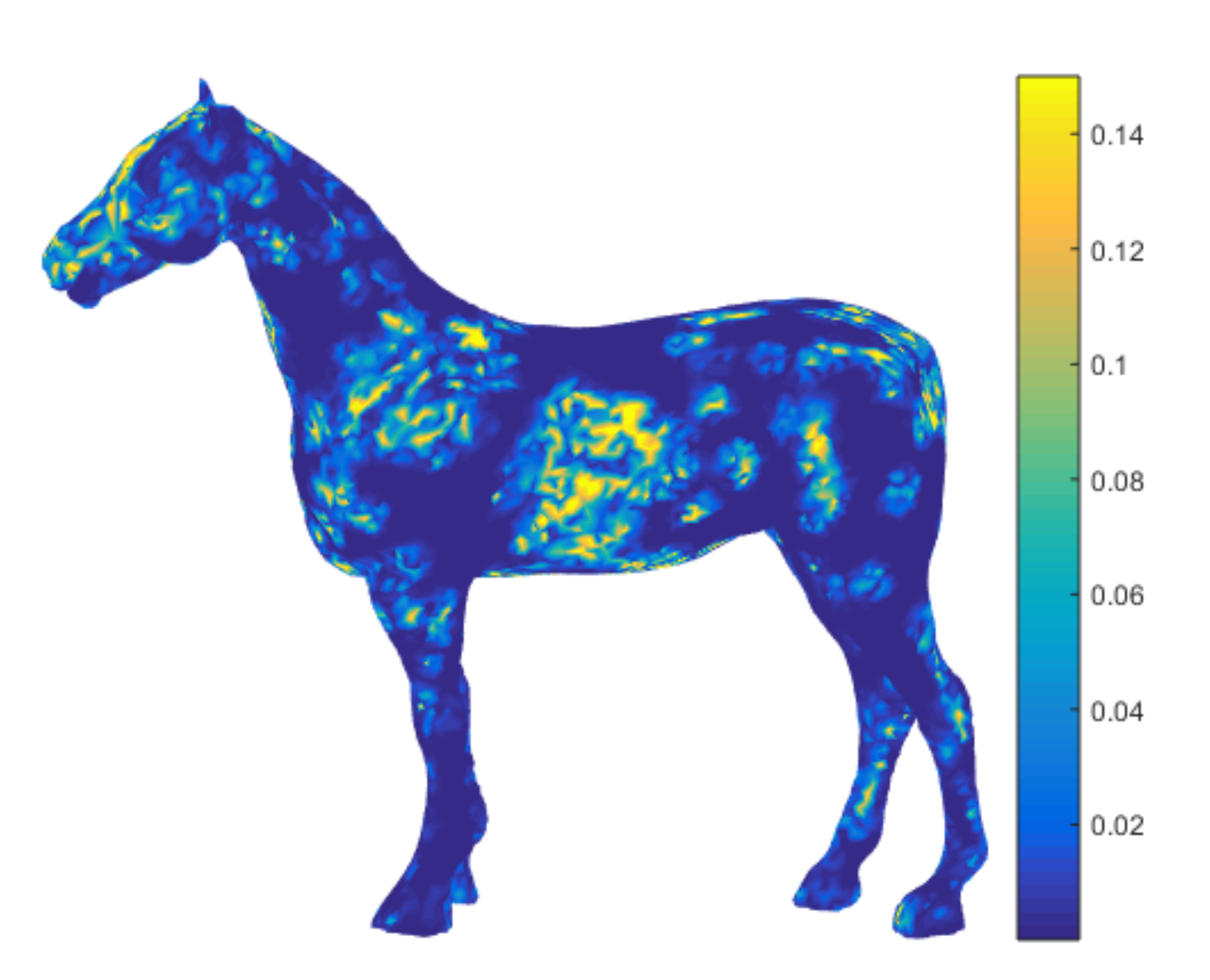}}
  \subfigure[]{
  \label{fig5:subfig:d} 
  \includegraphics[width=1.2in]{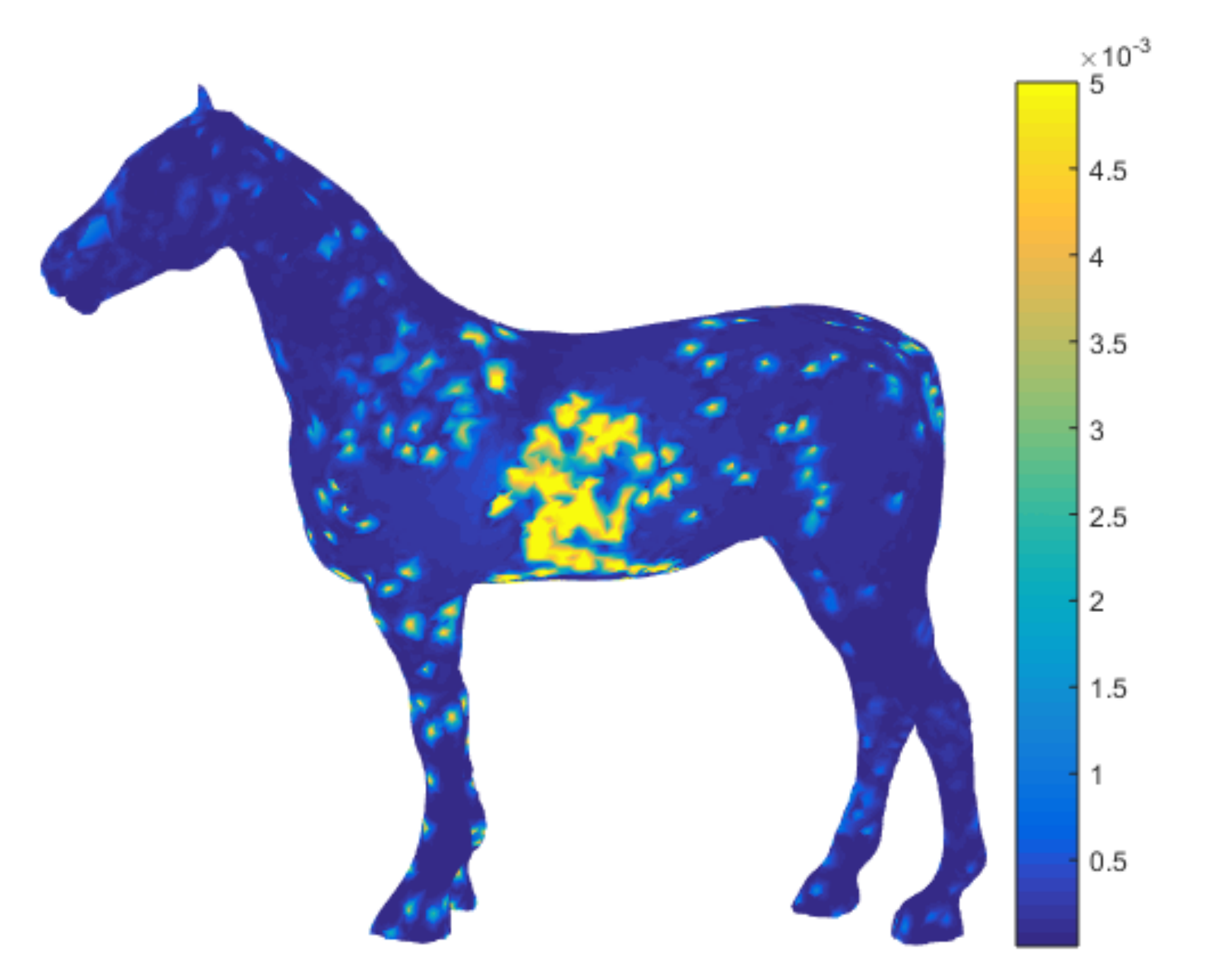}}
  \subfigure[]{
  \label{fig5:subfig:e} 
  \includegraphics[width=1.2in]{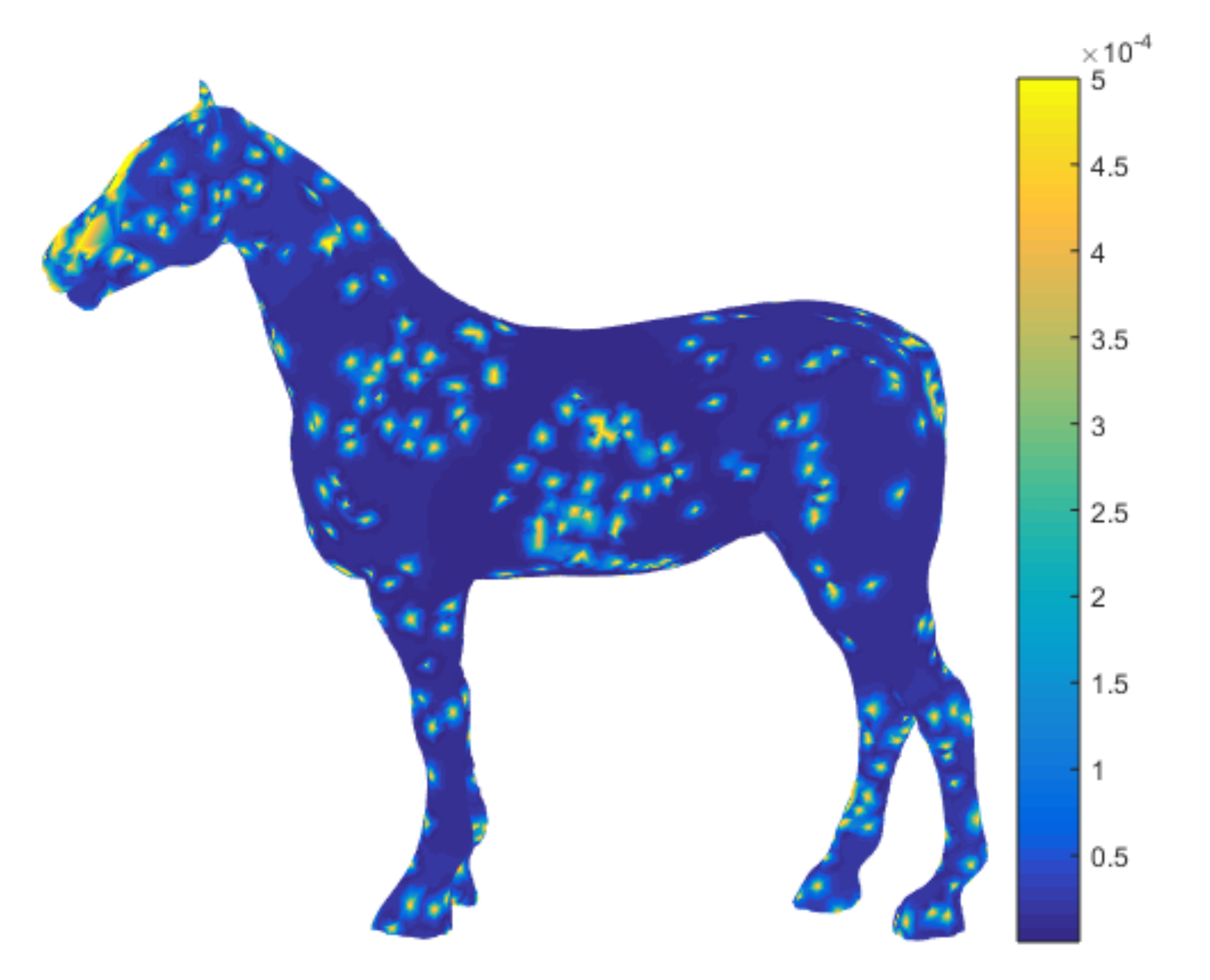}}
  \caption{Stego-object and the visualization of differences in the detection of features used for steganalysis.
  (a) The stego-object obtained after using the information hiding algorithms described in \cite{yang2016watermarking};
  (b) The absolute differences of vertex normals $\phi_{11}$ between the stego-objects and their corresponding cover-object; The
  absolute differences in (c) for the curvature
  ratios $\phi_{13}$; (d) for the azimuth angle $\phi_{14}$; (e) for the radial distance $\phi_{16}$.}
  \label{387-FeatureDiff}
\end{figure*}

During the generation of the stego-objects using the steganalysis-resistant watermarking method proposed in \cite{yang2016watermarking}, we consider multiple values for the parameter $K$ which determines the number of bins in the histogram of the radial distance parameter of the vertex where information is embedded. According to \cite{yang2016watermarking}, the upper bound of the embedding capacity is $\lfloor(K-2)/2\rfloor$. In our experiments we set the parameter $K=\{ 32, 64, 96, 128 \}$
and thus obtain multiple sets of stego-objects. Another parameter in the watermarking method from \cite{yang2016watermarking} is $n_{thr}$ which controls the robustness of the embedding method. In order to keep the distortion of the embedding to a relatively low level, we set the parameter $n_{thr}$ as 20. If the smallest number of the elements in
the bins from the objects is less than 20, we would choose $n_{thr}$ equal to the smallest nonzero number of the elements in the bins. Examples of stego-objects obtained using
the embedding method in \cite{yang2016watermarking} are shown in Figure~\ref{311-FeatureDiff}(a) and Figure~\ref{387-FeatureDiff}(a), where $K=128$. The absolute differences
of the vertex normals, the curvature ratios, the azimuth angles and the radial distances between the stego-object and its corresponding cover-object, representing the features
$\phi_{11}$, $\phi_{13}$, $\phi_{14}$ and $\phi_{16}$, detected on these stego-objects are shown in Figures~\ref{311-FeatureDiff} (b), (c), (d) and (e), and Figures~\ref{387-FeatureDiff} (b), (c), (d) and (e), for the two objects ``Head statue'' and ``Horse,'' respectively.

\begin{figure*}[htbp]
  \centering
  \subfigure[QDA for \cite{yang2016watermarking}]{
  \label{fig:subfig:a}
  \includegraphics[width=1.8in]{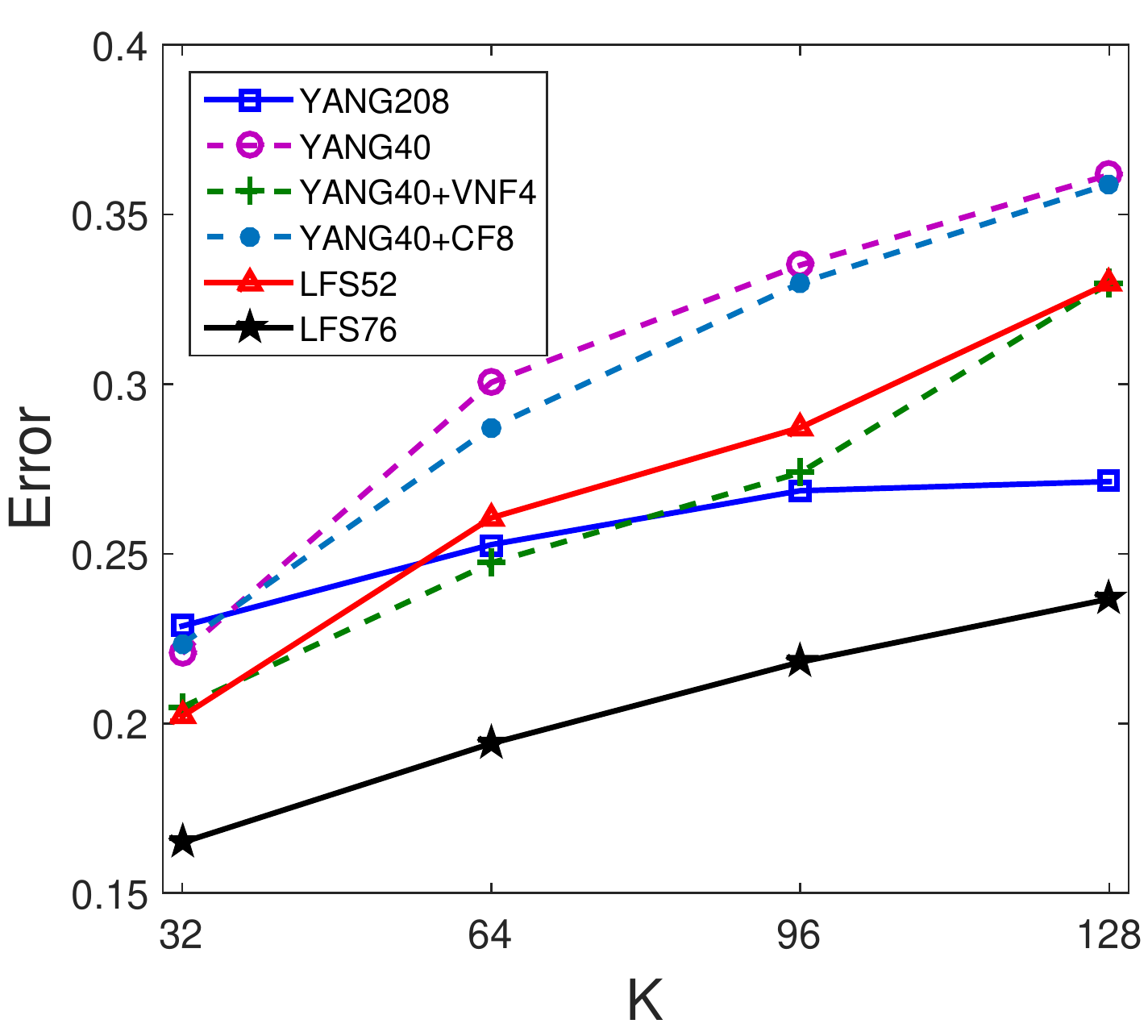}}
  \centering
  \subfigure[SVM for \cite{yang2016watermarking}]{
  \label{fig:subfig:b}
   \includegraphics[width=1.8in]{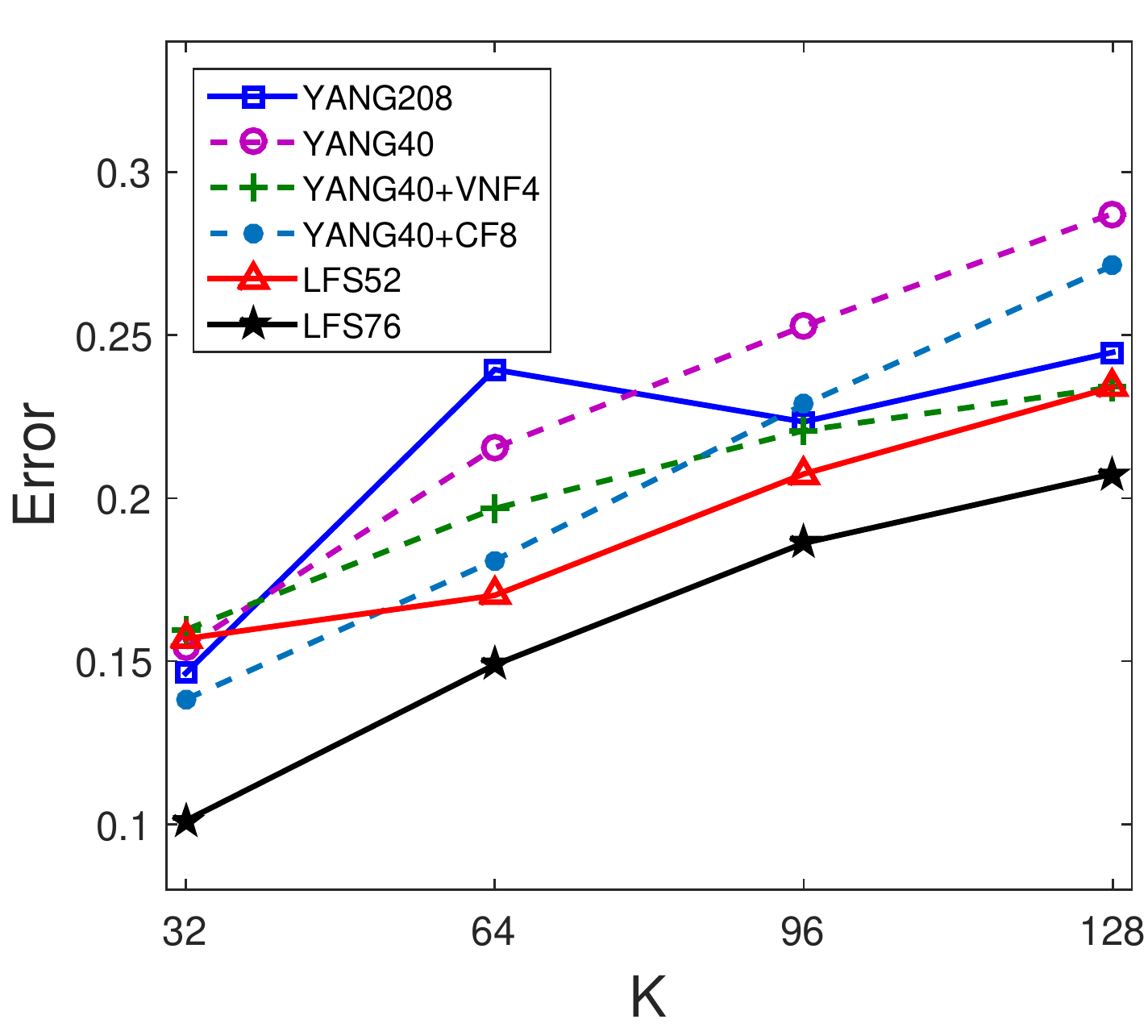}}
  \subfigure[FLD ensemble for \cite{yang2016watermarking}]{
  \label{fig:subfig:b}
  \includegraphics[width=1.8in]{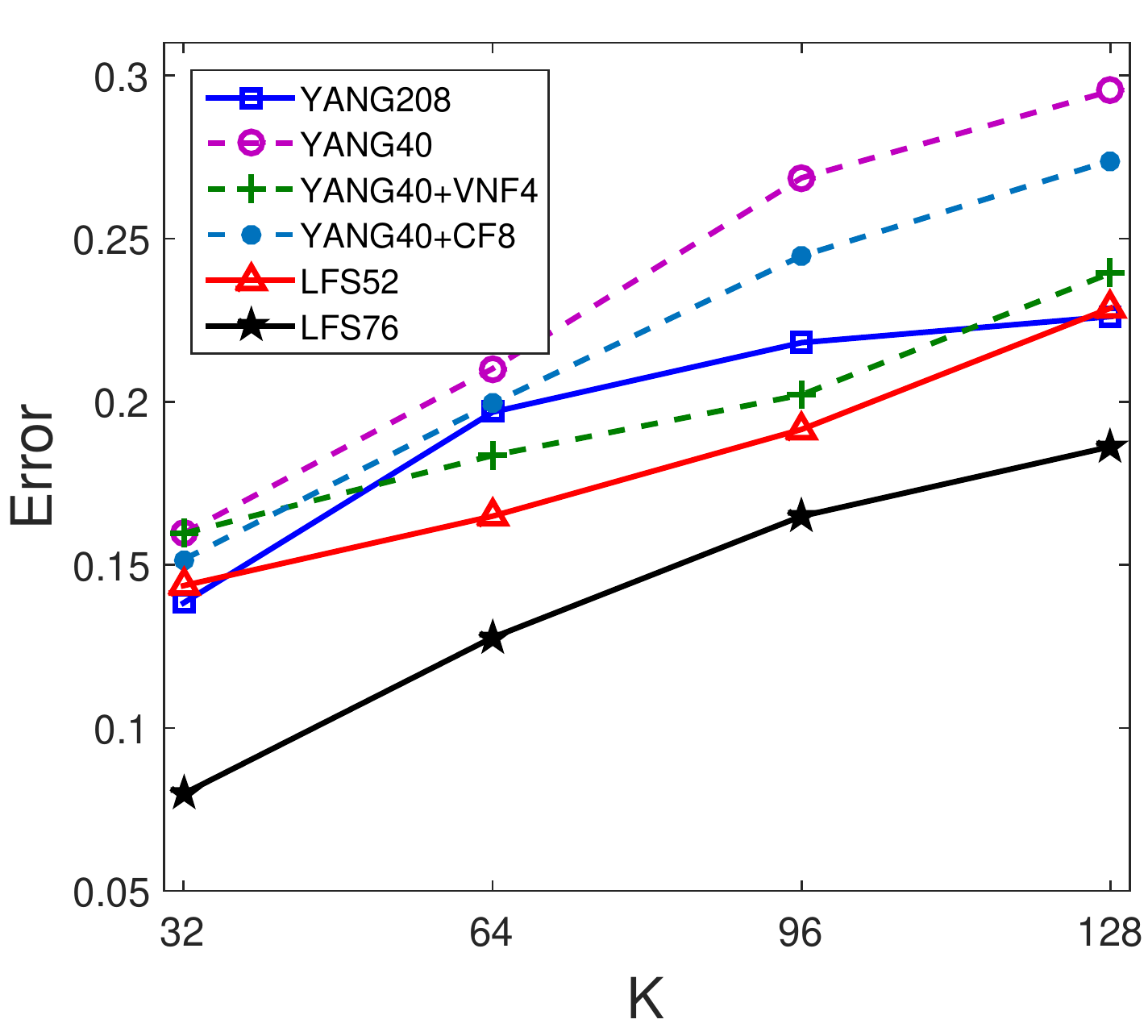}}
  \caption{Median value of detection errors of the steganalyzers trained as QDA classifiers, SVM classifiers and
  FLD ensembles on the testing set over 30 independent splits for the watermarking method proposed
  in \cite{yang2016watermarking} with different values for parameter K.}
  \label{yang16_K}
 \end{figure*}

Figure~\ref{yang16_K} shows the detection errors for the watermarking method \cite{yang2016watermarking} using the three steganalyzers, QDA, SVM and FLD ensembles, trained with the six combinations of feature sets, formed as mentioned above. It can be seen from Figure~\ref{yang16_K} that the LFS76 shows best performance among the six combinations, when using any of the three machine learning methods. We have observed that as the value of $K$ increases, the detection error tends to increase as well. This happens because the larger $K$ will lead to fewer elements in each bin,
so less vertices need to be changed for embedding a single bit.

For the mean-based watermarking method from \cite{cho2007oblivious}, we consider various values for the watermark strength
and message payload while fixing the incremental step size to $\Delta k=0.001$. An example of a stego-object obtained using
the watermarking method from \cite{cho2007oblivious} is shown in Figure~\ref{311-FeatureDiff} (f), where the watermark strength factor
is set as $\alpha=0.04$ and the message payload as 64 bits. The absolute differences of the features, $\phi_{11}$, $\phi_{13}$, $\phi_{14}$
and $\phi_{16}$, between the stego-object and its corresponding cover-object are shown in Figures~\ref{311-FeatureDiff} (g), (h), (i) and (j).
From these figures it can be observed that each feature identifies specific differences between the cover- and stego-objects, which usually do not overlap with each other.

 \begin{figure*}[htbp]
  \centering
  \subfigure[QDA for \cite{cho2007oblivious}]{
  \label{fig:subfig:a}
  \includegraphics[width=1.8in]{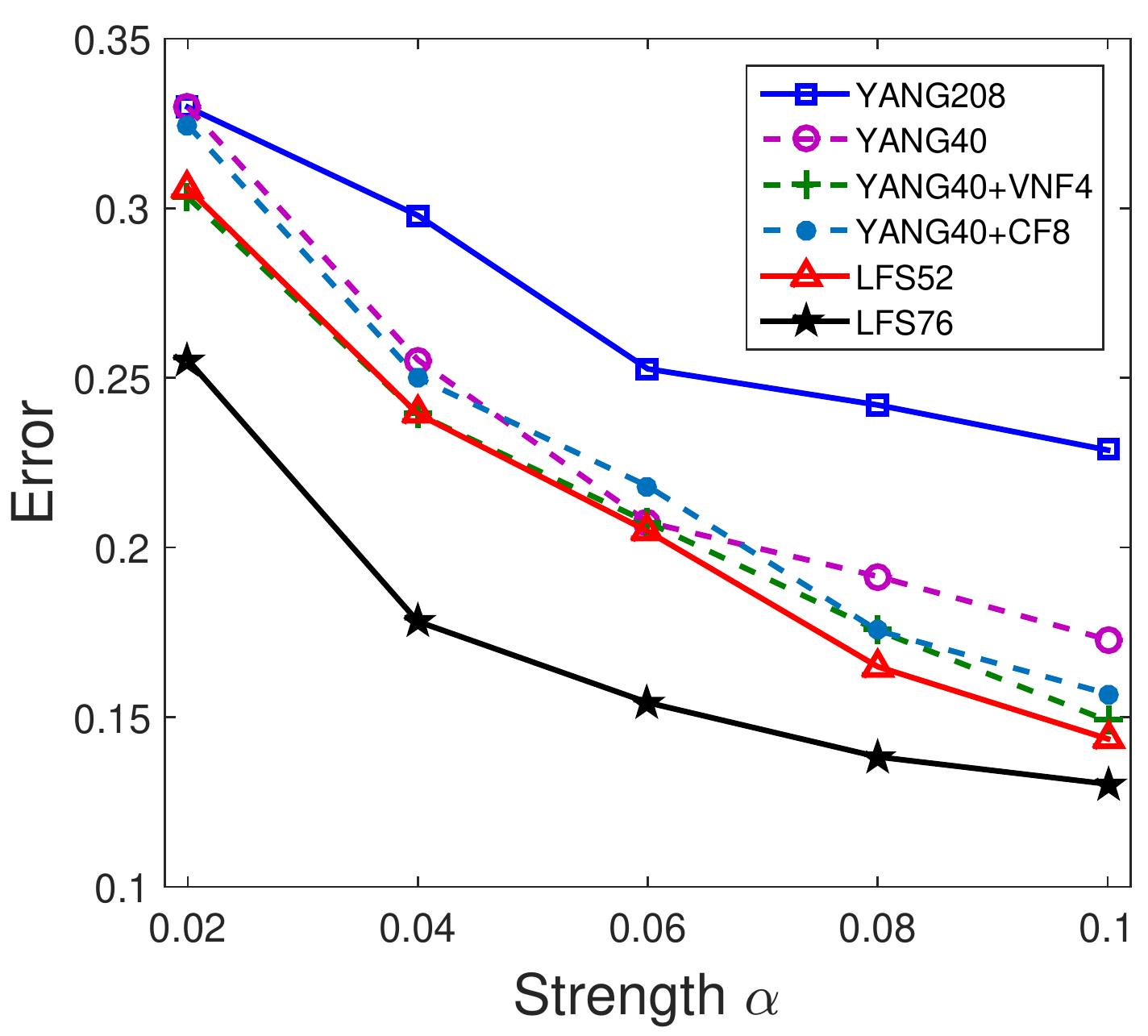}}
  \subfigure[SVM for \cite{cho2007oblivious}]{
  \label{fig:subfig:b}
  \includegraphics[width=1.8in]{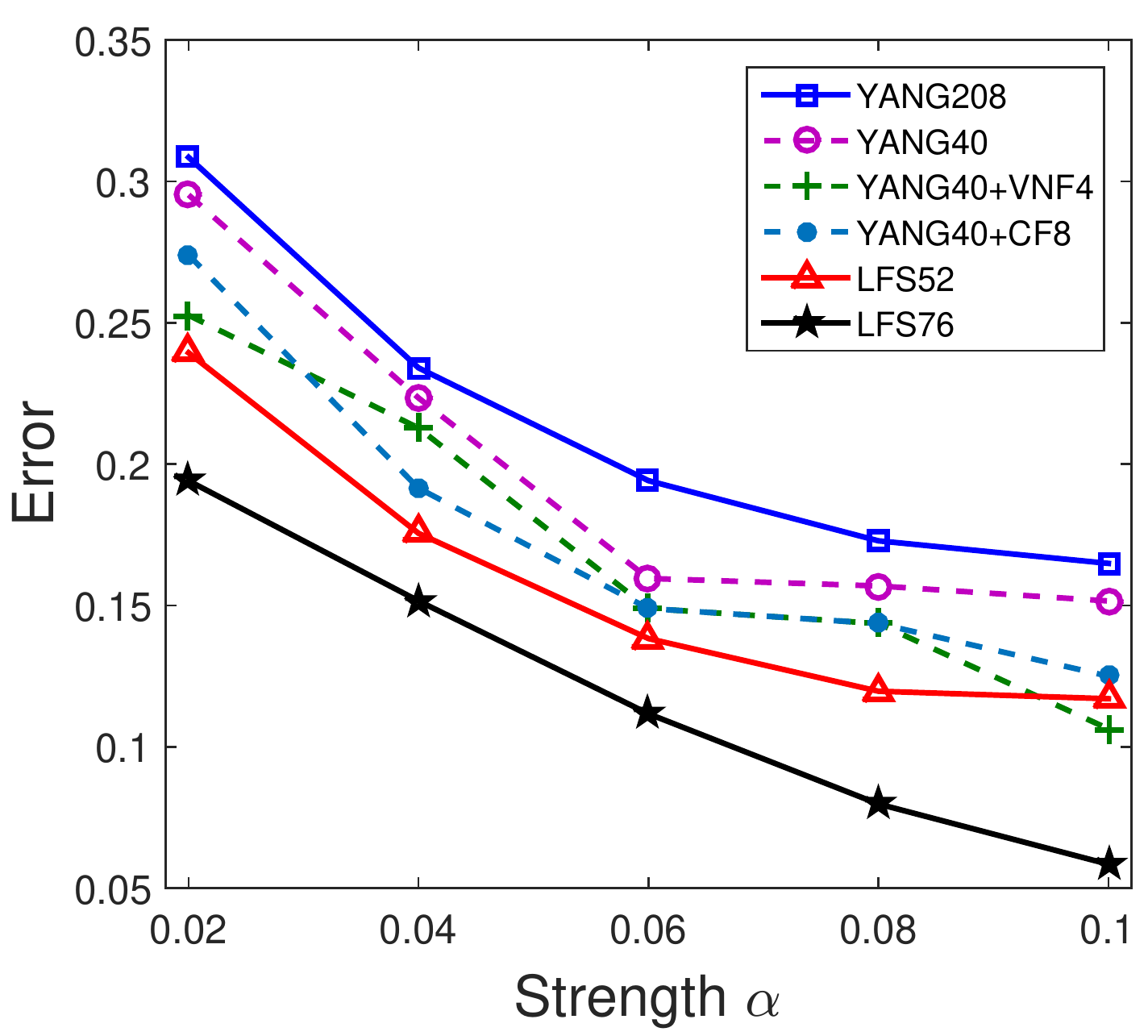}}
  \subfigure[FLD ensemble for \cite{cho2007oblivious}]{
  \label{fig:subfig:b}
  \includegraphics[width=1.8in]{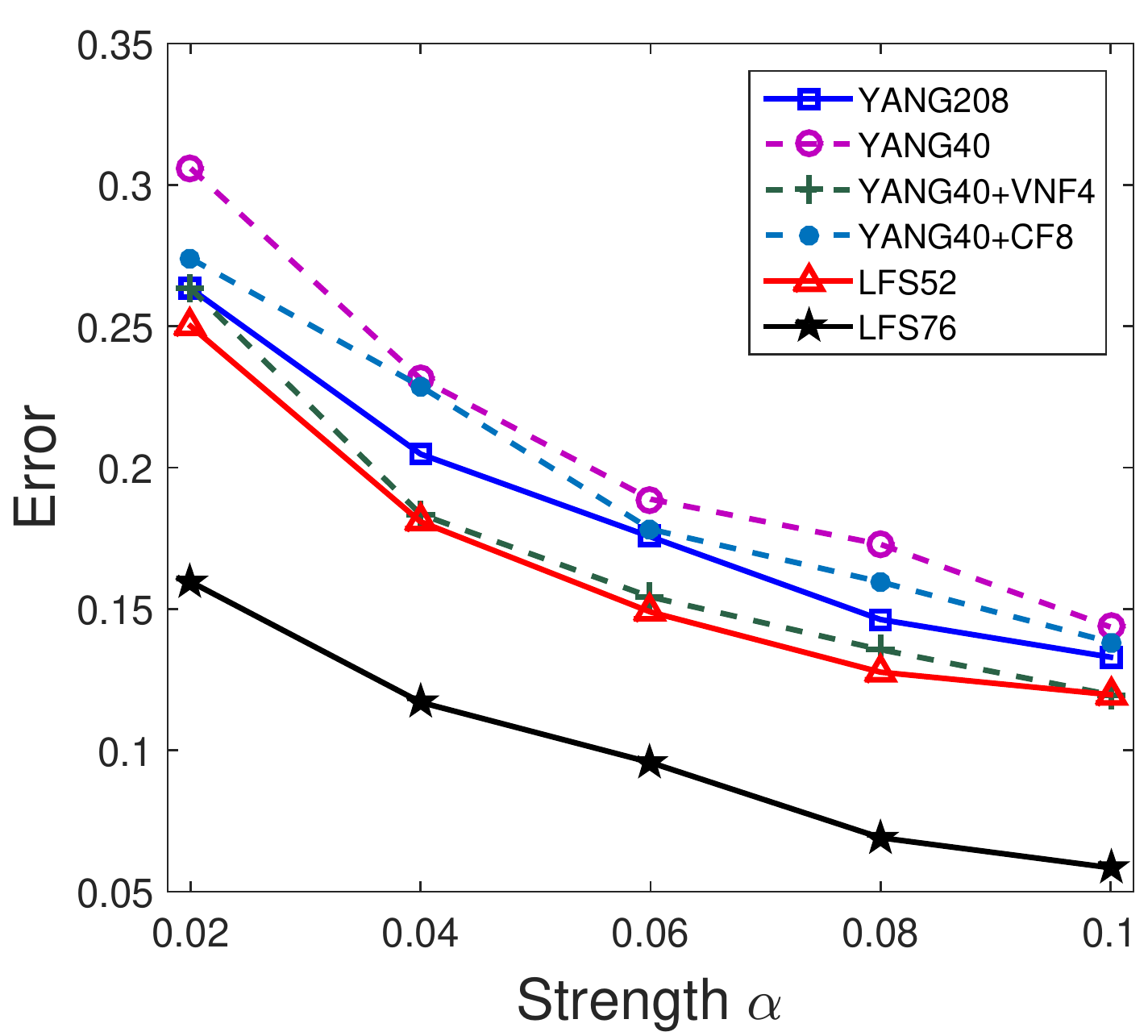}}\\
  \subfigure[QDA for \cite{cho2007oblivious}]{
  \label{fig:subfig:b}
  \includegraphics[width=1.8in]{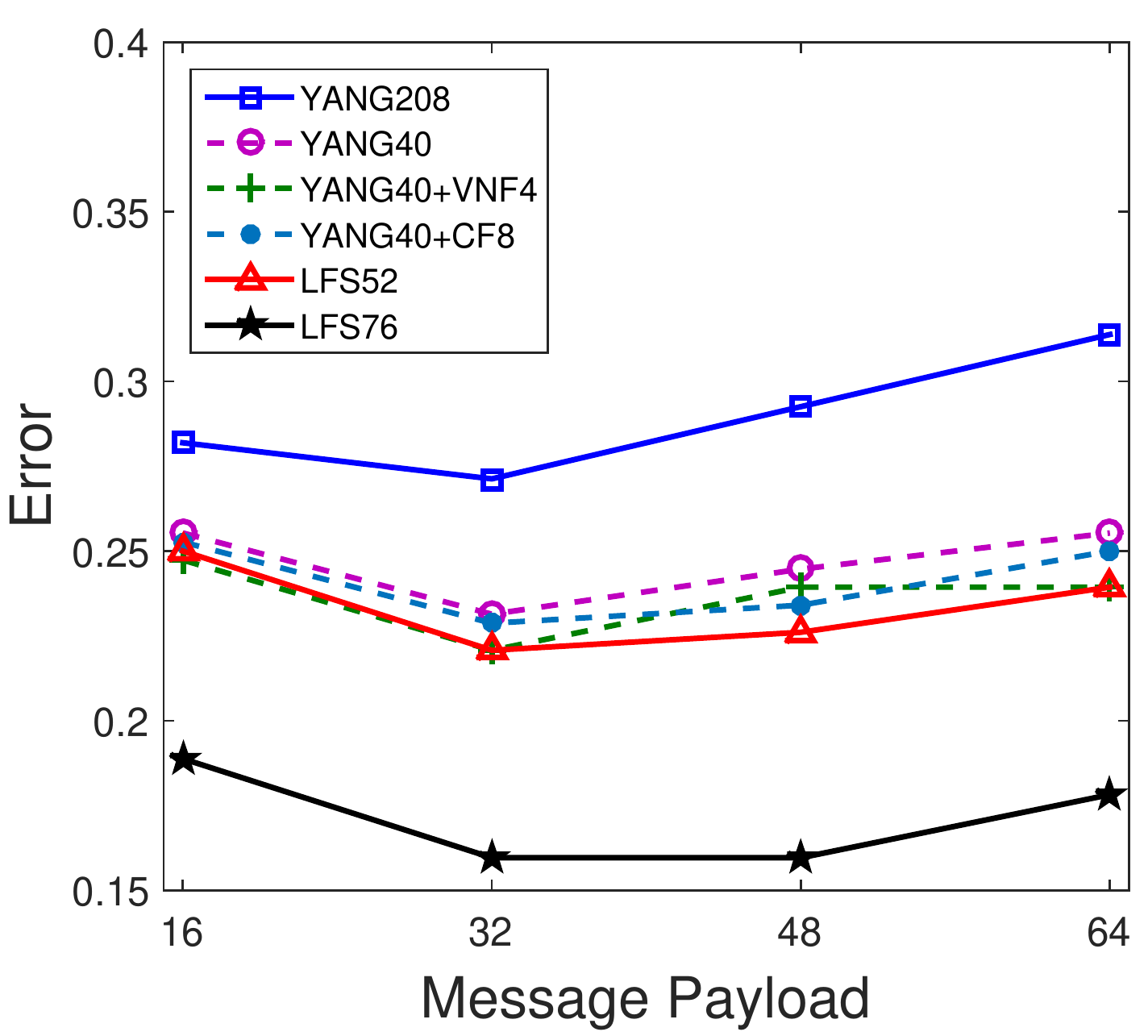}}
  \subfigure[SVM for \cite{cho2007oblivious}]{
  \label{fig:subfig:b}
  \includegraphics[width=1.8in]{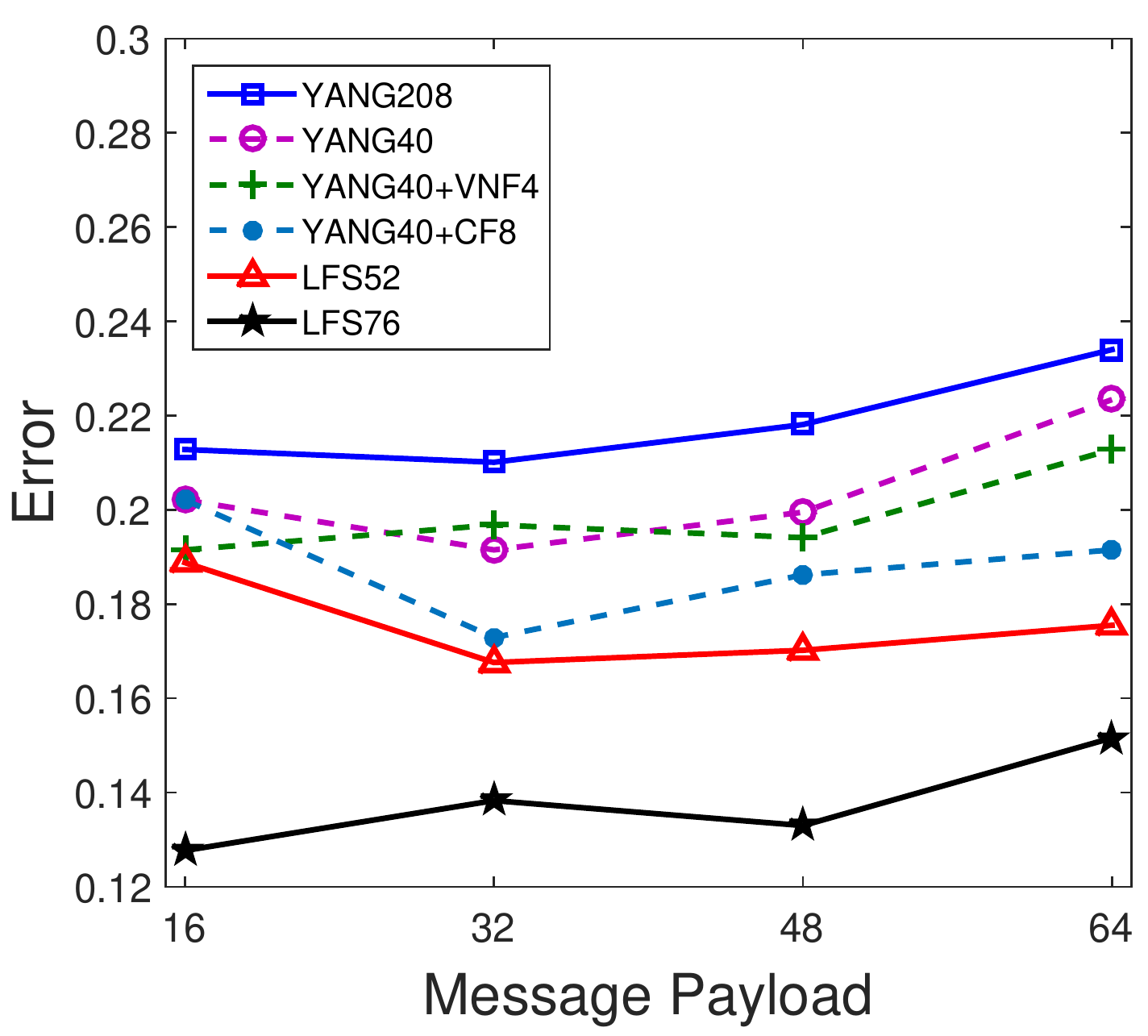}}
  \subfigure[FLD ensemble for \cite{cho2007oblivious}]{
  \label{fig:subfig:b}
  \includegraphics[width=1.8in]{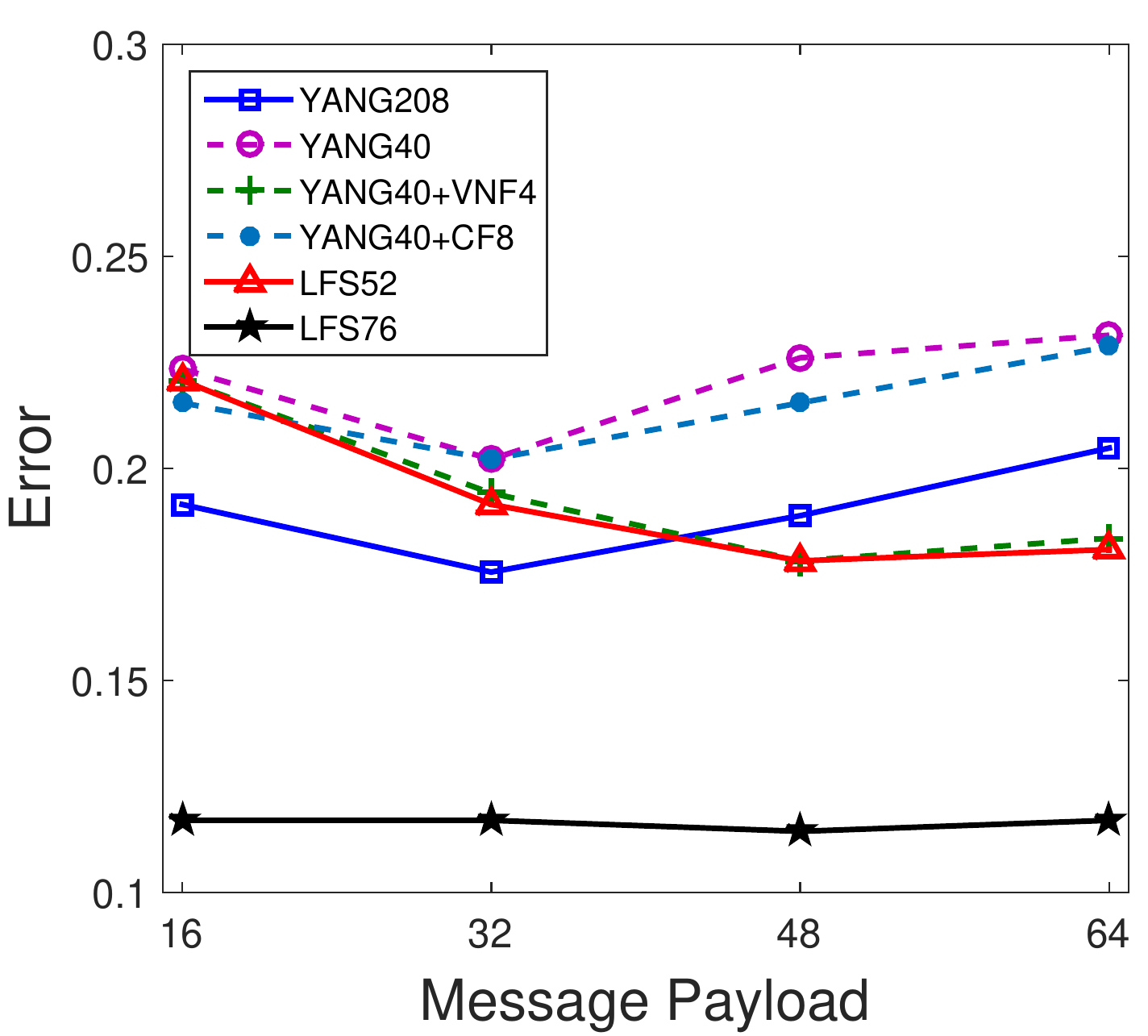}}
  \caption{Median value of the detection errors of the steganalyzers trained as QDA classifiers,
  SVM classifiers and FLD ensembles on the testing set over 30 independent splits for the watermarking
  method proposed in \cite{cho2007oblivious} when changing the watermarking strengths and the message length.}
  \label{Cho07-strength-length}
 \end{figure*}

\begin{figure*}[htbp]
  \centering
  \subfigure[QDA for \cite{chao2009high}]{
  \label{fig:subfig:a}
  \includegraphics[width=1.8in]{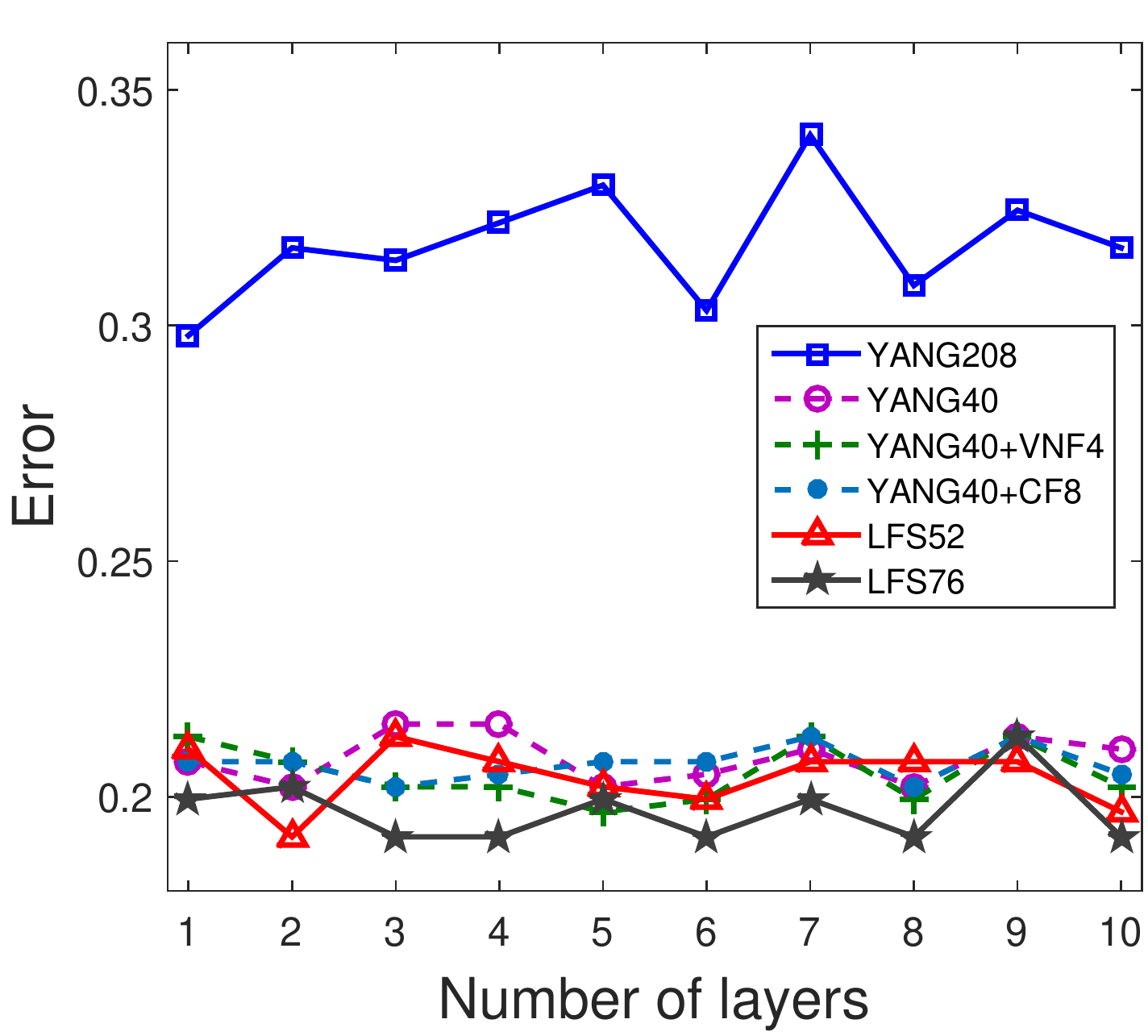}}
  \centering
  \subfigure[SVM for \cite{chao2009high}]{
  \label{fig:subfig:b}
   \includegraphics[width=1.8in]{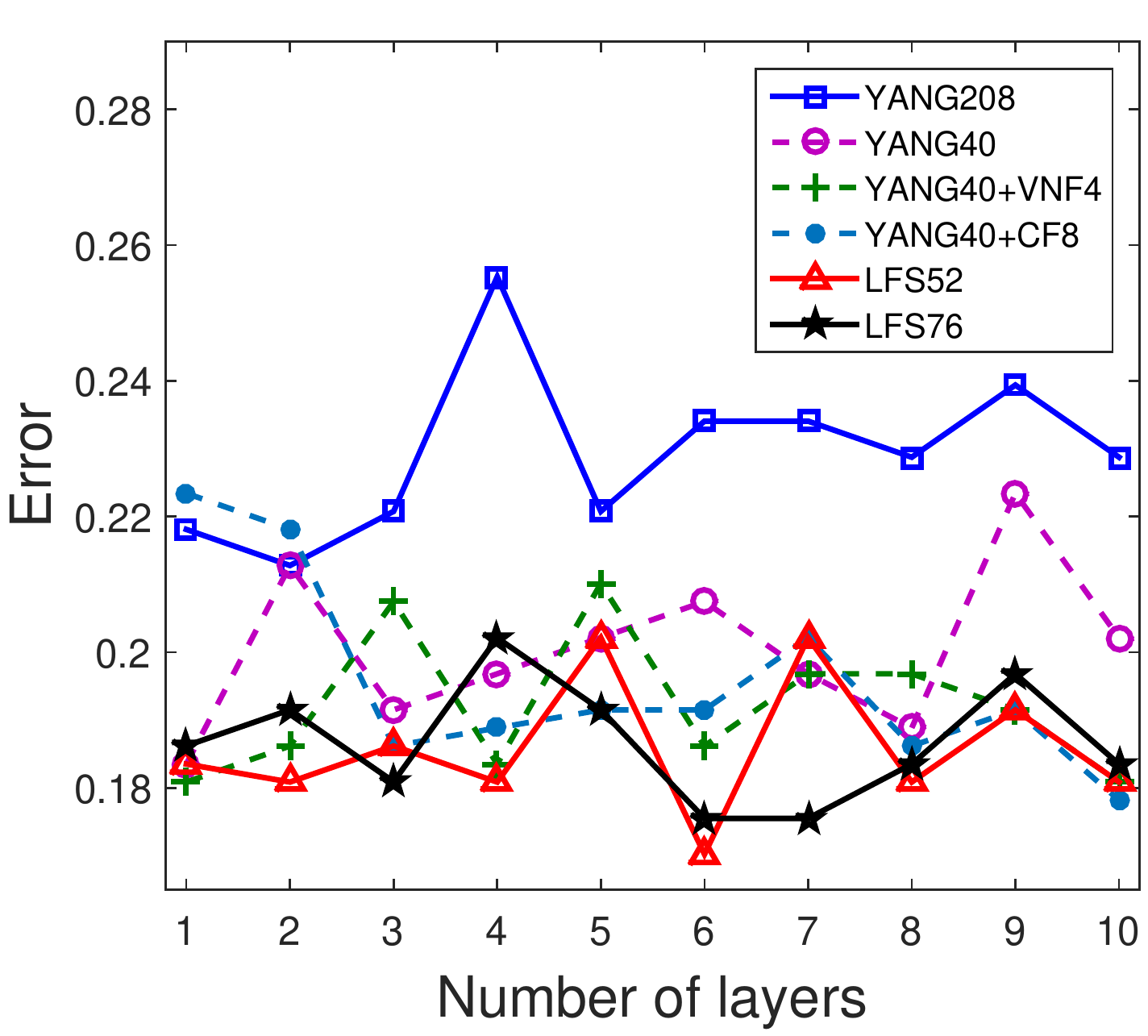}}
   \subfigure[FLD ensemble for \cite{chao2009high}]{
  \label{fig:subfig:b}
  \includegraphics[width=1.8in]{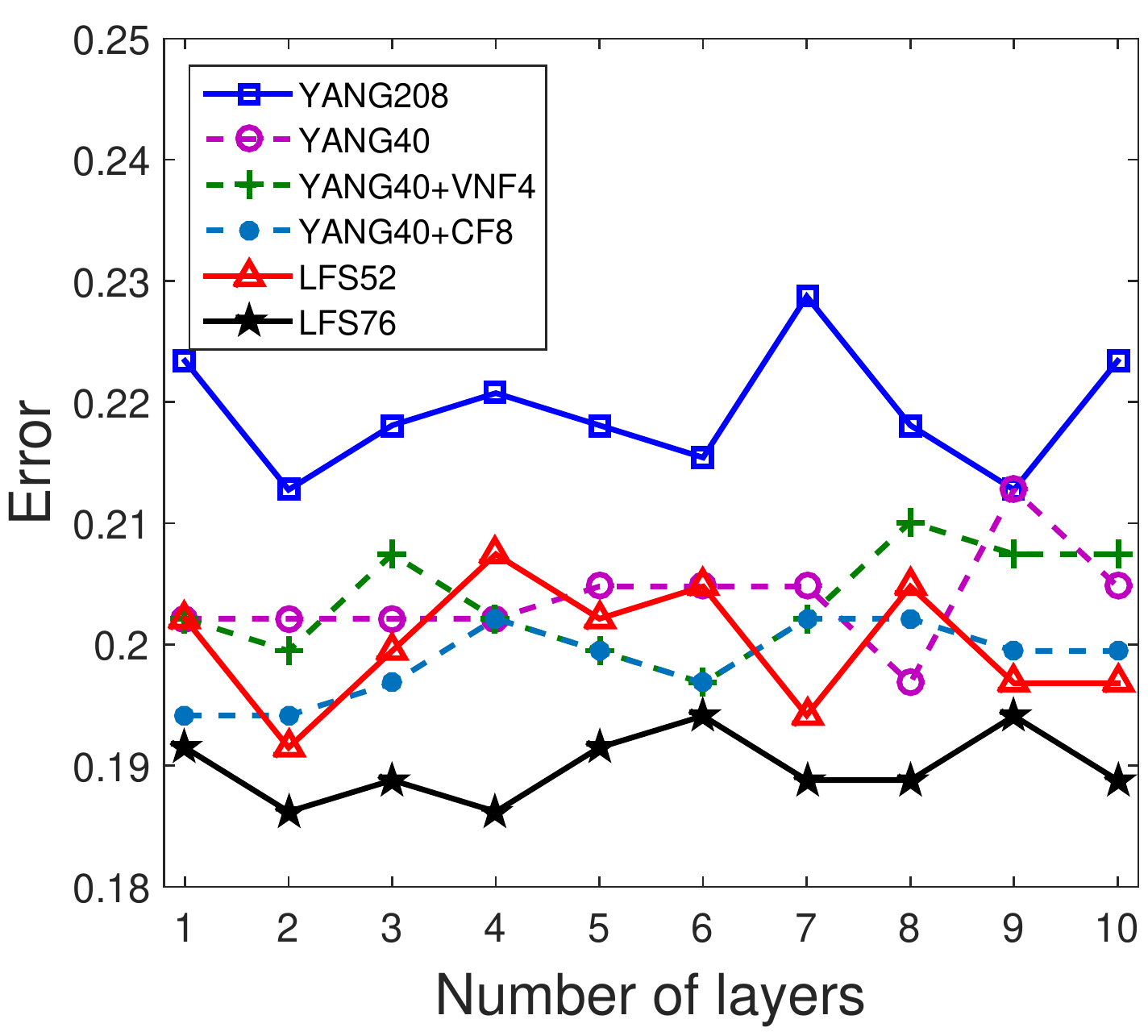}}
  \caption{Median value of detection errors of the steganalyzers trained as QDA classifiers, SVM classifiers and
  FLD ensembles on the testing set over 30 independent splits for the steganographic method proposed
  in \cite{chao2009high} when varying the number of layers.}
  \label{Chao09-1-10layers}
 \end{figure*}

Figure~\ref{Cho07-strength-length} depicts the median value of the detection errors of the watermarking
algorithm proposed in \cite{cho2007oblivious} using the steganalyzers trained as QDA classifiers, FLD ensembles and SVM classifiers over 6 different feature combinations, chosen as described above, and applied on the testing set for 30 independent data splits. In Figures \ref{Cho07-strength-length} (a), (b) and (c) we show
the results when the watermarking strengths are 0.02, 0.04, 0.06, 0.08 and 0.1, while the message length is fixed to 64 bits.
From these figures we can observe that as the watermarking strength increases,all steganalyzers provide better detection accuracy.
This is due to the fact that more significant changes are produced in the 3D object surface, by watermarks that have stronger embedding parameters.
Comparing the feature sets, it is evident that YANG40 has better performance than YANG208 when using the QDA and SVM
classifiers. Although YANG40 is a simplified version of YANG208, it preserves the most effective feature subsets in YANG208 and reduces the dimensionality in order to avoid overfitting. Combining either VNF4 or CF8 with YANG40 would get better performance than just by using YANG40. After adding the spherical
coordinate features to the features set LFS52, which was used in \cite{Li2016lfs}, the proposed LFS76 feature set achieves the best
steganalysis performance. In Figures~\ref{Cho07-strength-length} (b), (d) and (f) we show the results when increasing the message payload
from 16 to 32, 48 and 64 bits, while keeping the watermarking strength as 0.04.
The LFS76 feature set provides much better detection results than the other feature sets in all the cases when testing the steganalyzers.

When using the high-capacity steganographic method from \cite{chao2009high} we increase the number
of layers from 1 to 10, and we consider the number of intervals as 10000.
Increasing the number of embedding layers in this steganographic method corresponds to increasing
the payload capacity. During the embedding, all the vertices in the mesh are used as payload carriers,
except for three vertices which are used as references for the extraction process. Examples of a
stego-object obtained using the steganographic method from \cite{chao2009high} is shown in
Figure~\ref{311-FeatureDiff} (k), where the number of layers is 10.
Absolute differences of the features, $\phi_{11}$, $\phi_{13}$, $\phi_{14}$ and $\phi_{16}$, between the stego-object
and its corresponding cover-object are shown in Figures~\ref{311-FeatureDiff} (l), (m), (n) and (o).

The results provided by the three steganalyzers, using the QDA classifier, FLD ensemble and SVM, when
increasing the number of layers are provided in the plots from Figures~\ref{Chao09-1-10layers} (a), (b) and (c).
From these plots it can be observed that the proposed set of features LFS76 used for training FLD ensemble provides the best
results for any number of layers used for embedding. When the steganalyzers are trained as QDA classifiers, the LFS76
provides best results in most cases, except when the number of layers is 2, 5, or 9. When the steganalyzers are trained as SVM classifiers,
the feature set LFS76's performance is similar to
that achieved when using LFS52. It can be observed that the advantage of LFS76 with respect to LFS52 in detecting the steganographic method from \cite{chao2009high} is
not that high as that achieved in detecting the changes produced by the two watermarking methods from \cite{yang2016watermarking} and \cite{cho2007oblivious}.
This is because the embedding method from \cite{chao2009high} does not produce the modifications in the spherical coordinate system,
which makes the spherical coordinate features less useful for detecting the changes produced by embedding in this case. Another interesting
point is that the detection error for \cite{chao2009high} does not decline when the embedding capacity increases. The reason for this is that, according to
the multi-layer embedding framework applied in \cite{chao2009high}, the distortions produced to the objects are always controlled during the embedding.

\subsection{Analysing the efficiency of features for steganalysis}

In order to investigate the contribution of different categories of features from the set LFS76 to the steganalysis, we use the relevance between the feature vectors and the class label in order to assess each feature's importance. The measurement of the relevance is addressed by using the Pearson correlation coefficient,
\begin{equation}
  \rho ({\bf\Phi}_{i},\mathbf{Y}) =\frac{cov({\bf \Phi}_{i},\mathbf{Y})}{ \sigma^{}_{{\bf \Phi}_{i}}\sigma^{}_{\mathbf{Y}}}
  \label{relevance}
\end{equation}
where ${\bf\Phi}_{i}$ is the $i$-th feature of a given feature set, ${\bf \Phi}=\{{{\bf \Phi}_{i} }| i=1,2, \ldots ,N\}$, where $N$ is the dimensionality
of the input feature,  $\mathbf{Y}$ is the class label indicating whether the class corresponds to a cover or a stego object, $cov$ represents the covariance and
$\sigma^{}_{{\bf \Phi}_{i}}$ is the standard deviation of ${\bf \Phi}_{i}$. The Pearson correlation coefficient is well known as a measure of the linear dependence between two variables \cite{Hall99correlation-basedfeature}. Then we set $|\rho ({\bf\Phi}_{i},\mathbf{Y})|$ as the value of the relevance, where $|\rho ({\bf\Phi}_{i},\mathbf{Y})|=1$ indicates a highly linear relationship between the feature and the class label, corresponding to a better discriminant ability of that feature.

The analysis is conducted on the features extracted from the 354 cover objects used above and three sets of corresponding stego objects which are produced by the watermarking and steganographic algorithms
from  \cite{yang2016watermarking}, \cite{cho2007oblivious} and \cite{chao2009high}, respectively. We set the parameter $K$ in the steganalysis-resistant
watermarking algorithm from \cite{yang2016watermarking} as 128. For the watermarking method from \cite{cho2007oblivious}, in order to find a balance between the watermarking strength and its undetectability,
we set the watermarking strength as 0.04 and embed a payload of 64 bits. In the case when using
the steganographic method from \cite{chao2009high}, we consider ten layers of embedding.

\begin{figure*}[!htbp]
  \centering
  \subfigure[Result for \cite{yang2016watermarking} ]{
  \label{fig:subfig:a}
  \includegraphics[width=1.8in]{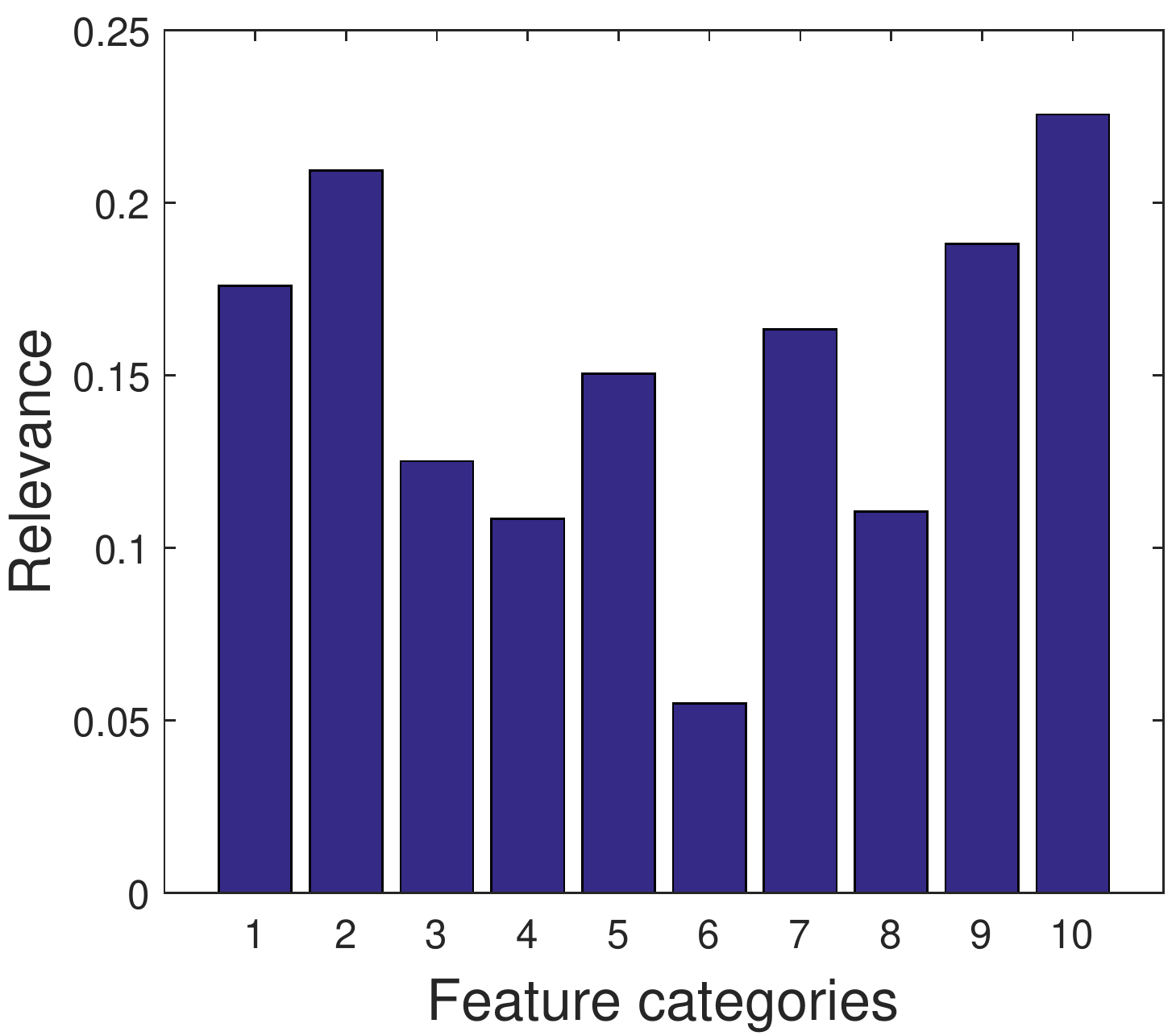}}
  \subfigure[Result for \cite{cho2007oblivious}]{
  \label{fig:subfig:b}
  \includegraphics[width=1.8in]{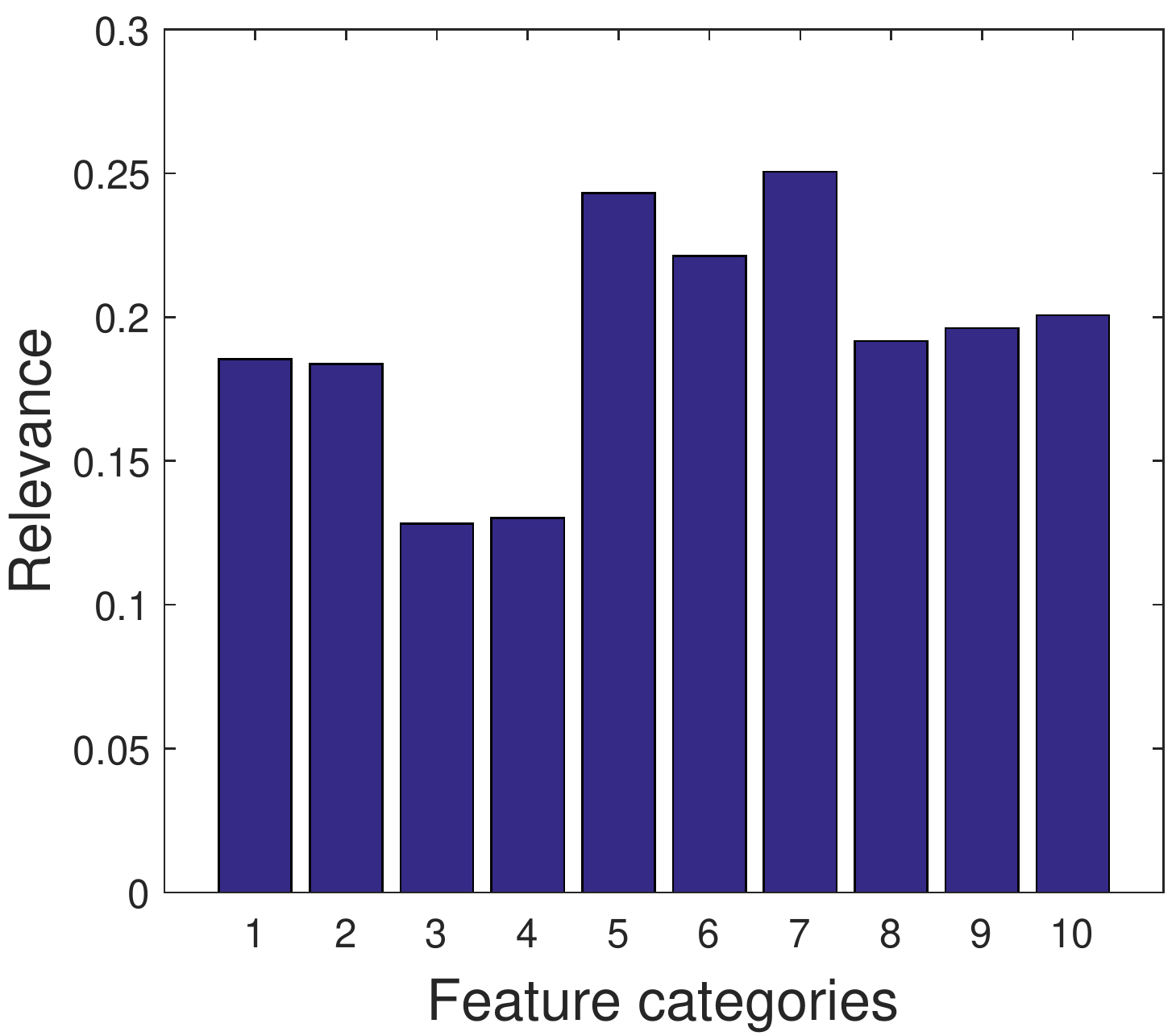}}
  \subfigure[Result for \cite{chao2009high}]{
  \label{fig:subfig:c}
  \includegraphics[width=1.8in]{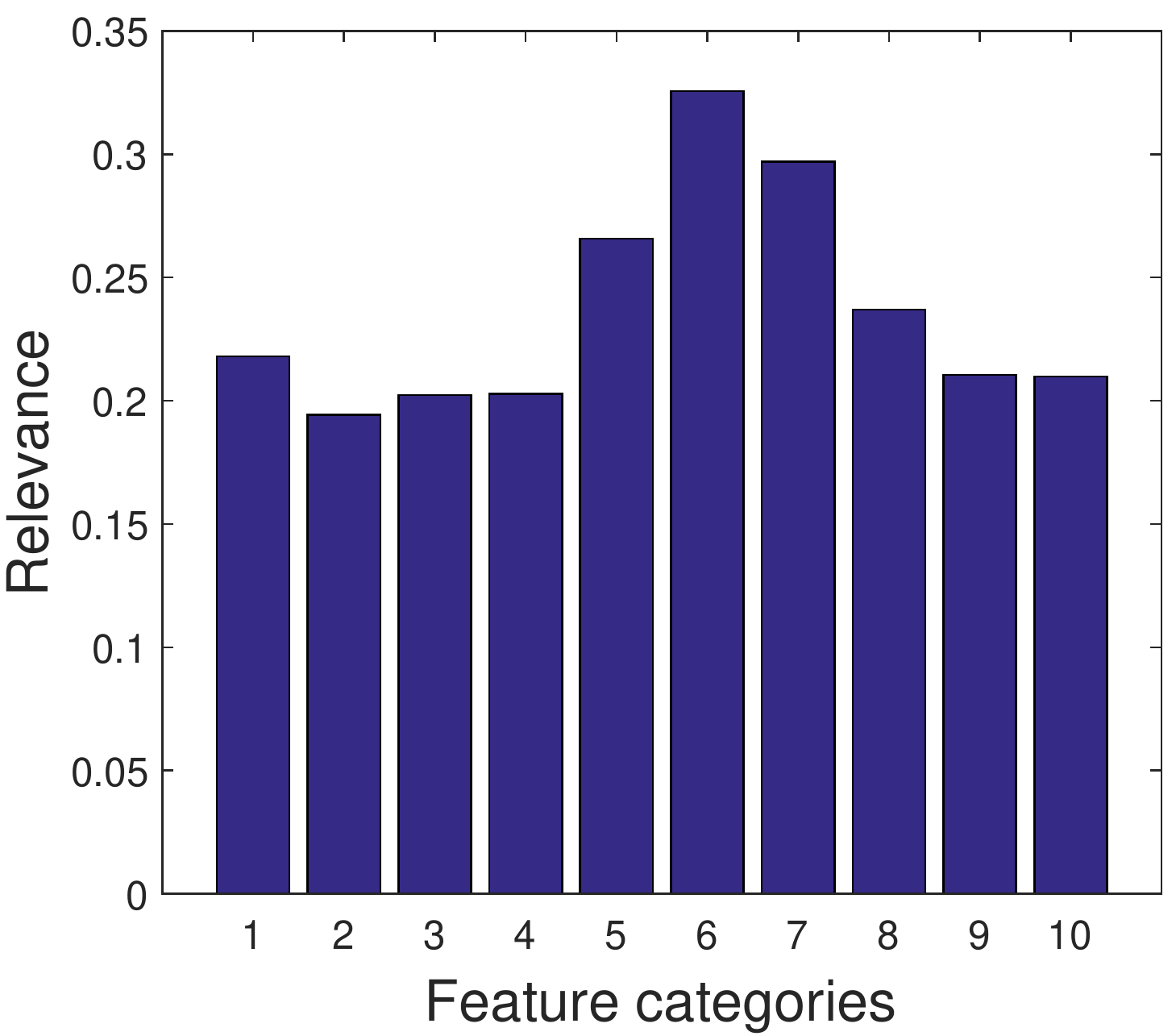}}
  \caption{The relevance between the features and the class label, cover (0) or stego (1), where the stego objects are generated by the three information hiding methods \cite{yang2016watermarking}, \cite{cho2007oblivious} and \cite{chao2009high}, respectively. The meaning of the category labels in (a), (b) and (c) are: 1, the vertex position in Cartesian coordinate system; 2, the vertex norm in Cartesian coordinate system; 3, the vertex position in Laplace coordinate system; 4, the vertex norm in Laplace coordinate system; 5, the face normal; 6, the dihedral angle; 7, the vertex normal; 8; the curvature; 9, the vertex position in spherical coordinates system; 10, the edge length in spherical coordinate system.}
  \label{rele-feature}
 \end{figure*}

We split the features from the set LFS76 into 10 categories according to their representation of the local shape geometry: 1, the vertex position in the Cartesian coordinate system ($\phi_1,\phi_2$ and $\phi_3$); 2, the vertex norm in the Cartesian coordinate system ($\phi_7$); 3, the vertex position in the Laplace coordinate system ($\phi_4,\phi_5$ and $\phi_6$); 4, the vertex norm in Laplace coordinate system ($\phi_8$); 5, the face normal ($\phi_{10}$); 6, the dihedral angle ($\phi_9$); 7, the vertex normal ($\phi_{11}$); 8, the curvature ($\phi_{12}$ and $\phi_{13}$); 9, the vertex position in spherical coordinates system ($\phi_{14},\phi_{15}$ and $\phi_{16}$); 10, the edge length in the spherical coordinate system ($\phi_{17}, \phi_{18}$ and $\phi_{19}$). The relevance of all the features in LFS76 Then, the relevance for all features from LFS72, is calculated according to (24), and the averaged relevances of the features in each category are shown in Figure~\ref{rele-feature}. From Figure~\ref{rele-feature} we can observe that the new proposed features, represented by labels 7, 8, 9, 10, have relatively high relevance to their class label. More specifically, in Figure~\ref{rele-feature} (a), the features characterizing the edge length in the spherical coordinate system (label 10) achieve the highest relevance. Meanwhile, in Figures~\ref{rele-feature} (b) and (c), the vertex normal feature (label 7) obtains highest and second highest relevances. It is interesting that the relevance of the dihedral angle (label 6) shows high relevance to class label in the cases of \cite{cho2007oblivious} and \cite{chao2009high}, but shows very low relevance when the stego objects are generated by the watermarking method from \cite{yang2016watermarking}. This may happen because all the vertices from a mesh are slightly changed by the 3D embedding methods from \cite{cho2007oblivious} and \cite{chao2009high}, while such changes are scattered among the vertices in the case of the method from \cite{yang2016watermarking}, as it can be observed from Figures~\ref{311-FeatureDiff} (b) (g) and (l). In addition, the watermarking method from \cite{yang2016watermarking} embeds information by changing the histogram of the radial coordinates of the vertices, then the neighboring vertices tend to be shifted in the same direction, so the dihedral angles may be relatively preserved after the embedding. The watermarking method from \cite{cho2007oblivious} shifts the vertices in a similar way to \cite{yang2016watermarking}, which explains why the relevance of dihedral angle in the case of \cite{cho2007oblivious} is lower than that of \cite{chao2009high}.

\begin{figure*}[!htbp]
  \centering
  \subfigure[QDA for \cite{yang2016watermarking}]{
  \label{fig:subfig:b} 
  \includegraphics[width=1.7in]{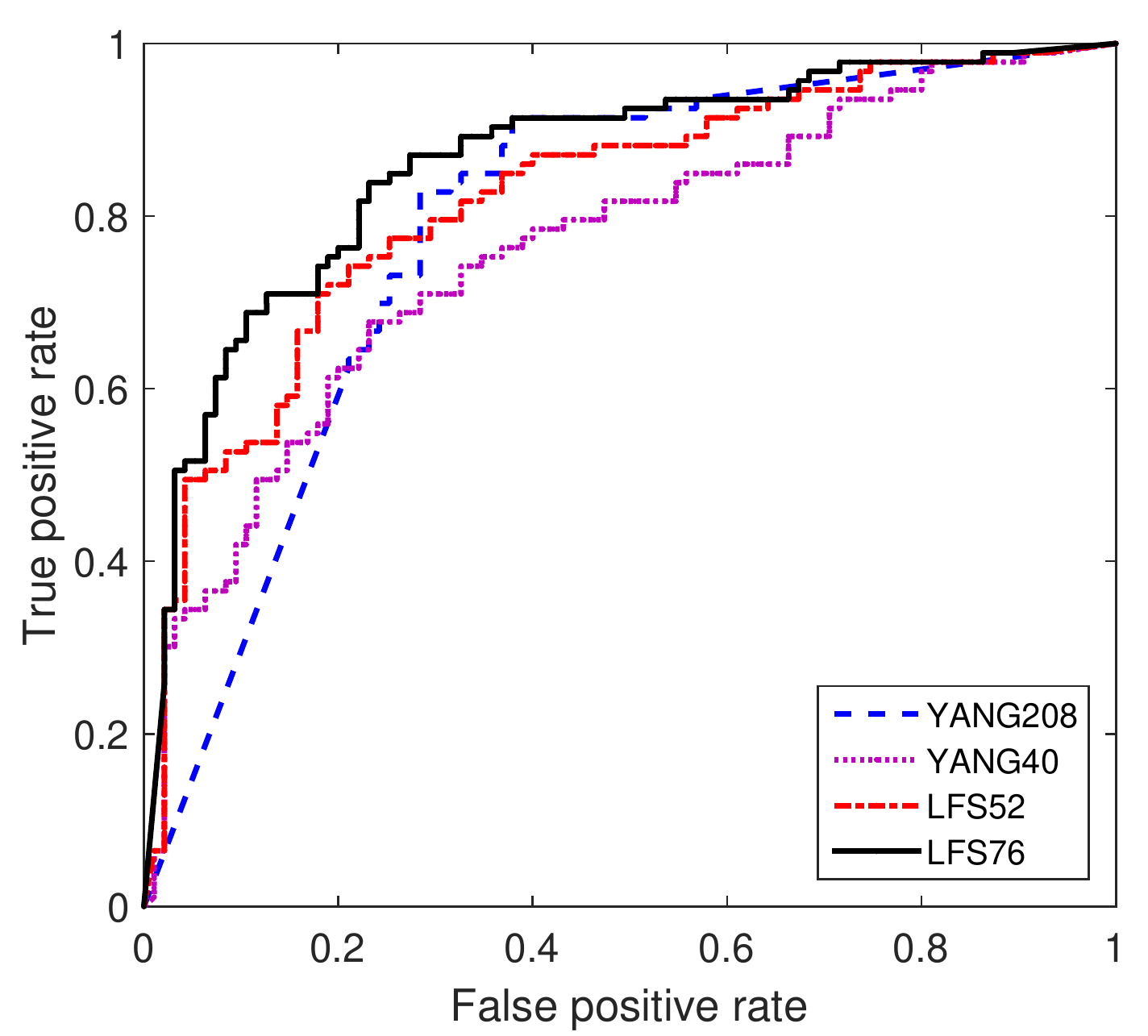}}
  \subfigure[SVM for \cite{yang2016watermarking}]{
  \label{fig:subfig:b} 
  \includegraphics[width=1.7in]{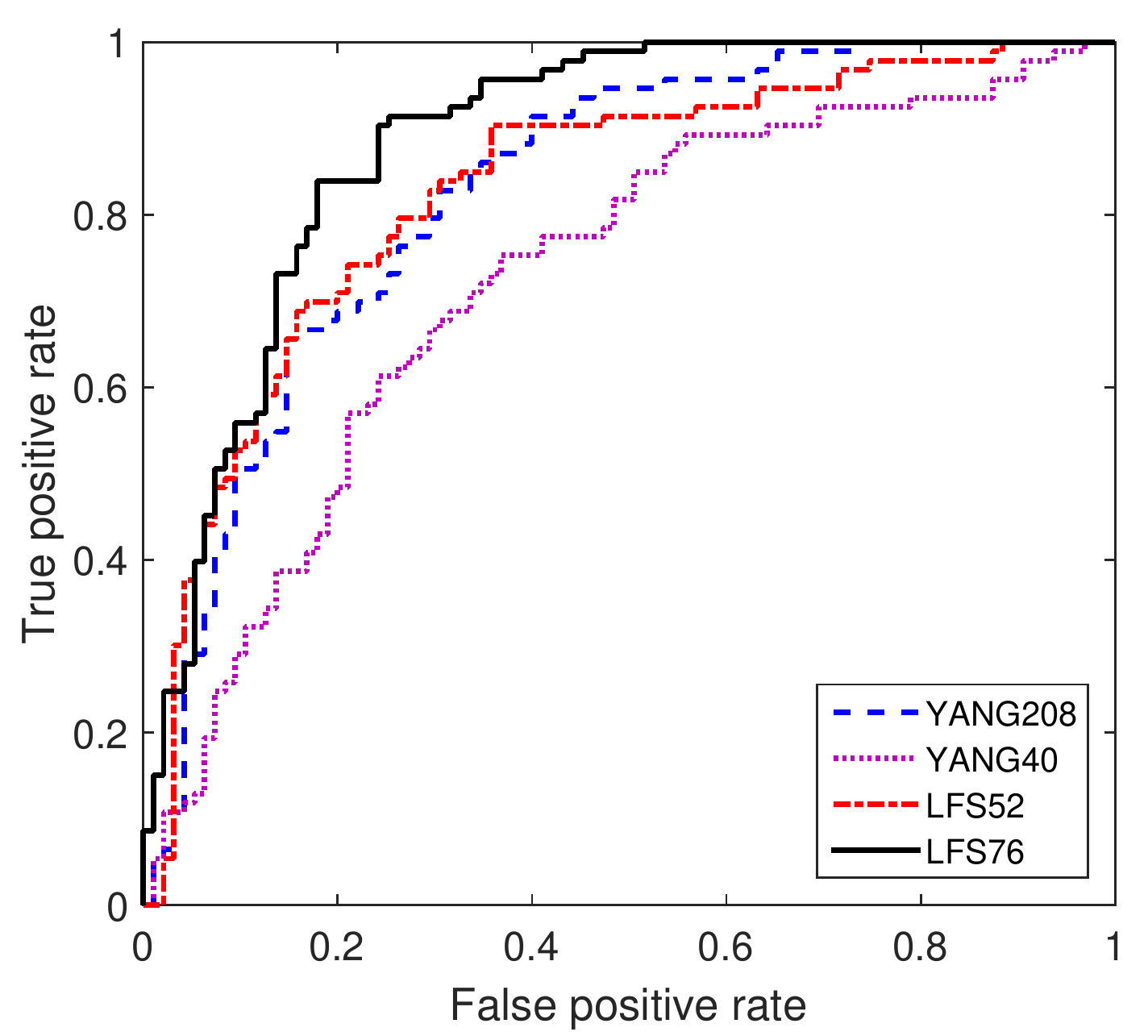}}
  \subfigure[FLD ensemble for \cite{yang2016watermarking}]{
  \label{fig:subfig:b} 
  \includegraphics[width=1.7in]{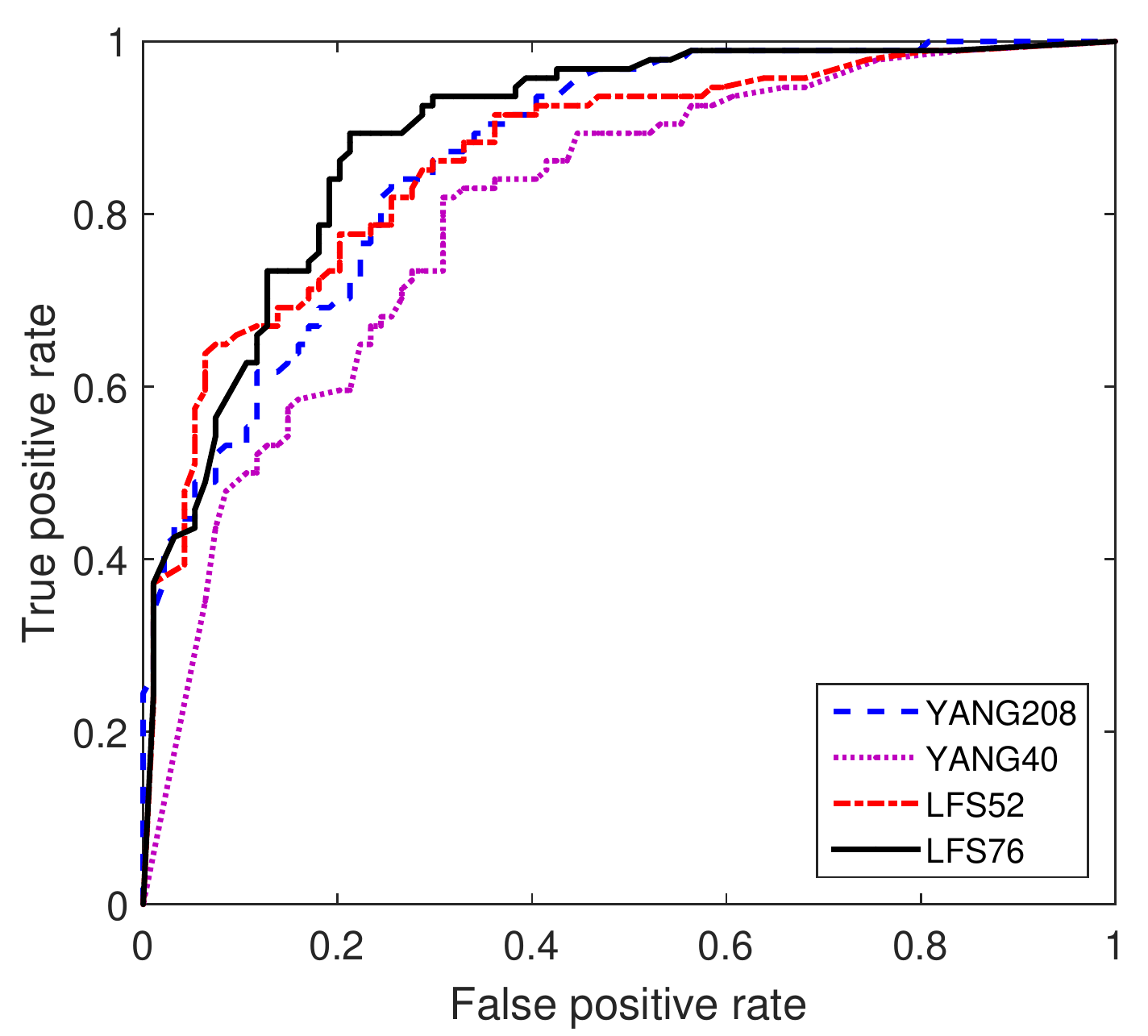}}\\
  \subfigure[QDA for \cite{cho2007oblivious}]{
  \label{fig:subfig:b} 
  \includegraphics[width=1.7in]{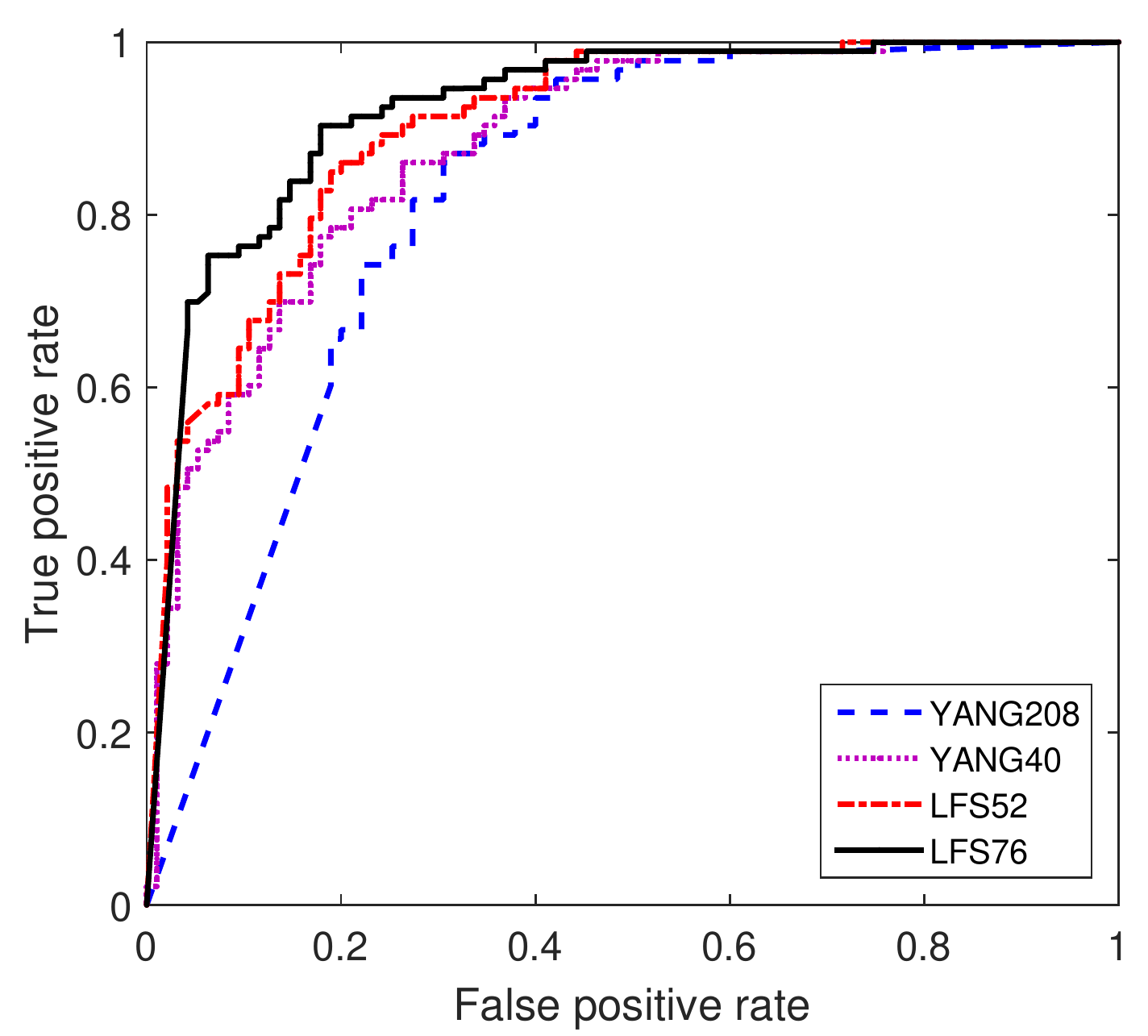}}
  \subfigure[SVM for \cite{cho2007oblivious}]{
  \label{fig:subfig:b} 
  \includegraphics[width=1.7in]{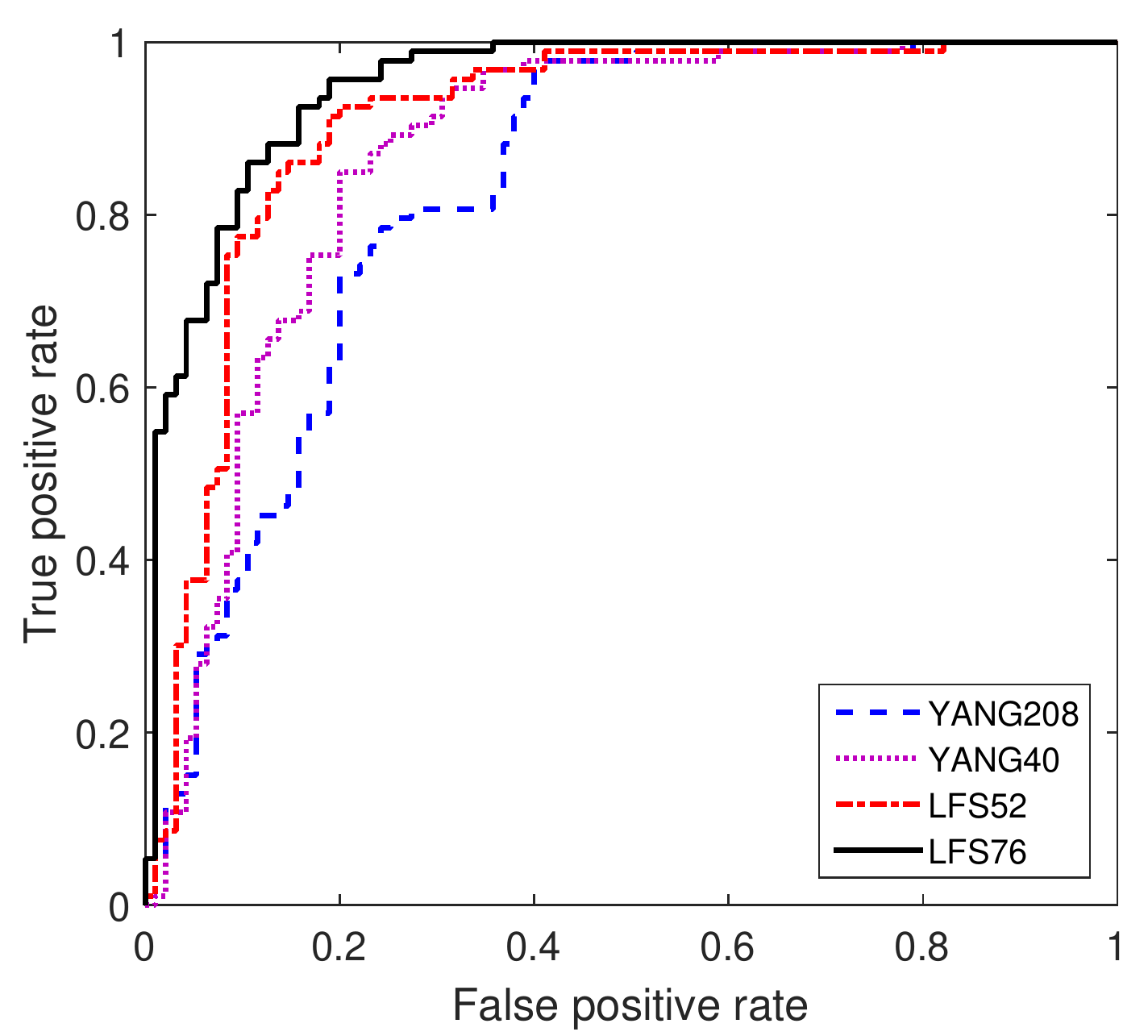}}
  \subfigure[FLD ensemble for \cite{cho2007oblivious}]{
  \label{fig:subfig:b} 
  \includegraphics[width=1.7in]{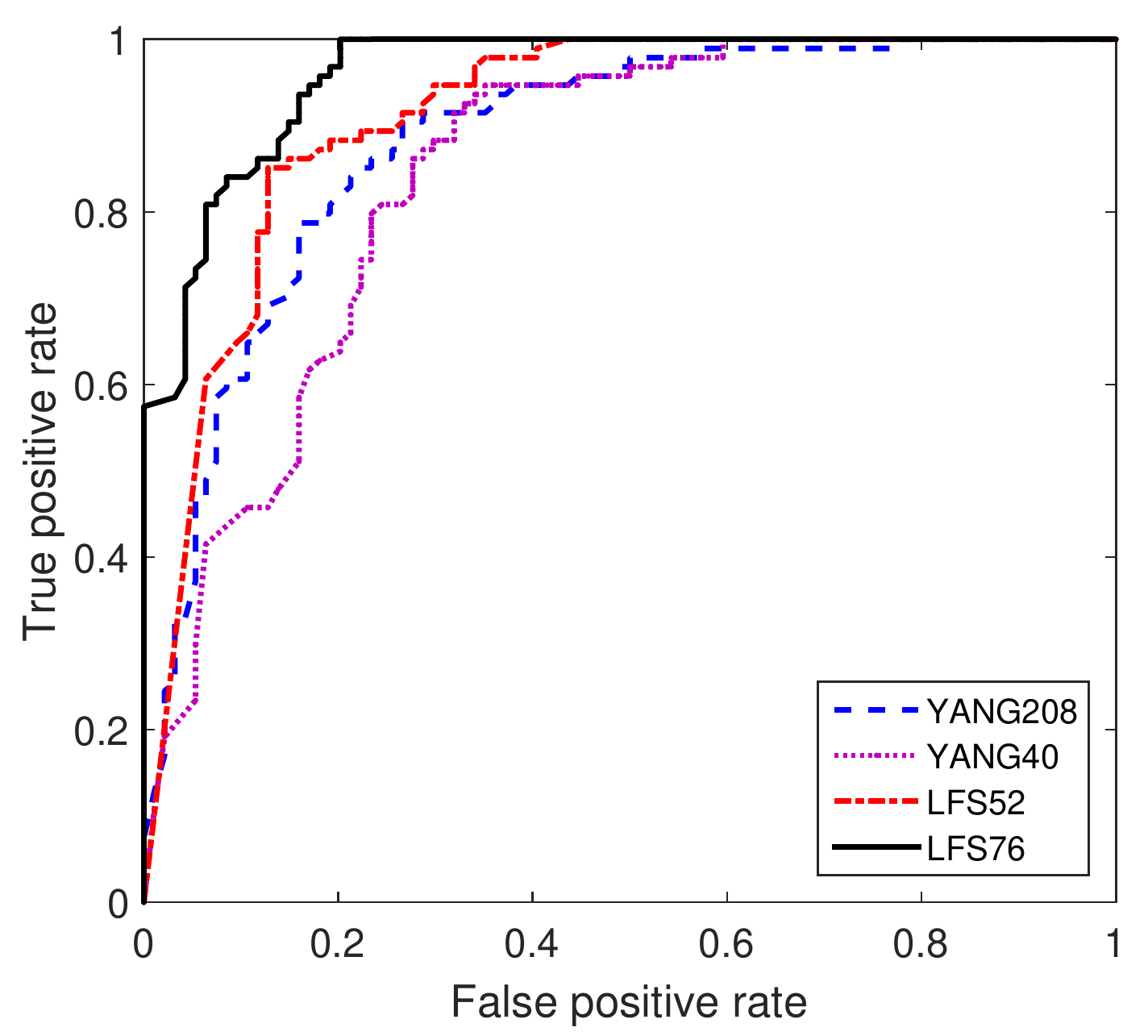}}\\
  \subfigure[QDA for \cite{chao2009high}]{
  \label{fig:subfig:a} 
  \includegraphics[width=1.7in]{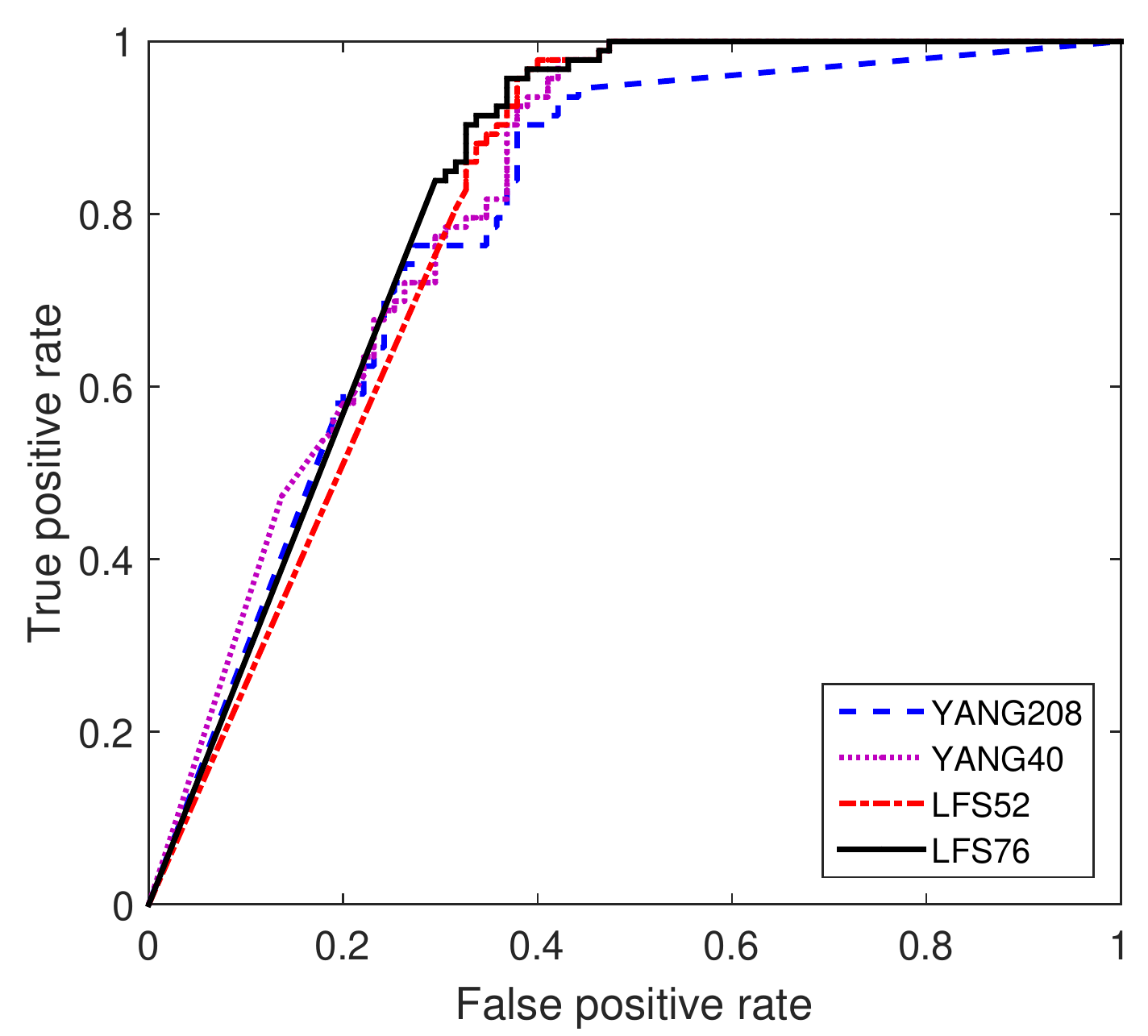}}
  \subfigure[SVM for \cite{chao2009high}]{
  \label{fig:subfig:b} 
  \includegraphics[width=1.7in]{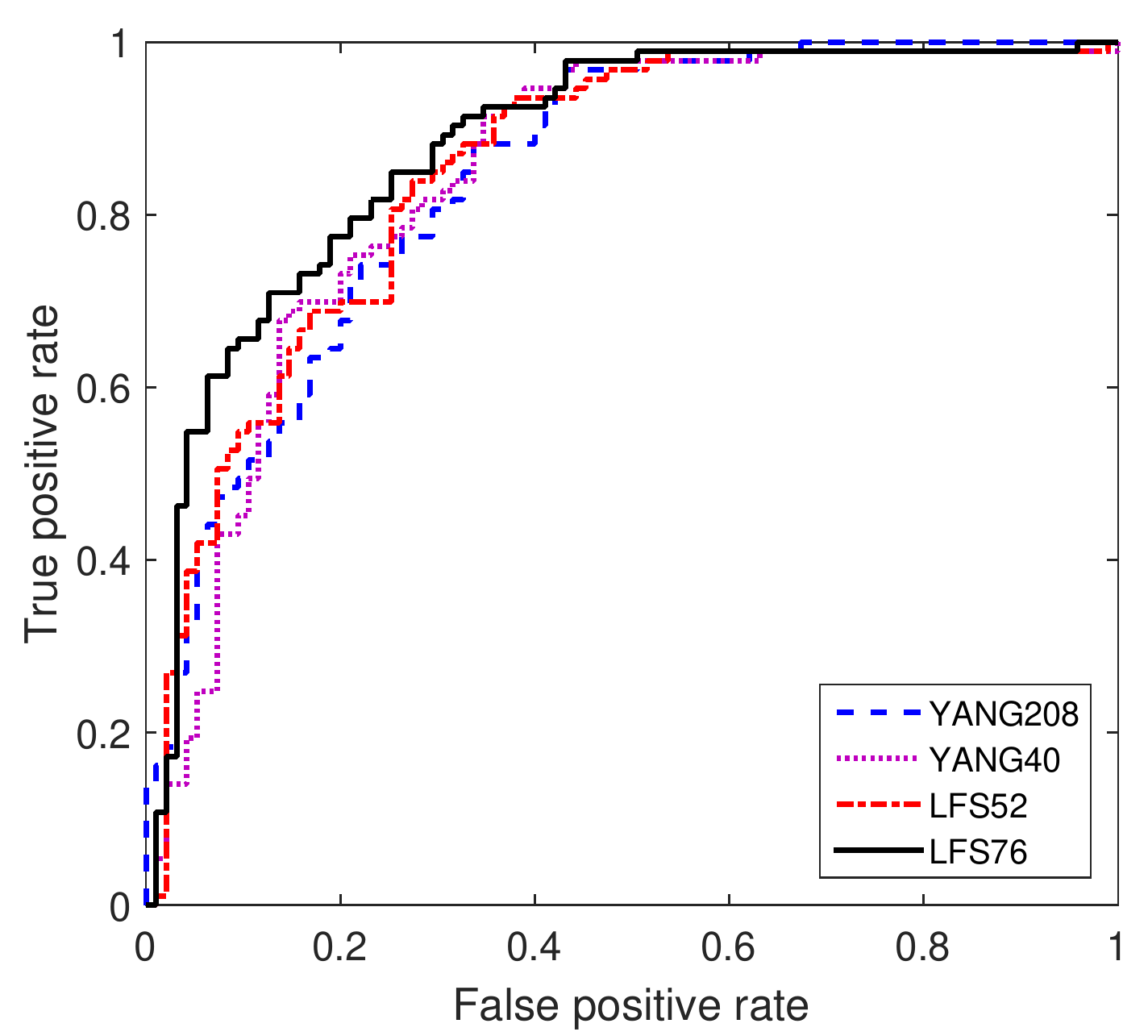}}
  \subfigure[FLD ensemble for \cite{chao2009high}]{
  \label{fig:subfig:b} 
  \includegraphics[width=1.7in]{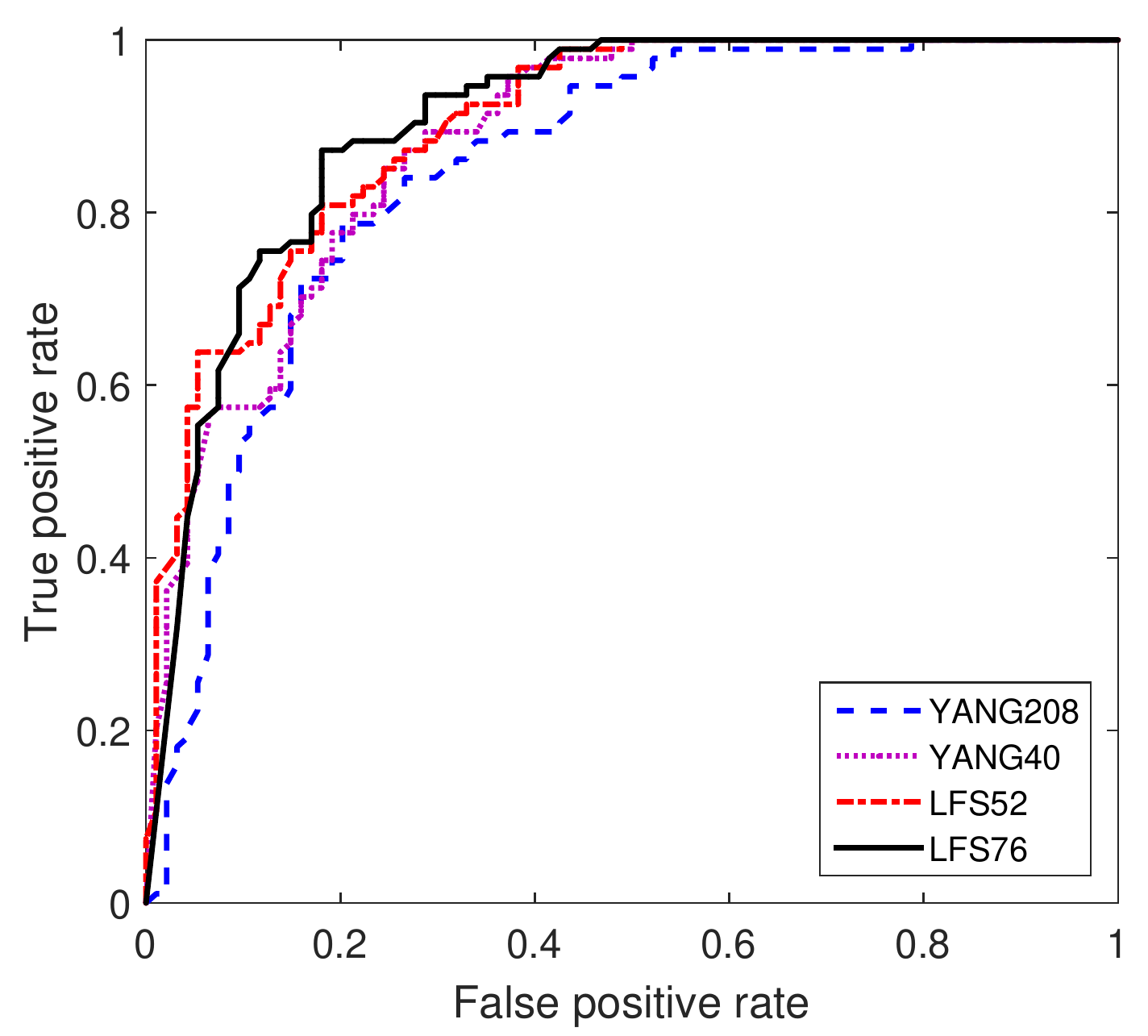}}\\
  \caption{ROC curves of steganalysis results for YANG208 \cite{yan2013steganalysis}, YANG40, LFS52 \cite{Li2016lfs} and the proposed feature set LFS76.}
  \label{ROC-curve}
\end{figure*}

In the following, we increase the feature set used for training steganalyzer gradually, from YANG40 to LFS52 and then to LFS76, and compare with YANG208. YANG40 includes the features represented by labels 1-6 in  Figure~\ref{rele-feature}. Features represented by labels 1-8 form LFS52 , while labels 1-10 correspond to LFS76.  In Figure~\ref{ROC-curve} we show the Receiver Operating Curves (ROC) results, considering the YANG208, YANG40, LFS52, and LFS76 feature sets
for training the three steganalyzers, when detecting the changes produced by the watermarking and steganographic algorithms
from  \cite{yang2016watermarking}, \cite{cho2007oblivious} and \cite{chao2009high}.   The parameters of the information hiding algorithms are the same as above when we calculate the features' relevance to the class label. The ROC curves in Figure~\ref{ROC-curve} show the true positives against the false positive rates, found for various threshold settings. A larger area under the ROC
curve means that the classifier has a better detection accuracy. It is shown in Figure~\ref{ROC-curve} that with the increase of the new features to YANG40, the feature sets LFS52 and LFS76 achieve better results and surpass the performance of YANG208. On the whole, the proposed feature set, LFS76, provides the best results in all nine cases with three different classifiers when assessing the changes
embedded by all three information hiding algorithms.

\begin{table}[!htbp]  \scriptsize
\centering
\caption{Median values and the standard deviations for the area under the ROC curves for the steganalysis results when considering the
information embedding by the algorithms proposed in \cite{yang2016watermarking}. }
\label{auc_yang}
\begin{tabular}{ c |c c c }
  \hline
  \multirow{2}{*}{Feature sets} & \multicolumn{3}{c}{Steganalyzers for method proposed in \cite{yang2016watermarking}} \\
   & QDA & SVM & FLD ensemble \\
  \hline
  YANG208 & 0.8002($\pm$0.0216) & 0.8222($\pm$0.0246) &	0.8690($\pm$0.0245) \\
  YANG40 &0.7547($\pm$0.0242)&0.7543($\pm$0.0287) &0.7740($\pm$0.0264) \\
  YANG40+VNF4 &0.7891($\pm$0.0240)&0.8346($\pm$0.0237)&0.8360($\pm$0.0240) \\
  YANG40+CF8 & 0.7709($\pm$0.0282)& 0.7920($\pm$0.0217)& 0.7977($\pm$0.0252)\\
  LFS52 &0.7976($\pm$0.0255)&0.8531($\pm$0.0234)&0.8559($\pm$0.0211) \\
  LFS76 & \textbf{0.8511($\pm$0.0276)}& \textbf{0.8726($\pm$0.0276)}& \textbf{0.8974($\pm$0.0227)}\\
  \hline
\end{tabular}
\end{table}

\begin{table}[!htbp]  \scriptsize
\centering
\caption{Median values and the standard deviations for the area under the ROC curves for the steganalysis results when considering the
information embedding by the algorithms proposed in \cite{cho2007oblivious}. }
\label{auc_cho}
\begin{tabular}{ c |c c c }
  \hline
  \multirow{2}{*}{Feature sets} & \multicolumn{3}{c}{Steganalyzers for method proposed in \cite{cho2007oblivious}} \\
   & QDA & SVM& FLD ensemble \\
  \hline
  YANG208 & 0.8002($\pm$0.0247) & 0.8140($\pm$0.0222)&	0.8765($\pm$0.0257) \\
  YANG40 &0.8476($\pm$0.0203)&0.8387($\pm$0.0196)&0.8083($\pm$0.0249) \\
  YANG40+VNF4 &0.8718($\pm$0.0278)&0.8513($\pm$0.0224)&0.8785($\pm$0.0226) \\
  YANG40+CF8 & 0.8459($\pm$0.0181)& 0.8612($\pm$0.0227)& 0.8570($\pm$0.0373)\\
  LFS52 &0.8613($\pm$0.0260)&0.8802($\pm$0.0222)&0.8917($\pm$0.0206) \\
  LFS76 & \textbf{0.8899($\pm$0.0215)}& \textbf{0.9366($\pm$0.0168)}& \textbf{0.9543($\pm$0.0115)}\\
  \hline
\end{tabular}
\end{table}

\begin{table}[!htbp] \scriptsize
\centering
\caption{Median values and the standard deviations for the area under the ROC curves for the steganalysis results when considering the
information embedding by the algorithms proposed in \cite{chao2009high}. }
\label{auc_chao}
\begin{tabular}{ c |c c c }
  \hline
  \multirow{2}{*}{Feature sets} & \multicolumn{3}{c}{Steganalyzers for method proposed in \cite{chao2009high}}\\
   & QDA classifier & SVM& FLD ensemble \\
  \hline
  YANG208 & 0.8035($\pm$0.0305)& 0.8426($\pm$0.0287) & 0.8573($\pm$0.0276) \\
  YANG40 & 0.8502($\pm$0.0273)&0.8707($\pm$0.0252)&0.8801($\pm$0.0211)\\
  YANG40+VNF4 & \textbf{0.8615($\pm$0.0172)}&0.8848($\pm$0.0229)&0.8928($\pm$0.0207) \\
  YANG40+CF8 &0.8570($\pm$0.0373)& 0.8903($\pm$0.0206)& 0.8926($\pm$0.0222)\\
  LFS52 &0.8520($\pm$0.0199)&0.8949($\pm$0.0240)&0.8925($\pm$0.0223) \\
  LFS76 &0.8462($\pm$0.0248)&\textbf{0.9023($\pm$0.0178)}&\textbf {0.8955($\pm$0.0253)} \\
  \hline
\end{tabular}
\end{table}

Tables~\ref{auc_yang}, \ref{auc_cho}, and \ref{auc_chao} provide the median values and the standard deviations of the area under the ROC curves for
the steganalysis methods when using six combinations of feature sets for 30 independent splits of the training/testing set.
It can be seen from Tables~\ref{auc_yang}, \ref{auc_cho} and \ref{auc_chao} that
the areas under the ROC curves of the steganalyzers increase with the addition of new features, such as VNF4 and CF4, to the YANG40 feature set. The benefit of adding VNF4 is slightly better than adding CF8 in general. After adding the features corresponding to the spherical coordinate system to LFS52, the usage of LFS76 feature set for 3D steganalysis results in larger areas under the ROC curves than any other combination of the features, in most of the cases. Meanwhile, the FLD ensemble acquire the best performance among the three kinds of classifiers in the detection of the two watermarking methods proposed in \cite{yang2016watermarking}
and \cite{cho2007oblivious}. Among different combinations of feature sets and machine learning methods used for 3D steganalysis, we conclude that the steganalyzer
using the feature set LFS76 and the SVM classifier produces the best results in the detection of the changes produced in 3D objects by the
steganographic algorithm from \cite{chao2009high}.

\section{Conclusion}
\label{Con}
The task of a 3D steganalyzer is very challenging because it has to find very small differences between stego-obejcts and
cover-objects. In this research study, we propose to use the statistics of some new shape features as inputs for 3D steganalyzers. We analyze various local features used for 3D steganalysis by evaluating their relevance to the class label and by testing their performance in the experiments.
The first four statistical moments in various 3D feature sets are used for training steganalyzers by three machine learning methods, namely, the quadratic discriminant, Fisher Linear Discriminant (FLD) ensemble, and the Support Vector Machine (SVM).
After training, these steganalyzers are used for differentiating the stego-objects from the cover-objects. The experimental results show that the proposed
3D feature sets, when used as inputs to the SVM and FLD ensemble, provide the best results for the steganalysis of 3D embeddings produced by three different
information hiding algorithms. In future studies we will assess the generalization ability of the proposed 3D steganalyzers.

\appendix
\section{}
When using the SVM with Gaussian kernel, two parameters need to be set prior to the training: the regularization parameter $C$ and
the radius of the Gaussian kernel $\gamma$. We apply a ``grid-search" on  $C$ and $\gamma$ using 5-fold cross-validation on the
training set. The training set is generated by a random split of the data set with 260 pairs of cover- and stego-meshes.
The search is firstly conducted in the grid defining the parameters:
\begin{equation}
\{ C,\gamma \} = (10^i,2^j) \in  {\cal S}  | i \in \{1,...,7\}, j \in \{-12,...,-1\}
\end{equation}
If the best parameters lay on the boundary of the grid, the search will continues on an expanded grid.

Table~\ref{yang16-svm-parameter} gives the optimal values for the parameters $\{ C,\gamma \}$ for the
detection of changes produced in 3D objects by the steganalysis-resistant watermarking method from \cite{yang2016watermarking} when
using various values of K.
Table~\ref{cho07-para-strength} and \ref{cho07-para-length} provide the best parameter combinations $\{ C,\gamma \}$,
when varying the embedding strength and increasing the payload, respectively, where the watermarks are
embedded by the watermarking method from \cite{cho2007oblivious}. While Table~\ref{chao09-para} provides the best choice of parameters,
when varying the number of layers for embedding the payload by the steganographic method from \cite{chao2009high}.

\begin{table}[!htbp] \scriptsize
\centering
\caption{Best parameter selection for SVM when is used as a steganalyzer for watermarks embedded by the watermarking method \cite{yang2016watermarking}
assuming various values of K.}
\label{yang16-svm-parameter}
\begin{tabular}{@{\extracolsep{3pt}}c@{\extracolsep{3pt}}|@{\extracolsep{3pt}}c@{\extracolsep{3pt}}c|@{\extracolsep{3pt}}c@{\extracolsep{3pt}}c|@{\extracolsep{3pt}}c@{\extracolsep{3pt}}c|@{\extracolsep{3pt}}c@{\extracolsep{3pt}}c}
  \hline
  K & \multicolumn{2}{c|}{32} & \multicolumn{2}{c|}{64} & \multicolumn{2}{c|}{96} & \multicolumn{2}{c}{128} \\
  \hline
   Feature sets & $C$ & $\gamma$ & $C$ & $\gamma$ & $C$ & $\gamma$ & $C$ & $\gamma$ \\
   \hline
  YANG208 & $10$ & $2^{-4}$ & $10^-13$ & $2^{-10}$ & $10$ & $2^{-9}$ & $10$ & $2^{-10}$  \\
  YANG40 & $10^4 $ & $ 2^{-8} $ & $ 10^2 $ & $ 2^{-4} $ & $ 10^6 $ & $ 2^{-11} $ & $ 10^4 $ & $ 2^{-9} $  \\
  YANG40+VNF4  &$ 10^7 $ & $ 2^{-9} $ & $10^4 $ & $ 2^{-9} $ & $ 10^5 $ & $ 2^{-10}$ & $ 10^5 $ & $ 2^{-8} $   \\
  YANG40+CF8 & $ 10^5 $ & $ 2^{-9} $ & $ 10^5 $ & $ 2^{-10} $ & $ 10^3 $ & $ 2^{-6} $ & $ 10^5 $ & $ 2^{-11} $  \\
  LFS52  & $ 10^5 $ & $ 2^{-12} $ & $ 10^6 $ & $ 2^{-9} $ & $ 10^7 $ & $ 2^{-12} $ & $ 10^7 $ & $ 2^{-12} $   \\
  LFS76  & $ 10^5 $ & $ 2^{-9} $ & $ 10^5 $ & $ 2^{-10} $ & $ 10^3 $ & $ 2^{-8} $ & $ 10^4 $ & $ 2^{-11} $   \\
  \hline
\end{tabular}
\end{table}

\begin{table}[!htbp] \scriptsize
\centering
\caption{Best parameter selection for SVM when is used as a steganalyzer for watermarks embedded by the watermarking method \cite{cho2007oblivious}
assuming various levels of watermark embedding strength.}
\label{cho07-para-strength}
\begin{tabular}{@{\extracolsep{3pt}}c@{\extracolsep{3pt}}|@{\extracolsep{3pt}}c@{\extracolsep{3pt}}c|@{\extracolsep{3pt}}c@{\extracolsep{3pt}}c|@{\extracolsep{3pt}}c@{\extracolsep{3pt}}c|@{\extracolsep{3pt}}c@{\extracolsep{3pt}}c|@{\extracolsep{3pt}}c@{\extracolsep{3pt}}c}
  \hline
   Strength & \multicolumn{2}{c|}{0.02} & \multicolumn{2}{c|}{0.04} & \multicolumn{2}{c|}{0.06} & \multicolumn{2}{c|}{0.08} & \multicolumn{2}{c}{0.1}\\
  \hline
   Feature sets & $C$ & $\gamma$ & $C$ & $\gamma$ & $C$ & $\gamma$ & $C$ & $\gamma$ &$C$ & $\gamma$ \\
   \hline
  YANG208 & $10^2$ & $2^{-13}$ & $10^2$ & $2^{-11}$ & $10^2$ & $2^{-9}$ & $10^3$ & $2^{-11}$ & $10$ &$ 2^{-9}$\\
  YANG40 & $10^5 $ & $ 2^{-11} $ & $ 10^6 $ & $ 2^{-9} $ & $ 10^2 $ & $ 2^{-2} $ & $ 10^2 $ & $ 2^{-6} $ & $ 10^3 $ & $ 2^{-1} $\\
  YANG40+VNF4  &$ 10^7 $ & $ 2^{-10} $ & $10^2 $ & $ 2^{-3} $ & $ 10^5 $ & $ 2^{-10}$ & $ 10^7 $ & $ 2^{-11} $ & $ 10^5 $ & $ 2^{-8}$ \\
  YANG40+CF8 & $ 10^7 $ & $ 2^{-11} $ & $ 10^4 $ & $ 2^{-7} $ & $ 10^3 $ & $ 2^{-5} $ & $ 10^3 $ & $ 2^{-6} $ & $ 10^6 $ & $ 2^{-11} $\\
  LFS52  & $ 10^4 $ & $ 2^{-7} $ & $ 10^6 $ & $ 2^{-10} $ & $ 10^6 $ & $ 2^{-13} $ & $ 10^5 $ & $ 2^{-10} $ & $ 10^4 $ & $ 2^{-8}$ \\
  LFS76  & $ 10^3 $ & $ 2^{-8} $ & $ 10^3 $ & $ 2^{-9} $ & $ 10^5 $ & $ 2^{-10} $ & $ 10^5 $ & $ 2^{-11} $ & $ 10^4 $ & $ 2^{-10}$ \\
  \hline
\end{tabular}

\end{table}

\begin{table}[!htbp] \scriptsize
\centering
\caption{Best parameter selection for SVM when is used as a steganalyzer for watermarks embedded by the watermarking method \cite{cho2007oblivious}
assuming various levels of message payload.}
\label{cho07-para-length}
\begin{tabular}{@{\extracolsep{3pt}}c@{\extracolsep{3pt}}|@{\extracolsep{3pt}}c@{\extracolsep{3pt}}c|@{\extracolsep{3pt}}c@{\extracolsep{3pt}}c|@{\extracolsep{3pt}}c@{\extracolsep{3pt}}c|@{\extracolsep{3pt}}c@{\extracolsep{3pt}}c}
  \hline
  Payload & \multicolumn{2}{c|}{16 bits} & \multicolumn{2}{c|}{32 bits} & \multicolumn{2}{c|}{48 bits} & \multicolumn{2}{c}{64 bits} \\
  \hline
   Feature sets & $C$ & $\gamma$ & $C$ & $\gamma$ & $C$ & $\gamma$ & $C$ & $\gamma$ \\
   \hline
  YANG208 & $10^2$ & $2^{-10}$ & $10$ & $2^{-8}$ & $10$ & $2^{-8}$ & $10^2$ & $2^{-11}$  \\
  YANG40 & $10^4 $ & $ 2^{-7} $ & $ 10^3 $ & $ 2^{-2} $ & $ 10^3 $ & $ 2^{-4} $ & $ 10^6 $ & $ 2^{-9} $  \\
  YANG40+VNF4  &$ 10^3 $ & $ 2^{-6} $ & $10^5 $ & $ 2^{-10} $ & $ 10^4 $ & $ 2^{-8}$ & $ 10^2 $ & $ 2^{-3} $   \\
  YANG40+CF8 & $ 10^5 $ & $ 2^{-11} $ & $ 10^5 $ & $ 2^{-9} $ & $ 10^6 $ & $ 2^{-12} $ & $ 10^4 $ & $ 2^{-7} $  \\
  LFS52  & $ 10^5 $ & $ 2^{-10} $ & $ 10^3 $ & $ 2^{-5} $ & $ 10^5 $ & $ 2^{-11} $ & $ 10^6 $ & $ 2^{-10} $   \\
  LFS76  & $ 10^5 $ & $ 2^{-12} $ & $ 10^7 $ & $ 2^{-10} $ & $ 10^4 $ & $ 2^{-5} $ & $ 10^3 $ & $ 2^{-9} $   \\
  \hline
\end{tabular}

\end{table}

\begin{table*}[!htbp] \scriptsize
\centering

\caption{Analysis of the parameter selection for the SVM steganalyzer when considering the steganographic method from \cite{chao2009high} for
embedding watermarks when assuming various numbers of embedding layers.}
\label{chao09-para}
\begin{tabular}{@{\extracolsep{3pt}}c@{\extracolsep{3pt}}|@{\extracolsep{3pt}}c@{\extracolsep{3pt}}c|@{\extracolsep{3pt}}c@{\extracolsep{3pt}}c|@{\extracolsep{3pt}}c@{\extracolsep{3pt}}c|@{\extracolsep{3pt}}c@{\extracolsep{3pt}}c|@{\extracolsep{3pt}}c@{\extracolsep{3pt}}c|@{\extracolsep{3pt}}c@{\extracolsep{3pt}}c|@{\extracolsep{3pt}}c@{\extracolsep{3pt}}c|@{\extracolsep{3pt}}c@{\extracolsep{3pt}}c|@{\extracolsep{3pt}}c@{\extracolsep{3pt}}c|@{\extracolsep{3pt}}c@{\extracolsep{3pt}}c}
  \hline
   Layer & \multicolumn{2}{c|}{1} & \multicolumn{2}{c|}{2} & \multicolumn{2}{c|}{3} & \multicolumn{2}{c|}{4} & \multicolumn{2}{c|}{5}& \multicolumn{2}{c|}{6} & \multicolumn{2}{c|}{7} & \multicolumn{2}{c|}{8} & \multicolumn{2}{c|}{9} & \multicolumn{2}{c}{10}\\
  \hline
   Feature sets & $C$ & $\gamma$ & $C$ & $\gamma$ & $C$ & $\gamma$ & $C$ & $\gamma$ &$C$ & $\gamma$ & $C$ & $\gamma$ & $C$ & $\gamma$ & $C$ & $\gamma$ & $C$ & $\gamma$ &$C$ & $\gamma$\\
   \hline
  YANG208 & $10^2$ & $2^{-11}$ & $10^2$ & $2^{-10}$ & $10$ & $2^{-6}$ & $10^2$ & $2^{-12}$ & $10^2$ &$ 2^{-10}$& $10^2$ & $2^{-10}$ & $10^2$ & $2^{-11}$ & $10^2$ & $2^{-10}$ & $10^2$ & $2^{-11}$ & $10^2$ &$ 2^{-11}$\\
  YANG40 & $10^5 $ & $ 2^{-10} $ & $ 10^3 $ & $ 2^{-9} $ & $ 10^5 $ & $ 2^{-10} $ & $ 10^6 $ & $ 2^{-10} $ & $ 10^7 $ & $ 2^{-11} $& $10^10$ & $2^{-11}$ & $10^6$ & $2^{-11}$ & $10^6$ & $2^{-10}$ & $10^7$ & $2^{-13}$ & $10^7$ &$ 2^{-10}$\\
  YANG40+VNF4  &$ 10^6 $ & $ 2^{-11} $ & $10^6 $ & $ 2^{-11} $ & $ 10^3 $ & $ 2^{-10}$ & $ 10^6 $ & $ 2^{-11} $ & $ 10^4 $ & $ 2^{-12}$ & $10^5$ & $2^{-9}$ & $10^7$ & $2^{-12}$ & $10^6$ & $2^{-11}$ & $10^5$ & $2^{-10}$ & $10^5$ &$ 2^{-10}$\\
  YANG40+CF8 & $ 10^4 $ & $ 2^{-12} $ & $ 10^4 $ & $ 2^{-12} $ & $ 10^5 $ & $ 2^{-10} $ & $ 10^4 $ & $ 2^{-8} $ & $ 10^5 $ & $ 2^{-9} $& $10^7$ & $2^{-12}$ & $10^3$ & $2^{-6}$ & $10^7$ & $2^{-11}$ & $10^5$ & $2^{-7}$ & $10^7$ &$ 2^{-9}$\\
  LFS52  & $ 10^6 $ & $ 2^{-11} $ & $ 10^5 $ & $ 2^{-10} $ & $ 10^7 $ & $ 2^{-12} $ & $ 10^7 $ & $ 2^{-12} $ & $ 10^7 $ & $ 2^{-9}$ & $10^5$ & $2^{-7}$ & $10^8$ & $2^{-8}$ & $10^6$ & $2^{-11}$ & $10^7$ & $2^{-11}$ & $10^6$ &$ 2^{-7}$\\
  LFS76  & $ 10^6 $ & $ 2^{-12} $ & $ 10^6 $ & $ 2^{-11} $ & $ 10^6 $ & $ 2^{-11} $ & $ 10^7 $ & $ 2^{-12} $ & $ 10^6 $ & $ 2^{-10}$ & $10^6$ & $2^{-11}$ & $10^5$ & $2^{-10}$ & $10^6$ & $2^{-10}$ & $10^7$ & $2^{-12}$ & $10^7$ &$ 2^{-12}$\\
  \hline
\end{tabular}

\end{table*}

\bibliographystyle{IEEEtran}
\bibliography{3Dsteganalysis}

\end{document}